\documentclass[11pt,a4paper]{article}
\pdfoutput=1
\usepackage[a4paper]{geometry}

\usepackage{amsmath,amssymb,amsfonts}
\usepackage{latexsym}
\usepackage{graphicx}
\usepackage{epsfig}

\usepackage{verbatim}
\usepackage{subfigure}

\usepackage{slashed}
\usepackage{dsfont}
\usepackage[small]{caption}

\relax

\def\be{\begin{equation}}
\def\ee{\end{equation}}
\def\bea{\begin{eqnarray}}
\def\eea{\end{eqnarray}}

\newcommand\fverb{\setbox\pippobox=\hbox\bgroup\verb}
\newcommand\fverbdo{\egroup\medskip\noindent%
                        \fbox{\unhbox\pippobox}\ }
\newcommand\fverbit{\egroup\item[\fbox{\unhbox\pippobox}]}

\newcommand{\ba}{\begin{aligned}}
\newcommand{\ea}{\end{aligned}}
\newcommand{\bear}{\begin{eqnarray}}

\newcommand{\eear}{\end{eqnarray}}

\newcommand{\dau}{\partial}
\newcommand{\xtp}{x_{\text{tp}}}
\newcommand{\dagh}{^{\dagger}}
\newcommand{\Tr}{\text{Tr}}
\newcommand{\I}{\,\textrm{Im}\,}
\newcommand{\R}{\,\textrm{Re}\,}
\newcommand{\DD}{\slashed{D}}
\newcommand{\om}{\omega}
\newcommand{\kv}{\vec{k}}
\newcommand{\eff}{_{\text{eff}}}
\newcommand{\bes}{\begin{equation*}}
\newcommand{\ees}{\end{equation*}}

\newbox\pippobox

\def\lab{\label}
\def\6{\partial}

\def\a{\alpha}

\def\half{\frac12}

\def\le{\left}
\def\ri{\right}
\def\cO{{\cal O}}

\def\C0{{\bf C_0}}
\def\Y0{{\bf Y_0}}
\def\G0{{\bf G_0}}

\def\m{\mu}

\def\t{\theta}

\def\sq
\def\a{\alpha}

\def\l{\lambda}
\def\g{\gamma}

\def\o{\omega}

\def\bz{\begin{itemize}}
\def\ez{\end{itemize}}
\def\bn{\begin{enumerate}}
\def\en{\end{enumerate}}

\def\ben{\begin{enumerate}}
\def\een{\end{enumerate}}

\def\bk{\bar{k}}

\def\bV{V}
\def\sik{\vec{\sigma}\cdot\vec{k}}

\allowdisplaybreaks[2]
\numberwithin{equation}{section}

\newcommand{\eq}[1]{\begin{equation}
                     \begin{split} #1 \end{split}
                     \end{equation}}

\newcommand{\ov}[1]{\overline{#1}}
\newcommand{\ul}[1]{\underline{#1}}

\def\a{\alpha}
\def\o{\omega}
\def\6{\partial}
\def\lab{\label}
\def\half{\frac12}
\def\le{\left}
\def\ri{\right}
\def\tk{\tilde{k}}
\def\tw{\tilde{\o}}
\def\be{\begin{equation}}
\def\ee{\end{equation}}
\def\bea{\begin{eqnarray}}
\def\eea{\end{eqnarray}}
\def\bz{\begin{itemize}}
\def\ez{\end{itemize}}

\def\cO{{\cal O}}

\begin{document}

\DeclareGraphicsExtensions{.jpg,.JPG,.jpeg,.pdf,.png,.eps}
\graphicspath{{./graphics/}}

\begin{flushright} \small
 ITP--UU--12/27 \\ SPIN--12/25 \\ CERN-PH-TH/2012-240
\end{flushright}

\vskip1cm

\begin{center}
{\large\bfseries Holographic models for undoped Weyl semimetals}\\[8mm]
Umut G\"ursoy$^{\dag}$, Vivian Jacobs$^*$, Erik Plauschinn$^{*\hspace{1pt}\ddag}$, Henk Stoof$^*$, Stefan Vandoren$^*$ \\[5mm]
{\small\slshape \textsuperscript{\dag} Theory Group, Physics Department, CERN, CH-1211 Geneva 23, Switzerland\\[3mm]
\textsuperscript{*} Institute for Theoretical Physics \emph{and} Spinoza Institute,
 Utrecht University \\ 3508 TD Utrecht, The Netherlands \\[5mm]
{\upshape\ttfamily Umut.Gursoy@cern.ch; V.P.J.Jacobs, E.Plauschinn, H.T.C.Stoof, S.J.G.Vandoren@uu.nl}\\[3mm]}
\end{center}

\vspace{1.5cm} \hrule\bigskip \centerline{\bfseries Abstract}
\medskip

We continue our recently proposed holographic description of single-particle correlation functions for four-dimensional chiral fermions with Lifshitz scaling at zero chemical potential, paying particular attention to the dynamical exponent $z=2$. We present new results for the spectral densities and dispersion relations at non-zero momenta and temperature. In contrast to the relativistic case with $z=1$, we find the existence of a quantum phase transition from a non-Fermi liquid into a Fermi liquid in which two Fermi surfaces spontaneously form, even at zero chemical potential. Our findings show that the boundary system behaves like an undoped Weyl semimetal.
\renewcommand*{\thefootnote}{\fnsymbol{footnote}}
\footnotetext[3]{Affiliation per 1 September 2012:
Dipartimento di Fisica e Astronomia, Universit\`a degli Studi di Padova, Italy.
}
\renewcommand*{\thefootnote}{\arabic{footnote}}
\bigskip
\hrule\bigskip


\newpage

\tableofcontents


\section{Introduction}\lab{sec:intro}

Over the past years, the AdS/CFT correspondence has become a more and more popular and widespread tool which offers the opportunity to apply ideas from string theory to realistic materials studied in condensed-matter physics, see e.g. \cite{Hartnoll:2009sz,Herzog:2009xv,McGreevy:2009xe,Sachdev:2010ch,Hartnoll:2011fn} and references therein. More specifically, it is potentially very useful when considering strongly coupled or critical condensed matter, which is generically not possible to describe within perturbation theory. However, in the context of the AdS/CFT correspondence these systems can be understood and investigated more conveniently using a dual theory in a higher-dimensional curved spacetime, i.e., in the framework of general relativity. Most commonly, the duality exists between a weakly coupled, classical gravity theory in a curved anti-de-Sitter (AdS) spacetime on the one hand, and a strongly coupled, conformal field theory (CFT) living on the flat boundary of the AdS spacetime on the other hand. Such a conformal field theory describes for example a quantum critical point of the condensed-matter system under consideration, and the observables that are most readily available from the AdS/CFT correspondence are usually correlation functions of the {\em composite} operators that classify the conformal field theory. For our purposes, most relevant  are correlators of fermionic operators, since they may show Fermi or non-Fermi liquid-like behavior most easily \cite{Lee:2008xf,Faulkner:2009wj,Liu:2009dm,Zaanen,Cubrovic:2010bf}.

However, in condensed-matter physics it is more natural to think in terms of fermionic {\em single-particle} operators, i.e., creation and annihilation operators in a Fock space, that satisfy the following (equal-time) anti-commutation relations
\eq{
\lab{fock}
\bigl[ \psi_{\alpha}(\vec{x},t), \psi_{\alpha'}\dagh(\vec{x}',t) \bigr]_+ = \delta(\vec{x}-\vec{x}')\delta_{\alpha,\alpha'}\;,
}
where $\alpha$ labels the spin of the electron. Furthermore, in experiments on condensed-matter systems, the quantity that is measured is essentially always related to \emph{single-particle} or \emph{two-particle} correlation functions and not to the above-mentioned composite operators. For example, the retarded single-particle correlation or Green's function $G_R$ is measurable in electronic systems using angle-resolved photoemission spectroscopy (ARPES). For this reason, we are interested in finding  $G_R$ of a strongly interacting condensed-matter system using holographic methods. Inspired by previous work on this topic, such as \cite{Pomarol} and \cite{Faulkner:2010tq,Hartnoll:2009ns}, we have shown recently how the usual holographic prescription can be modified in such a manner that it allows for the construction of the retarded \emph{single-particle} Green's function \cite{Gursoy:2011gz}. In a sense, our method can be seen as a bulk derivation of the semi-holographic description advocated in \cite{Faulkner:2010tq}.

A crucial consequence of the anti-commutation relations in equation \eqref{fock} is the existence of a sum rule for the corresponding single-particle two-point correlation function, given by
\eq{
\frac{1}{\pi} \int_{-\infty}^{+\infty} \textrm{d}\o \,
                \I\left[G_{R;\alpha,\alpha'}(\vec{k},\o)\right] = \delta_{\alpha,\alpha'} \;.
}
This sum rule is essential in determining whether the quantity under consideration is the correlator of a single-particle or a composite operator, and therefore  plays a central role in our construction.

The quantum critical points described by the AdS/CFT correspondence are  characterized by their dynamical scaling exponent $z$. In the usual correspondence, an anti-de-Sitter background leads to a quantum critical point with relativistic scaling, i.e., $z=1$. But from a condensed-matter perspective we are also interested in quantum critical points that have a dynamical scaling exponent different from one. To achieve this, the usual anti-de-Sitter background is generalized into a so-called Lifshitz background, which leads to a non-relativistic, i.e., $z\neq1$ scaling on the boundary \cite{Kachru}. Non-zero temperature effects can then be studied by placing a black brane in the Lifshitz spacetime \cite{Taylor}, see also \cite{Keranen:2012mx} for a more recent discussion. The pure Lifshitz geometry without a black brane develops singularities due to diverging tidal forces \cite{Horowitz:2011gh}; as a result the zero-temperature limit might become ill-defined\footnote{See however \cite{Harrison:2012vy,Bao:2012yt} for possible resolutions within string theory.}. While this could be of possible concern for our analysis, our findings show that the fermionic Green's function is well defined at zero temperature and the limit $T\rightarrow 0$ is smooth, so at least the spectral-weight function for fermions does not suffer from any singularities.

In \cite{Gursoy:2011gz} we have described how to construct the retarded Green's function of a strongly interacting, but particle-hole symmetric system of chiral fermions with an arbitrary dynamical exponent $z$ using a Lifshitz black-brane background. The aim of the present paper is to analyze the physics that follows from our prescription. In particular, we calculate the retarded single-particle Green's function in various cases, at zero and non-zero temperatures, and present the corresponding spectral-weight function and dispersion relations. Although we give some results for $z=1$, we mostly focus on the case $z=2$ and four boundary spacetime dimensions. With this number of dimensions, the boundary fermions are Weyl fermions and the boundary system behaves like an interacting Weyl semimetal. A semimetal is a gapless semiconductor. In addition, a Weyl semimetal \cite{Weyl1,Weyl2} is a semimetal with touching valence and conduction bands based on chiral two-component fermions that in the non-interacting limit and at low energies, satisfy the Weyl equation $\pm \vec\sigma\cdot \vec k\,\psi=E\,\psi$. Here, the $\pm$ denotes the chirality of the fermion. Because the total system has to be chirality-invariant, the most simple realizations of such a Weyl semimetal considered in the literature usually contain two of such linear-dispersion cones with opposite chirality, separated in momentum-space \cite{Weyl1}. The single-particle propagator presented here represents the physics of one of these chiral cones. Since the holographic boundary theory is that of Weyl fermions, and one cannot write down mass terms for Weyl fermions, the system will automatically be gapless at zero chemical potential. Due to the Weyl character of the holographic boundary fermions, the topology of the band structure is therefore protected, just like in Weyl semimetals.

We would like to stress that holography is employed in a bottom-up manner in this paper. That is, the holographic model is an effective low-energy theory, satisfying the above definition of a Weyl semimetal, whereas the microscopic structure of the actual condensed-matter system remains hidden. Of course, all holographic AdS/CMT models have this latter feature in common. We just assume that the Weyl semimetal under consideration leads to a low-energy theory that is strongly coupled, chiral and scale-invariant. The advantage of a bottom-up approach is that we can explore a range of possible situations.

In this paper, we are studying the properties of the ground state of such a holographic model for a semimetal at zero doping, so there is a zero total charge density. But as will be discussed, this ground state has in fact very interesting properties. In particular, zero chemical potential corresponds to particle-hole symmetric systems, and there are many-body correlations due to the possibility of creating particle-hole pairs.  This indicates that we are not describing a single fermionic excitation, but a fermionic many-body problem. By computing the momentum distribution of the particles and holes, we find for $z\neq1$ that our holographic model for a Weyl semimetal contains a quantum phase transition between a non-Fermi-liquid phase and a Fermi-liquid phase with two Fermi surfaces, one for the particles and one for the holes, even at zero doping. In this phase, there is a non-zero density of both particles and holes, and conduction can take place.

The scale set by the Fermi surface is determined by an appropriate combination of dimensionless parameters $\lambda$ and $g$ that enter in our prescription for the Green's function on the holographic boundary \cite{Gursoy:2011gz}, and the quantum critical point is at $\lambda=0$. One can think of $1/\l$ as a spin-orbit coupling constant and of $g$ as an effective interaction parameter coupling the elementary fermion to the conformal field theory. These parameters are not present in the standard formulation of holography, but enter our holographic prescription for the fermionic single-particle Green's function. We discuss the nature of these parameters in more detail in the next section.

\begin{figure}[t]
\centering
\subfigure[$\l >0$]{
\includegraphics[width=0.25\textwidth]{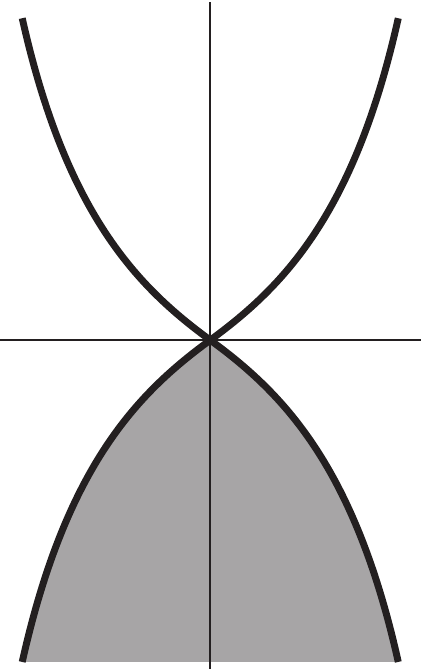}
\begin{picture}(0,0)
\put(-59,167){\small $\om$}
\put(-3,79){\small $k$}
\end{picture}
}
\hspace{2cm}
\subfigure[$\l<0$]{
\includegraphics[width=0.25\textwidth]{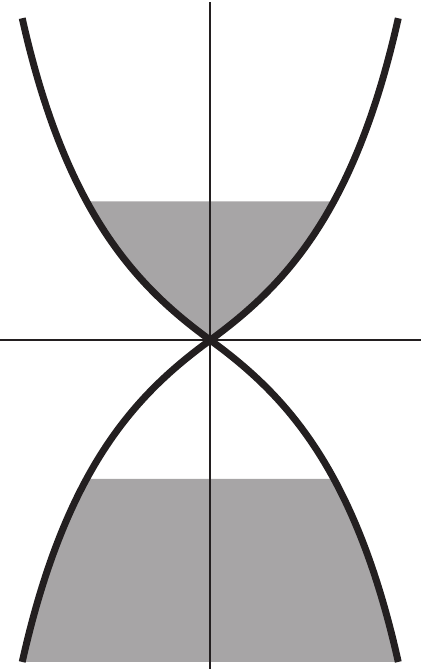}
\begin{picture}(0,0)
\put(-59,167){\small $\om$}
\put(-3,79){\small $k$}
\end{picture}
}
\caption{An idealized picture of the ground state that we find for $\l>0$ and $\l<0$. Grey areas refer to filled energy levels. In this picture, we only show the effect of the interactions on the occupation numbers and not on the dispersion relations. }
\label{dispcone}
\end{figure}
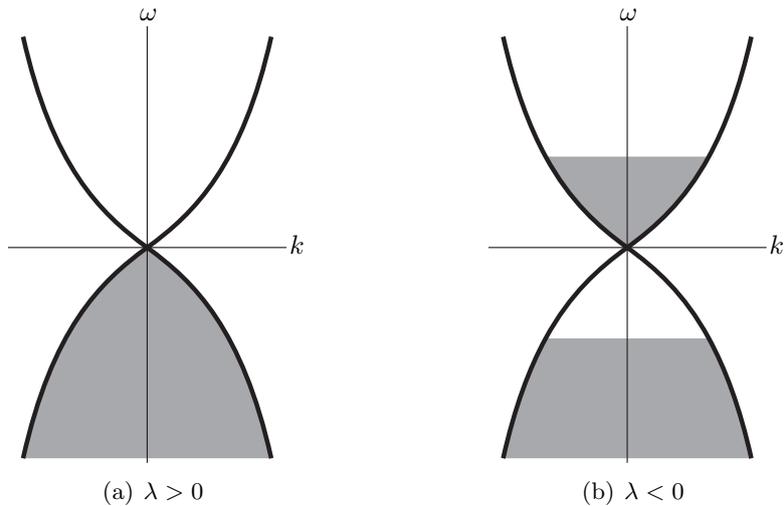

The phase transition that takes place is schematically illustrated in figure \ref{dispcone}, where we show the non-interacting band structure and how it is populated in the ground state of each phase. In particular, for $\l>0$ the conduction band is empty and the valence band is completely filled, whereas for $\l<0$ the conduction band contains a Fermi sea of particles and the valence band a Fermi sea of holes. Note that we present here an idealized picture that neglects the renormalization of the Fermi surfaces involved, which is discussed at length in this paper.

\medskip
The layout of the paper is as follows. In section \ref{sect:pres}, we briefly summarize some basic formulas from \cite{Gursoy:2011gz} that are needed here. In section \ref{sect:results}, we present analytic and numerical results for the single-particle spectral function for the cases $z=1$ and $z=2$. In section \ref{sect:QPT} we consider the single-particle momentum distribution, and describe  properties of the quantum phase transition and the Fermi liquid involved, that does allow for well-defined but strongly renormalized quasi-particles and quasi-holes near the two Fermi surfaces. The main part of the paper ends with conclusions in section \ref{sect:concl}. Finally, there are a number of appendices which contain our conventions and give more details about the calculations.


\section{Single-particle Green's function for Lifshitz fermions}\label{sect:pres}

In this section, we determine  solutions to the Dirac equation in a Lifshitz spacetime, from which we construct the retarded fermionic single-particle Green's function. To keep the paper self-contained, we start by  defining our notation and conventions. Readers who are only interested in the result of the calculation may skip directly to  equation~\eqref{GR2}.
Further details on the derivation can be found in the appendices, in particular, for remarks on how to switch from natural to SI units we would like to refer to appendix \ref{app:units}. Some of the results in this section have already appeared in more detail in our previous work \cite{Gursoy:2011gz}, other related work on fermions in Lifshitz backgrounds can be found in \cite{Korovin:2011kw,Alishahiha:2012nm,Fang:2012pw}.

\bigskip
We begin by specifying the gravitational background metric of a Lifshitz black brane \cite{Taylor}, i.e.,
\eq{
  \label{metric_02}
  {\rm d}s^2 =  - V^2(r) \, r^{\,2z} {\rm d}t^2+\frac{{\rm d}r^2}{r^2\, V^2(r)}  + r^{\,2} {\rm d} {\vec x}^{\:2} \ ,\hspace{40pt}
  V^2(r) = 1 - \Big(\frac{r_h}{r}\Big)^{\,d+z-1} \;,
}
where $d$ denotes the number of spatial dimensions.
The extra spatial coordinate $r$ runs from the horizon at $r=r_h$ to the boundary at $r=\infty$, and the temperature of the black brane is obtained by demanding the absence of a conical
singularity at $r_h$, leading to \cite{Taylor}
\eq{
\lab{T}
T = \frac{d+z-1}{4\pi} \: (r_h)^{z} \;.
}
The metric \eqref{metric_02} enjoys the Lifshitz isometry, which will be inherited by the boundary theory as we will show shortly, and which reads
\eq{
\lab{scaling}
r\to \l r\;, \hspace{30pt}
t\to \l^{-z} t\;,  \hspace{30pt}
x\to \l^{-1} x \;,  \hspace{30pt}
T \to \l^z T\;.
}

Next, we introduce a Dirac fermion $\Psi$ into the above gravitational background. Following \cite{Pomarol,Gursoy:2011gz}, we include a boundary action on an ultra-violet (UV) cut-off surface at $r=r_0$. While many of the results  can be phrased for arbitrary dimensions, for later purposes we specify to $d=4$ spatial dimensions already here. In the conventions of \cite{Gursoy:2011gz}, the total action for the bulk fermion that we consider is
\eq{
\label{totac}
 S_{\text{total}}[\Psi] =
  i\hspace{1pt}g_f &\int {\rm d}^{5}x \hspace{1pt}\sqrt{-g} \: \left(
  \frac{1}{2}\, \ov \Psi \overrightarrow{ \slashed{\mathcal{D}}} \Psi
  -  \frac{1}{2}\, \ov \Psi \overleftarrow{ \slashed{\mathcal{D}}} \Psi
  -
  M \,\ov
  \Psi \Psi \right)\\
  -&\int_{r=r_0} {\rm d}^{4}x \hspace{1pt}\sqrt{-h} \left( \Psi_+\dagh Z \DD_z(r,x) \Psi_+ + i \, g_f
\sqrt{g^{rr}}\: \Psi_+\dagh\Psi_- \right)\;,
}
where the first term on the right-hand side is the Dirac action in the bulk and the second term is the boundary action.
Our notation is such that $\ov \Psi = \Psi^{\dagger} \Gamma^{\ul 0}$ with  $\Gamma^{\ul 0}$ being anti-hermitian, and a natural choice for the Dirac matrices employed in this paper is given in appendix \ref{app:spinors}.
Furthermore, we have decomposed $\Psi$ into chiral components according to its
eigenvalue under $\Gamma^{\ul r}$ \cite{Liu:2009dm,Zaanen} in the following way
\eq{
\lab{psipm}
\Psi \equiv \binom{\Psi_+}{\Psi_-} \;,
\hspace{40pt}
 \Gamma^{\ul r}\,\Psi =\binom{+\Psi_+}{-\Psi_-}  \;.
}
The normalization constants $g_f$ and $Z$ appearing in \eqref{totac} are left unspecified for the moment. The coupling of the fermions to the gravitational field is through the vielbeins $e_{\ul a}{}^{\mu}$ and the spin connection $(\Omega_{\mu})_{\ul a \ul b}$, which are determined from the  geometry and whose explicit form can be found in appendix \ref{app:spinors}.
The symbol  $\slashed D_z$ appearing in the boundary action is the usual Dirac operator for a fermion in a Lifshitz background with arbitrary dynamical exponent $z$, and $h$ denotes the determinant of the induced metric on the boundary.

Following \cite{Liu1}, we also define Fourier-transformed spinors on each constant $r$ slice as
\eq{
\label{FTspinor}
\Psi_{\pm}(r,x)=\int \frac{{\rm d}^4p}{(2\pi)^4}\,\psi_{\pm}(r,p)\,e^{ip_\mu x^\m}\; ,\qquad p_\mu=(-\omega,\vec k)\ ,}
where $\omega$ and $\vec k$ denote the frequency and momenta of the plane wave.
The Dirac equation $\bigl( \slashed{\mathcal{D}} - M \bigr) \Psi = 0 $ resulting from \eqref{totac} (see \cite{Gursoy:2011gz} for more details) and  in-falling boundary conditions that we shall employ at the horizon imply a relation between $\psi_+$ and $\psi_-$. It can be expressed as
\eq{
\lab{xi1}
\psi_-(r,p) = -i \hspace{1pt}\xi(r,p)\, \psi_+(r,p)\ ,
}
and can be used to integrate out $\psi_-$ from the action. Together with (\ref{totac}) and \eqref{FTspinor} this results in a holographic effective action for the field $\psi_+$ on the cut-off surface as
\eq{
\lab{totac1}
S_{\rm eff}[\Psi_+]  = - \int_{r=r_0}\frac{{\rm d}^4p}{(2\pi)^4}\:\sqrt{-h}\:  \psi_+^\dagger \le[
  Z \slashed D_z(p) +g_f \sqrt{g^{rr}} \xi (r_0,p) \ri] \psi_+\;.
}
The Green's function of $\psi_+$ that follows from \eqref{totac1} for our geometry \eqref{metric_02} reads
\eq{
\lab{GR1}
G_R(r_0,p) = - \le(r_0^z V(r_0) \slashed D_z(p) + \frac{g_f}{Z} r_0^{1+z}V^2(r_0)\: \xi(r_0,p) \ri)^{-1}\ ,
}
where we have rescaled $\psi_+$ so that it acquires a canonically normalized kinetic term. The final step is now to take a double-scaling limit
\eq{
\lab{dslim}
r_0\to\infty \;, \hspace{40pt}
 g_f\to 0\;, \hspace{40pt}
 g_f r_0^{1+z-2M} = {\rm const.}
}
As we take this limit, we make sure that the resulting effective action has a kinetic term expected from a theory with dynamical scaling $z$. In the case $z\neq1$, this can be achieved by renormalizing away a relativistic kinetic term and adding appropriate counter-terms to the action \cite{Gursoy:2011gz}. The result for $z\neq1$ then is
\begin{equation}\label{GR2}
G_R(\vec k,\omega)= - \le(\o - \:\frac{1}{\lambda} \vec{\sigma}\cdot\vec{k}\: |\vec k|^{z-1} -\: \Sigma(p) \ri)^{-1}\;,
 \end{equation}
where $\lambda$ is an arbitrary real number in our approach and where in the following we denote $k=|\vec k|$. Before we discuss the self-energy $\Sigma(p)$, we would like to spend some words on the nature and interpretation of the parameter $\lambda$, both from a holographic point of view, and from a condensed matter point of view of the boundary.

The parameter $\lambda$ appeared first in \cite{Gursoy:2011gz}, where it was denoted $\lambda=-1/\eta$ (see the discussion between (3.18) and (3.19) in that reference). It is a coefficient that is determined by the holographic renormalization procedure for Lifshitz geometries in the presence of bulk fermions that has not been not carried out. In our set-up, it is the parameter that multiplies the counterterm for the fermions on the UV brane involving spatial derivatives, after sending the UV cut-off to infinity. Given a particular bulk model, $\lambda$ is a fixed number. For instance, for a dynamical exponent $z=1$, and relativistic symmetry, $\lambda =\pm 1$ (or $\pm 1/c$ in natural units, $c$ being the speed of light). For other values of the dynamical exponent, we do not know the magnitude of $\lambda$, neither do we know its sign. It is fixed and not tunable for a given bulk action. Nevertheless, we treat it here as a variable, and allow it to vary in the phase diagram of the boundary system, in much the same way as the bulk fermion mass $M$. In this manner we can explore all the possible physical properties of the system, even though at present we do not precisely know its value for the background that we are using. Moreover, as explained in Section 4.5, the same phase diagram can be obtained from a fixed value of $\lambda$, but by varying the bulk mass $M$ over a broader range, see also \eqref{defsigspec}. From this point of view, treating $\lambda$ as a variable is perfectly viable, as long as one has enough bulk models (or string compactifications) with sufficiently many different choices for $\lambda$. This is completely consistent within our approach, as we have not specified the bulk action. We only used the metric that couples to the bulk fermions.

Forgetting about holography, and looking at the boundary system from a condensed-matter point of view, the parameter $\lambda$ is the strength of the spin-orbit coupling $\vec{\sigma} \cdot \vec{k}\, k^{z-1}$. What the value of the spin-orbit coupling is depends on the underlying microscopic model for the fermions, and we cannot compute it without specifying the model. In fact, it may depend on the properties of the material.  Hence we treat it as a parameter that encodes this ignorance, and we allow to vary it. In other words, we study how the physics changes as a function of $\lambda$, and the result of this analysis is the main content of our paper. In particular, we found that the sign of $\lambda$ appears to be crucial in determining whether the system displays Fermi-liquid behavior or not. In a certain sense, we are doing model building, and thus we checked that for all values of $\lambda$, important physical consistency conditions such as sum rules and Kramer-Kronig relations are satisfied.

After having discussed the significance of the parameter $\lambda$, we return to the self-energy appearing in the Green's function \eqref{GR2}. In our set-up, the holographic self-energy $\Sigma(p)$ is by construction an effective description of the interactions between the elementary chiral fermions. This interaction term arises from coupling the elementary field $\psi_+$ to the conformal field theory encoded in the Lifshitz background. As can be seen by comparing with \eqref{GR1}, it is related to the quantity $\xi(r,p)$ introduced in \eqref{xi1} by
\eq{ \lab{defsig}
\Sigma(p) =  -g\lim_{r_0\to\infty} r_0^{2M}  \xi(r_0,p) \;,
\hspace{50pt}
 -\tfrac{1}{2}< M<\tfrac{1}{2}\;,
}
where $g$ is the coupling constant  which stays finite in the limit (\ref{dslim}) and reads
\begin{equation}\label{g}
g = \frac{g_f}{Z} \,r_0^{1+z-2M}.
\end{equation}
The definition of the self-energy $\Sigma(p)$ in \eqref{defsig} is valid for all values of the momentum $\kv$, however, for $\kv=0$ the allowed range of $M$ is extended to $-z/2 < M < z/2$. This
 can be derived from the asymptotic behavior of $\xi(r,p)$ near the boundary, and for more details we refer the reader to appendix \ref{app:asymp}.
Furthermore, it is possible to extend the relation between $\xi(r,p)$ and $\Sigma(p)$ even to $|M|>1/2$ for non-zero momentum, which requires introducing certain counter-terms on the cut-off surface at $r_0$ before taking the limit $r_0\to\infty$. Since we will also be interested in the range $1/2<M<z/2$, using the results in appendix \ref{app:asymp}, we can show that the relation \eqref{defsig} in this case should be modified to
\eq{ \lab{defsigspec}
\Sigma(p) =  -g\lim_{r_0\to\infty} \left(r_0^{2M}  \xi(r_0,p) - \frac{\sik}{2M-1}\, r_0^{2M-1}\right)\;,
\hspace{40pt}
\tfrac{1}{2}<M<\tfrac{z}{2}\; .
}
Note that the second term in this expression, which is divergent for $M>1/2$, removes the divergence in $\xi(r,p)$ and  yields a finite result for $\Sigma(p)$.

The transfer matrix $\xi(r,p)$ defined in equation \eqref{xi1} is a complex two-by-two matrix. In the case we are interested in, that is, $d=4$, it can be diagonalized by choosing the Weyl basis of gamma matrices \eqref{even}. A first-order differential equation for the eigenvalues $\xi_{\pm}(r,p)$ of $\xi(r,p)$ can be derived, which was achieved in the anti-de-Sitter case ($z=1$) in \cite{Liu1} and generalized to arbitrary $z$ in
\cite{Gursoy:2011gz}. The resulting differential equations read
\eq{
\lab{xieq}
r^2V \dau_r\xi_\pm + 2Mr \xi_\pm = \frac{\omega}{r^{z-1}V} \mp k_3 + \left( \frac{\omega}{r^{z-1}V} \pm k_3\right)
\xi_\pm^2 \;,
}
where we used rotational invariance to set $\kv=(0,0,k_3)$.
Imposing in-falling boundary conditions at the black-brane horizon, which corresponds to considering the retarded Green's function, leads to the boundary condition
\eq{\lab{bc} \xi_\pm(r_h) = i \;.
}
The functions $\xi_{\pm}(r,p)$ as well as the resulting self-energy $\Sigma(p)$ have various symmetries, which are discussed in more detail in appendix \ref{app:symm}.

From the imaginary part of the retarded Green's function $G_R$, the total spectral weight, or spectral function, can be obtained, which is of great importance in condensed-matter physics and is also directly observable in experiments. It is defined as the trace of the two-by-two matrix $G_R$, i.e.,
\eq{
 \lab{imspec}
\rho(\kv,\om) = \frac{1}{2\pi} \I \Tr\hspace{1pt} \bigl[G_R(\kv,\om) \bigr] \;.
}
It was shown in \cite{Gursoy:2011gz} that the spectral-weight function corresponding to \eqref{GR2} satisfies the sum rule
\eq{
\lab{sumrule}
\int_{-\infty}^{\infty} \text{d} \om\,\rho(\kv,\om)= 1
}
in the physically allowed range for  the scale dimensions of the CFT operator coupled to $\psi$.
In  \cite{Gursoy:2011gz}, this range was found to be $-z/2<M<z/2$, where the lower bound is due to unitarity in the field theory, and the upper bound is a consequence of the requirement that the coupling of the elementary fermion to the CFT is irrelevant in the ultra-violet. Within this allowed physical range for $M$, we have checked numerically that the Kramers-Kronig relations are satisfied and the sum rule \eqref{sumrule} is obeyed. The latter is a consequence of the fact that $G_R$ is the Green's function for an elementary field $\psi$ and its hermitian conjugate $\psi\dagh$, that satisfy anti-commutation relations. In conclusion, the prescription \eqref{GR2} thus allows us to holographically compute   \emph{single-particle} Green's functions, which are important from a condensed-matter perspective.

\begin{figure}[t]
\centering
\scalebox{0.8}{\includegraphics[width=0.7\textwidth]{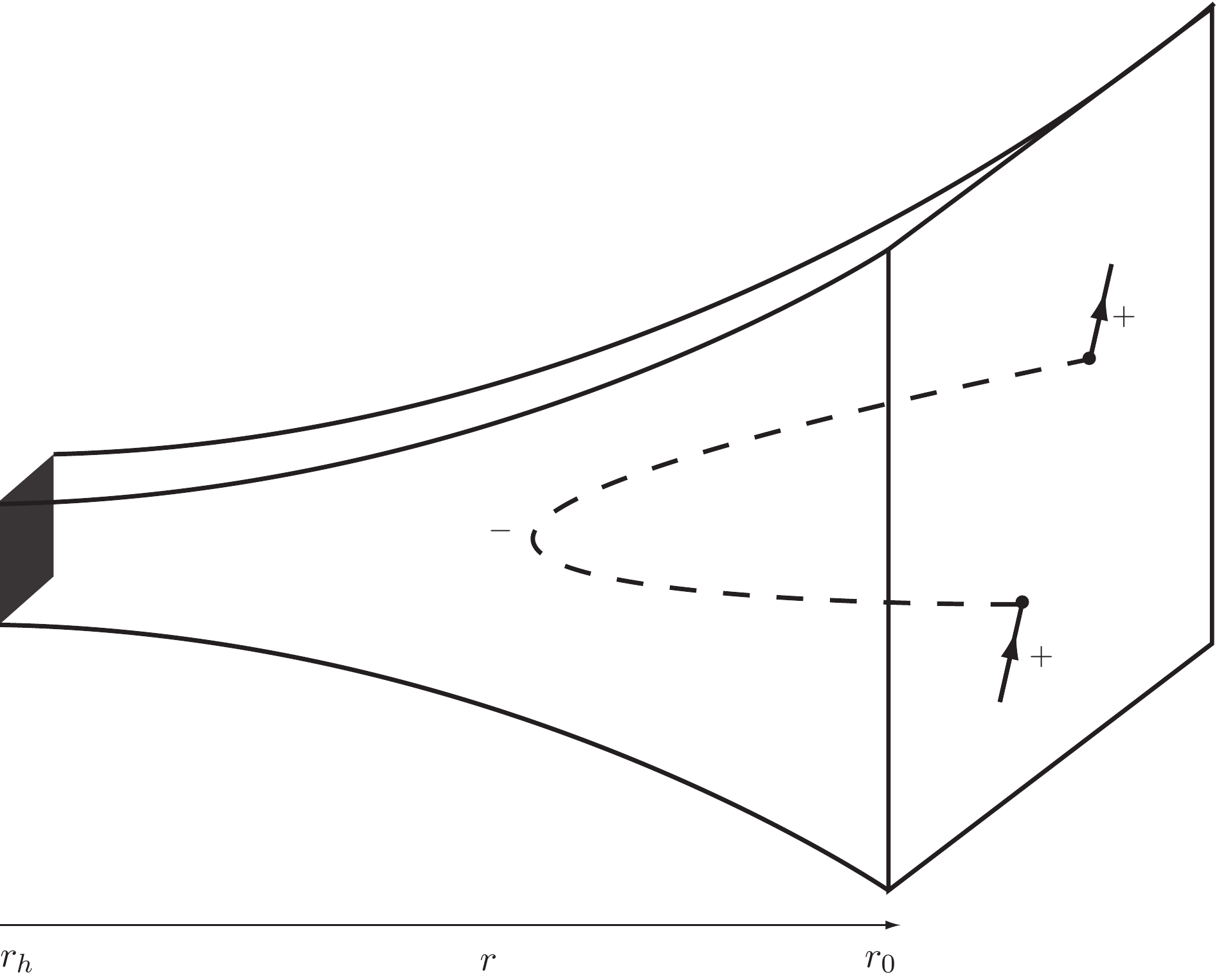}}
\caption{Illustration of our holographic construction of the self-energy.}
\label{figA}
\end{figure}

\medskip
To close this section, we summarize and illustrate our holographic construction in figure \ref{figA}. In particular, equation $\eqref{GR2}$ describes chiral single-particle propagators on the boundary of the spacetime. These single-fermion excitations are modeled as dynamical sources coupled to the fermionic composite operators in the conformal field theory. These chiral single-fermions interact with each other, which is described via the coupling with the conformal field theory due to a fermionic excitation of opposite chirality that travels into the bulk spacetime, feels the bulk gravitational effects classically, and comes back to the boundary, forming the self-energy of the single fermions. In the ultra-violet (UV), the points where the chiral dynamical source emits and reabsorbs the fermion of opposite chirality are very close together and the latter fermion cannot travel far into the bulk. Hence, it only notices the flat background on the cut-off surface in the bulk spacetime, which leads to a free fermionic propagator in the UV. For the infra-red (IR) case, however, the fermion of opposite chirality can indeed travel far into the curved part of the bulk spacetime, and as a result the IR dynamics will be dominated by interactions and will generally not be free. This corresponds physically to the usual ``energy-scale'' interpretation of the extra spatial direction.
For clarity, we recite the three different couplings that one should distinguish here.
\begin{enumerate}
\item The first coupling is the analog of the 't Hooft coupling $\lambda_H$ that describes the effective coupling in the conformal field theory. This is assumed to be large in our model, hence the corresponding string states in the bulk are assumed to decouple.
\item The second coupling is the inverse of Newton's constant, $G_5^{-1}$, that is proportional to $N^2$, where $N$ is the degree of the gauge group governing the conformal field theory. This is assumed to be large, allowing us to treat the fermion bulk action as a perturbation on the action of general relativity. In particular, we may ignore the back-reaction of the bulk fermions on the spacetime in the large-$N$ limit. This corresponds to the fact that the connected parts of four-point functions of composite operators in the conformal field theory vanish.
\item Finally there is what we call $g$, the coupling between the dynamical source and the dominant channel in the conformal field theory.
\end{enumerate}
The first two coupling constants are implicit in our model, as the gravitational action is not specified. A strongly interacting conformal field theory is characterized by non-trivial scaling dimensions, different than the engineering dimensions. This is indeed the case modeled here. In order to produce a non-trivial self-energy for the single fermion, the first and last coupling constants should be large, at least of order 1. As a consequence, there is always a momentum-space region in the IR where the self-energy is dominant over the kinetic term in the single-fermion propagator $\eqref{GR2}$. This is what we mean when saying that the single fermions are "strongly interacting". In the context of the Weyl semimetal, the single-particle propagator from $\eqref{GR2}$ models the excitations of one of the chiral cones.


\section{Single-particle spectra}\lab{sect:results}

In the previous section, we have outlined the construction of the retarded single-particle Green's function. In the present section, we now study this Green's function  for both the relativistic $z=1$ and non-relativistic $z=2$ case in more detail, and work out its physical properties. The zero-temperature results were first obtained in \cite{Gursoy:2011gz}, and here we consider non-zero temperatures as well. However, let us stress that the systems we describe are always at zero chemical potential.


\subsection{Relativistic case $z=1$}

In the case that the elementary fermion $\psi_+$ interacts with a relativistic CFT, the background is given by an AdS black brane and is described by the metric shown in equation \eqref{metric_02} for $z=1$. The mass range for the bulk Dirac fermion is then restricted to lie within the interval $-1/2<M<1/2$.


\subsubsection{Zero temperature}

For vanishing temperature and $z=1$, the self-energy $\Sigma(p)$  can be computed analytically for arbitrary frequencies $\o$ and momenta $\vec{k}$. The result for the full retarded Green's function has been worked out in \cite{Gursoy:2011gz} and reads
\begin{equation}
\label{GRAdS2}
G_R(\vec k,\omega)=-\frac{1}{p^2\le(1-g \hspace{1pt}c_1\hspace{1pt} e^{-i\pi(M+\half)}p^{2M-1}\ri)}
\Big(\omega+\vec \sigma \cdot \vec k\Big) \ ,
\end{equation}
where we defined $p\equiv\sqrt{\omega^2-\vec{k}\cdot\vec{k}}$ as well as the constant (for arbitrary values of $z$)
\eq{
\lab{cz}
c_z = (2z)^{-\frac{2M}{z}} \frac{\Gamma\le(\half-\frac{M}{z} \ri)}{\Gamma\le(\half+\frac{M}{z}\ri)} \;.
}
Note that \eqref{GRAdS2} can be extended into the complex plane by allowing for complex momenta $p$, which is important when determining the pole structure of the Green's function. Because of the branch point at $p=0$, we need to introduce a branch cut which is taken to run from $p=0$ to $p=-i\infty$ for later convenience. Furthermore, we find that in order for \eqref{GRAdS2} to be free of singularities in the upper half $\o$ plane, i.e., to satisfy the Kramers-Kronig relations, we have to demand
\eq{
\lab{posgot}
g>0 \;,
}
which is derived in detail in appendix \ref{app:sum3}.
It is then straightforward to show that the Green's function \eqref{GRAdS2} also satisfies the sum rule \eqref{sumrule}, as it was done in \cite{Gursoy:2011gz} and summarized in appendices \ref{app:sum1} and \ref{app:sum2}. Furthermore, as mentioned above,  \eqref{GRAdS2} is valid on the complex $p$ plane, with the prescription that on the real $\o$ line the self-energy is found  by using
\eq{
\label{realline}
p^{2M-1}e^{-i\pi(M+\half)}=  \left\{\begin{array}{l@{\hspace{20pt}}l@{\hspace{20pt}}r@{}}
+p^{2M-1}e^{-i\pi(M+\half)} \;, & p\equiv\sqrt{\omega^2-\vec{k}\cdot\vec{k}}\;, & \omega> +|\vec{k}| \;, \\
+p^{2M-1}e^{+i\pi(M+\half)}\;,  & p\equiv\sqrt{\omega^2-\vec{k}\cdot\vec{k}}\;, & \omega< -|\vec{k}|\;, \\
-p^{2M-1} \;, & p\equiv\sqrt{\vec{k}\cdot\vec{k} - \omega^2}\;, & - |\vec{k}|<\omega< +|\vec{k}| \;.
\end{array} \right.
}
This directly follows from \eqref{GRAdS2} by noting that the region $\o<-|\vec{k}|$  for real frequencies is obtained from the region  $\o>|\vec{k}|$ by $p \to e^{i\pi} p$, that is $\o \to e^{i\pi}\o$ and $\vec{k} \to e^{i\pi} \vec{k}$, while the region
$|\o|<|\vec{k}|$ is obtained via $p \to e^{i\pi/2} p$. Note also that the Green's function for $- |\vec{k}|<\omega< +|\vec{k}|$ has no imaginary part.


\subsubsection{Non-zero temperature}\label{sec:z1spec}

We have studied the effects of non-zero temperature on the structure and form of the Green's function by  numerical integration of the Dirac equation.
To illustrate our results, we have included figures \ref{figImTrGRz1disp3D} and \ref{figImTrGRz1disp3DT2} which show the total spectral-weight function \eqref{imspec} for $T=1/30$ and $T=2$, respectively.
Furthermore, we made a distinction between positive and negative masses $M$, and we employed the rotational symmetry to set $\vec k = (0,0,k_3)$. Let us discuss these figures in some more detail:
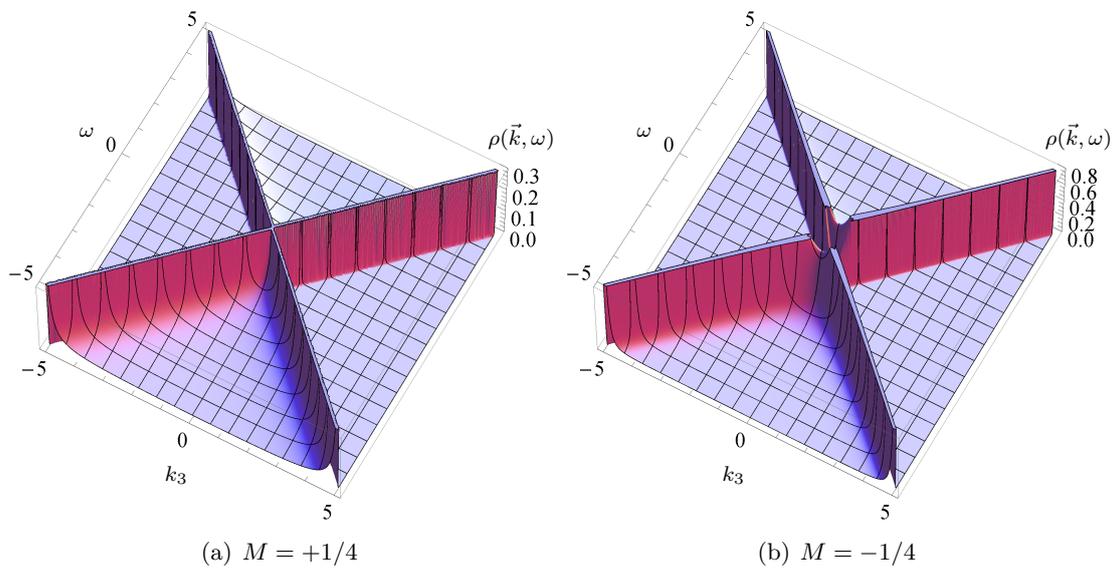
\begin{figure}[p]
\centering
\subfigure[$M=+1/4$]{
\includegraphics[width=0.475\textwidth]{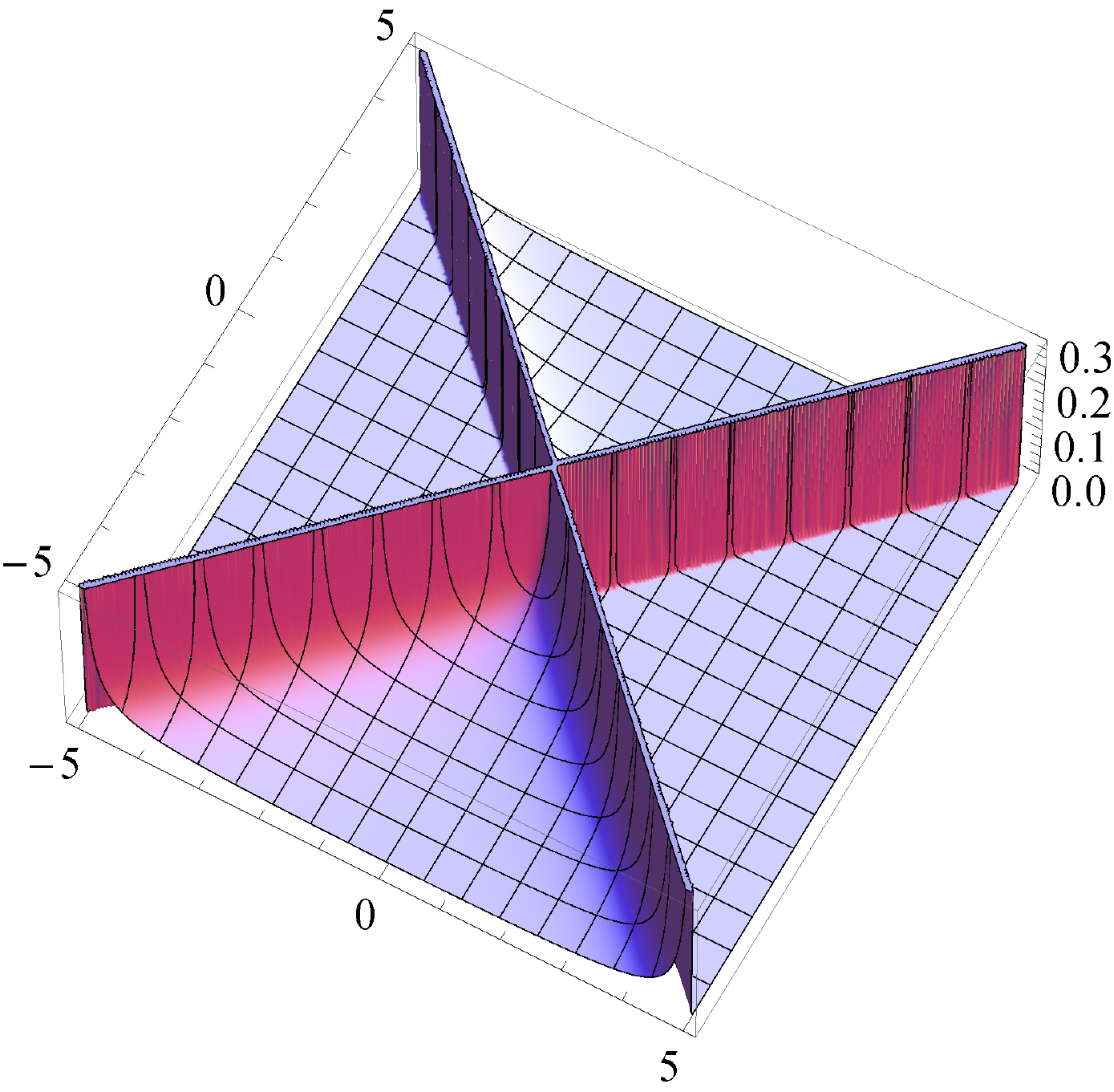}
\begin{picture}(0,0)
\put(-144,16){\scriptsize $k_3$}
\put(-176,146){\scriptsize $\omega$}
\put(-23,143){\scriptsize $\rho(\vec k, \omega)$}
\end{picture}
}
\hspace{-8pt}
\subfigure[$M=-1/4$]{
\includegraphics[width=0.475\textwidth]{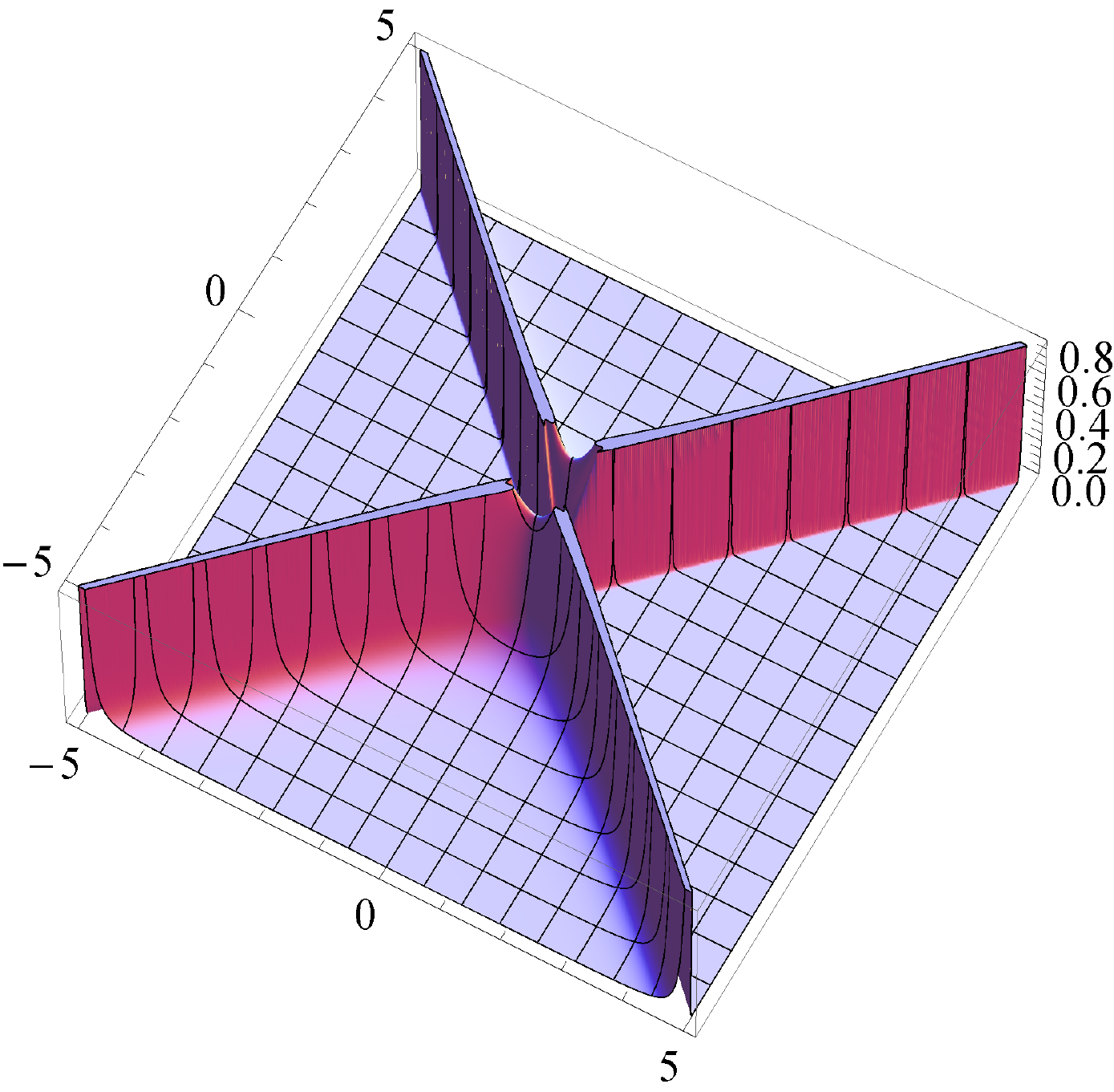}
\begin{picture}(0,0)
\put(-144,16){\scriptsize $k_3$}
\put(-176,146){\scriptsize $\omega$}
\put(-23,143){\scriptsize $\rho(\vec k, \omega)$}
\end{picture}
}
\caption{Total spectral function for $T=1/30$, $g=1$ and  $M=\pm1/4$.}
\label{figImTrGRz1disp3D}
\end{figure}
\begin{figure}[p]
\centering
\subfigure[$M=+1/4$]{
\includegraphics[width=0.475\textwidth]{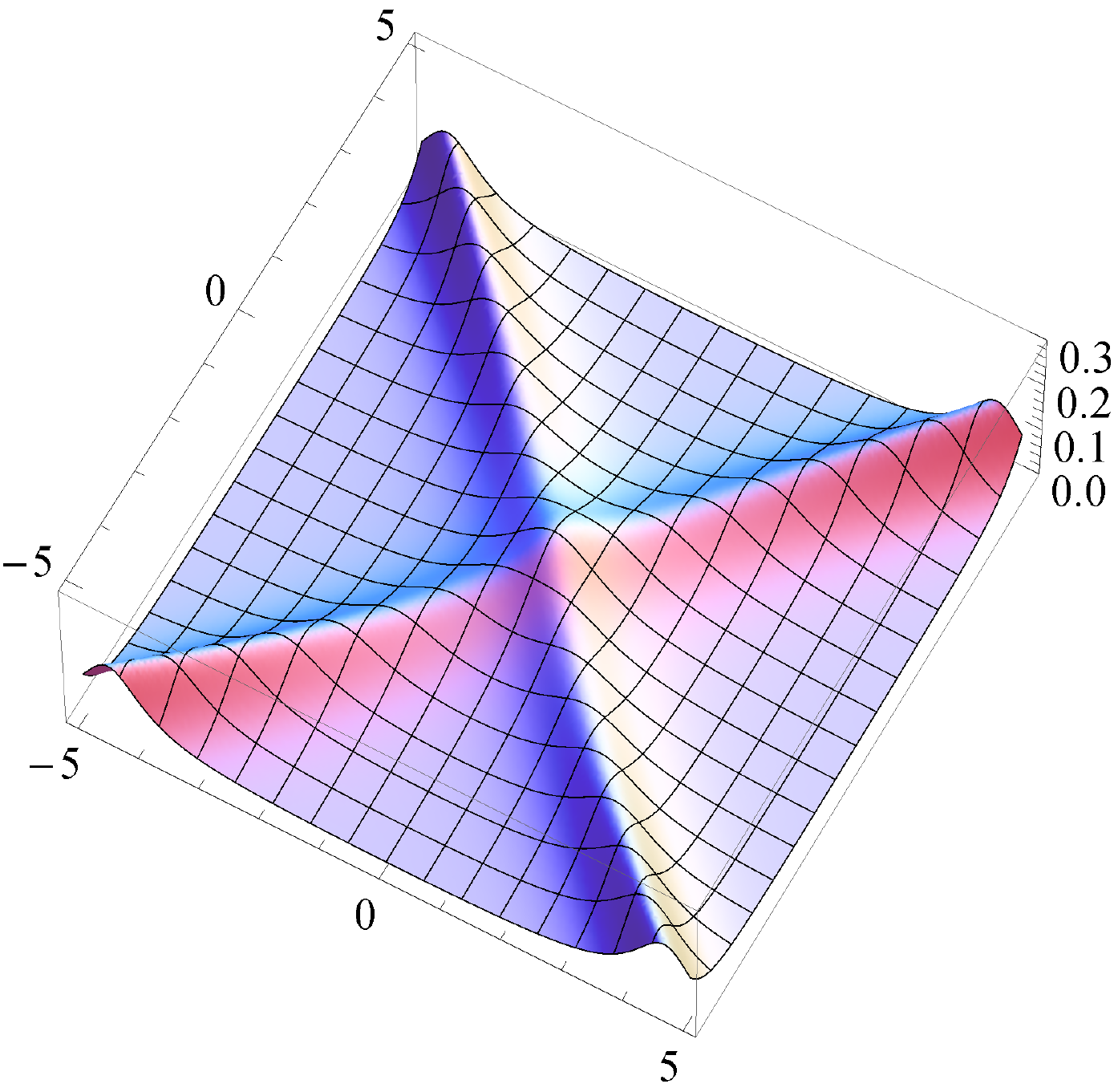}
\begin{picture}(0,0)
\put(-144,16){\scriptsize $k_3$}
\put(-176,146){\scriptsize $\omega$}
\put(-23,143){\scriptsize $\rho(\vec k, \omega)$}
\end{picture}
}
\hspace{-8pt}
\subfigure[$M=-1/4$]{
\includegraphics[width=0.475\textwidth]{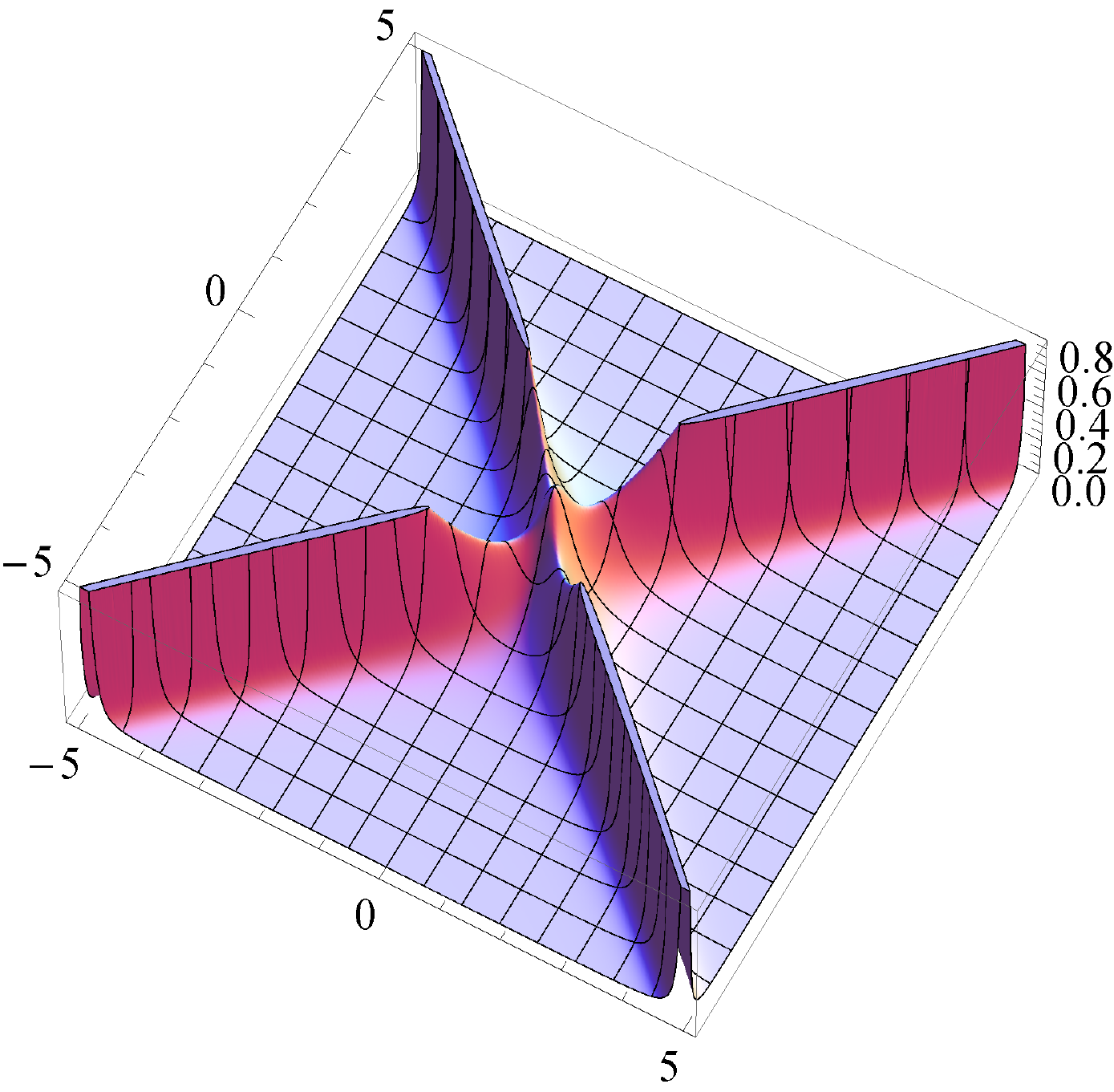}
\begin{picture}(0,0)
\put(-144,16){\scriptsize $k_3$}
\put(-176,146){\scriptsize $\omega$}
\put(-23,143){\scriptsize $\rho(\vec k, \omega)$}
\end{picture}
}
\caption{Total spectral function for $T=2$, $g=1$ and  $M=\pm1/4$.}
\label{figImTrGRz1disp3DT2}
\end{figure}
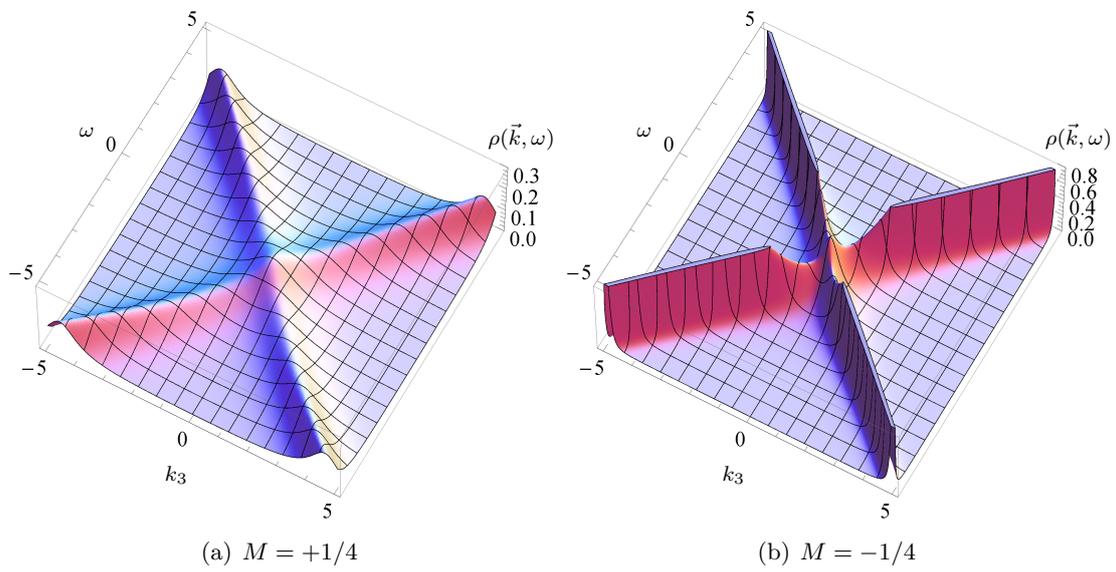
\begin{itemize}

\item First, we note that a non-vanishing temperature results in a smearing out of the features of the Green's function, as can be seen by comparing figures \ref{figImTrGRz1disp3D} and \ref{figImTrGRz1disp3DT2}.

\item For $M>0$ we observe a large spectral weight approximately at $\om = \pm k_3$, which is due to the  ``light cone'' of relativistic physics. In the zero-temperature case \eqref{realline}, the imaginary part of the Green's function is strictly zero outside of the light cone in the region $- |\vec{k}|<\omega< +|\vec{k}| $. For non-zero temperatures, there is a small contribution in this region, which increases with temperature.

Note that the peak in the spectral weight is not caused by a pole in the Green's function, but by the aforementioned branch cut. Therefore, it has no interpretation as a well-defined quasi-particle excitation, but it is still possible to approximately determine the dispersion relation, as it is shown in the next subsection.

\item For negative masses $M<0$, the spectrum has a similar form. Indeed, there is again a large peak at $\om = \pm k_3$. However, for low $T$, the imaginary part of the Green's function  goes to zero at $k_3=0$ as $\om \rightarrow 0$. It has a maximum in between, which leads to a broad maximum in the spectrum. Its approximate location is determined analytically in section \ref{sec:z1disp}.

We have also investigated the ratio $\mathcal{R}$ of the location of the maximum  and its width as a function of $g$. This ratio is approximately constant, which means that its shape becomes broader as we increase its position, and vice versa. The spectral function at $k_3=0$ looks very similar to what is plotted in figure \ref{figmassbump} for the non-relativistic case with $z=2$. It is not possible to tune the parameters such that the peak is located sufficiently far away from the origin at $\omega=0$ while its maximum remains sharp. Therefore, we cannot interpret this feature in the spectral function as a massive particle. Again, we give some further analytic arguments of this statement at zero temperature in section \ref{sec:z1disp}.

\end{itemize}


\subsubsection{Dispersion relation}\label{sec:z1disp}

In the last subsection \ref{sec:z1spec}, we have illustrated that the spectral weight is  peaked at particular functions $\om(\vec{k})$. In this section, we now investigate the corresponding dispersion relation of the theory. For vanishing temperature this can be done analytically via the single-particle Green's function by solving
\eq{
\lab{Disp}
\textrm{Re}\, G_R^{-1}(\vec k, \omega) =0 \;.
}
By definition, the solutions to this equation determine the dispersion relation of the would-be (quasi-)particle. The interpretation as a particle only becomes justified if the width is small compared to the energy of the particle.  Indeed, a large peak in the spectral density function is only obtained when both Re($G_R^{-1}$) and $({\rm Im}\,\Sigma)^2$ are minimal, as follows from the identity
\begin{equation}
{\rm Im}\, G_R=-\frac{{\rm Im}\,\Sigma}{({\rm Re}\,G_R^{-1})^2+({\rm Im}\,\Sigma)^2}\ .
\end{equation}

Let us first solve \eqref{Disp}. Using the explicit expression \eqref{GRAdS2}, equation \eqref{Disp} generically gives two possible dispersion relations, namely
\eq{
\lab{disp1}
 \omega = \pm |\vec k|
 \hspace{30pt}{\rm and}\hspace{30pt}
  g\,c_1 \textrm{Re}\le[e^{-i\pi(M+\half)} p^{2M-1} \ri]= 1 \;.
}
The first solution describes  a free, massless relativistic excitation which is always present.
The second equation in \eqref{disp1} should be studied separately in the two regions $|\o|<|\vec{k}|$ and $|\o|>|\vec{k}|$. In the first case, using \eqref{realline} we obtain the solution
\eq{
\lab{disp2}
  p^2 = \le(-\frac{1}{g\hspace{1pt}c_1}\ri)^{\frac{2}{2M-1}} \;.
}
Since $g \hspace{1pt}c_1 > 0$ and the left-hand side of (\ref{disp2}) should be real, we conclude that  there is no solution for $|\o|<|\vec{k}|$.
In the second case,  $|\o|>|\vec{k}|$, the solution reads
\eq{
\lab{disp3}
  p^2 = \Bigl(g \,c_1 \cos\big[\pi(M+\tfrac{1}{2})\big]\Bigr)^{\frac{2}{1-2M}} \;.
}
For $0 < M < 1/2$, we do not find a solution since $g \hspace{1pt} c_1 >0$.
On the other hand, for  $-1/2<M<0$, we obtain
\eq{
\lab{disp4}
\o = \pm \sqrt{\vec k\cdot \vec k + m^2}\;, \hspace{40pt}
  m^2=  \Bigl(g \hspace{1pt}c_1 \cos\big[\pi(M+\tfrac{1}{2})\big]\Bigr)^{\frac{2}{1-2M}} \;,
}
which is indeed close to the locations of the maxima in the spectral-weight function that were found for non-zero temperature. For zero temperature, equation \eqref{disp4} yields the exact result.

Now we can look at the width, by computing the imaginary part of the self-energy at the values \eqref{disp4}. A straightforward calculation shows that, at zero temperature, we have
\begin{equation}
\I \Sigma=-(\o-\sik) \tan\left[\pi\left(M+\frac{1}{2}\right)\right]\ ,
\end{equation}
with $\o$ given by \eqref{disp4}. However, for momenta small compared to the gap $m$  (in the restframe of the would-be particle), we can approximate
\begin{equation}
\I \Sigma \approx m \tan\left[\pi\left(M+\frac{1}{2}\right)\right]\ .
\end{equation}
The width is then comparable to the gap, and therefore the peak in the spectral density does not have a quasi-particle interpretation. One might try to make the width smaller by taking the value of $M$ close to the unitarity bound, $M= -1/2$. However, notice that then the gap also narrows down. We conclude therefore that no true quasi-particles exist.


\subsection{Non-relativistic case $z\neq1$}

In principle, every value of the dynamical exponent $z$ can be considered using the prescription shown in \eqref{GR2}. However, for our purposes we are mostly interested in  $z=2$ where the elementary fermion interacts with a CFT exhibiting $z=2$ Lifshitz scale invariance.


\subsubsection{Zero temperature}

Unlike the relativistic case studied above, for an arbitrary dynamical exponent  $z$ we were not able to  obtain $G_R(\omega,\vec k)$ analytically for the both $\o$ and $\vec{k}$ non-vanishing. But it is possible to determine the zero-temperature result for $\vec{k}=0$ and $\o=0$ separately, and then restrict the general expression  to a great extent.

Let us therefore start with the case $\vec{k}=0$ and $T=0$. From \cite{Gursoy:2011gz} we recall the expression for the Green's function as
\eq{
\lab{Gz}
G_{R}( \vec 0,\omega) = -\frac{1}{\omega -  g \,c_z\:  \omega^{\frac{2M}{z}} e^{-i \pi (\frac{M}{z}+\half)}}\;,
}
where we employed the definition \eqref{cz}.
In the  case with $\o=0$ and $T=0$ we instead find \cite{Gursoy:2011gz}
\eq{
\lab{Gzk}
G_{R}( \vec k,0) = \frac{1}{ \frac{1}{\lambda}\, \sik\; k^{z-1} + g \,c_1\; k^{2M-1} \sik }\;,
}
where again $k$ denotes $|\vec{k}|$.
The generic case at vanishing temperature when both $\vec{k}$ and $\o$ are non-zero  can then be restricted as follows. Using the scaling and rotational symmetries of the self-energy (see appendix \ref{app:sca}), and defining $u = \o/k^z$ for notational simplicity, we have
\eq{
\lab{Gzko}
G_{R}( \vec k,\o) = -\frac{1}{\omega -\frac{1}{\lambda}\, \sik\; k^{z-1} - g\le( k^{2M} s_{1,M}\le(u\ri) + k^{2M-1} s_{2,M}\le(u\ri) \sik \ri)}\;,
}
where  $s_{1,M}$ and $s_{2,M}$ are complex functions of $u$ and $M$ only.
 Furthermore, we can derive conditions on the functions $s_{1,2}$ using the symmetry of $\Sigma(p)$ under the change $M\to -M$, as discussed in appendix \ref{app:Msym}.
 The asymptotic behavior of the functions $s_{1,M}(u)$ and $s_{2,M}(u)$ in the limits  $u\to 0$ and $u\to\infty$ is  fixed by the expressions \eqref{Gz} and \eqref{Gzk} given above. We therefore have
\eq{
\lab{flims1}
\begin{array}{lcllcl}
s_{1,M}(u) &\xrightarrow{\;u\to\infty\;}& u^{\frac{2M}{z}} e^{-i\pi\le(\frac{M}{z}+\half\ri)} \,c_z\;,  \hspace{30pt} &
s_{2,M}(u) &\xrightarrow{\;u\to\infty\;}& 0\;, \\
s_{1,M}(u) &\xrightarrow{\;u\to0\;}& 0\;, &
s_{2,M}(u) &\xrightarrow{\;u\to0\;}& c_1\;.
\end{array}
}
Unfortunately, these conditions do not seem to be sufficient to determine the analytic form of \eqref{Gzko}. Therefore we have to study the Green's function numerically. Yet, we can determine the qualitative form of the dispersion relation by analytic arguments as we describe below in section \ref{sec:z2disp}.


\subsubsection{Non-zero temperature}

We have studied the retarded Green's function and the corresponding spectral-weight function for $z=2$  numerically as a function of $\om$, $\kv$, $T$ and $\lambda$.
Using the symmetries summarized in appendix \ref{app:symm}, that is chirality and particle-hole symmetry, we observe that the components of the spectral-weight function obey the following relations
\eq{
\rho^{\pm}(\kv,\om) = \rho^{\mp}(-\kv,\om) =\rho^{\pm}(-\kv,-\o) \;.
}
Consequently, it suffices to consider separate components instead of the trace over chiral components.

Next, we recall that for $z\neq1$ the Green's function \eqref{GR2} contains the parameter $\lambda$, which we have not determined analytically. Treating $\lambda$ as a free parameter, we observe that under a change of sign of $\lambda$, the  far UV behavior of one component of the spectral weight function
asymptotes the UV behavior of the other component with the original sign, that is,
\eq{
\rho_{-\lambda}^{\pm}(\kv,\om)
\quad\xrightarrow{\;|\om|\rightarrow \infty,\;|k|\rightarrow \infty\;}\quad
 \rho_{+\lambda}^{\mp}(\kv,\om)=\rho_{+\lambda}^{\pm}(-\kv,\om) \;.
}
Since $\lambda$ can be interpreted as a spin-orbit coupling constant, we use the following convention for plotting components of the spectral-weight function: when $\lambda$ is positive (negative), the plus(minus) component of the spectral-weight function is shown. In this manner we make sure that we always compare spectra with equal group velocity in the UV, also when $\lambda$ changes sign. We make this choice because qualitatively the UV physics is then, apart from the topology of the band structure, independent of the sign of $\lambda$. This allows us to compare more clearly the physics for positive and negative values of $\lambda$ as we will see shortly.

In figures \ref{figz2displambda} and \ref{figz2displambdaMneg} we show numerical results for separate components of the corresponding spectral-weight function. The parameters are chosen as $M=\pm1/4$, $g=1$ and $T=1/30$, for a number of different values for $\lambda$.
We discuss some of the features in turn:
\begin{itemize}

\item In figure \ref{figz2displambda}, showing the results for positive $M$, we see that the spectral-weight function is very sharply peaked in the UV and behaves as a free chiral fermion with a quadratic dispersion. In the IR, the self-energy becomes dominant which changes  and smears out the form of the free-fermion dispersion due to strong interactions. When $\lambda$ switches sign, the spectral-weight function changes in the IR, with the convention such that the far UV behavior stays the same.
For $\lambda>0$, the band structure is similar to that of a Weyl semimetal, as can be seen by combining it with the density plot of $\rho^{-}(\kv,\om)$ for $\lambda>0$ (not shown) and comparing it with figure 1. Indeed, the observed $\vec{\sigma} \cdot \kv$ dependence and the absence of a gap are defining properties of the Weyl semimetal. For $\lambda<0$, the system is gapless too, and similar to a Weyl semimetal in the presence of Fermi surfaces.

\item In figure \ref{figz2displambdaMneg}, where $M$ is negative, we observe a phenomenon similar to the relativistic case with $M<0$. In particular,  the spectral weight has to vanish for $k=0$ as $\om \rightarrow 0$, so there is a maximum in the spectral-weight function at non-zero $\om$. However, the ratio of the location and the width of the peak, $\mathcal{R}$, again remains constant as we change $g$, so there is no gap generation with a quasi-particle interpretation. This is further illustrated in figure  \ref{figmassbump}.

\end{itemize}
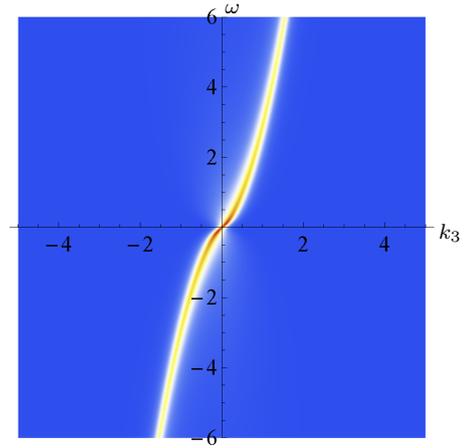
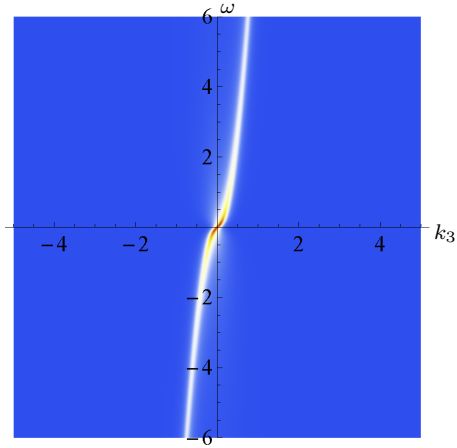
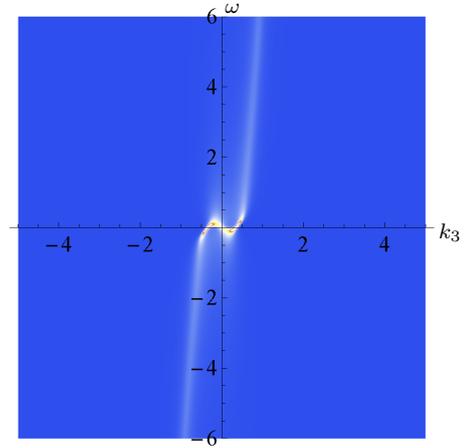
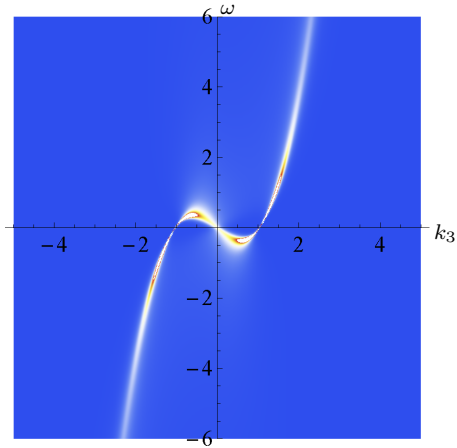
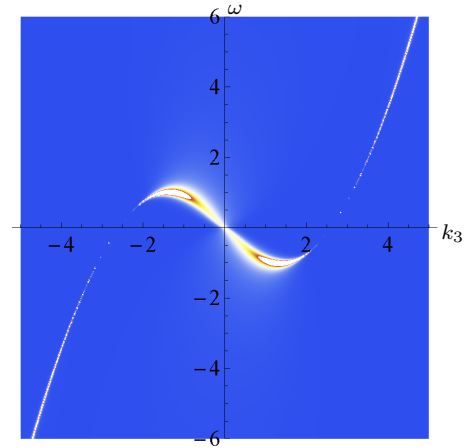
\begin{figure}[p]
\centering
\vskip-2pt
\subfigure[Density plot of $\rho^+(\vec k,\omega)$ at $\lambda=2$]{
\includegraphics[width=0.38\textwidth]{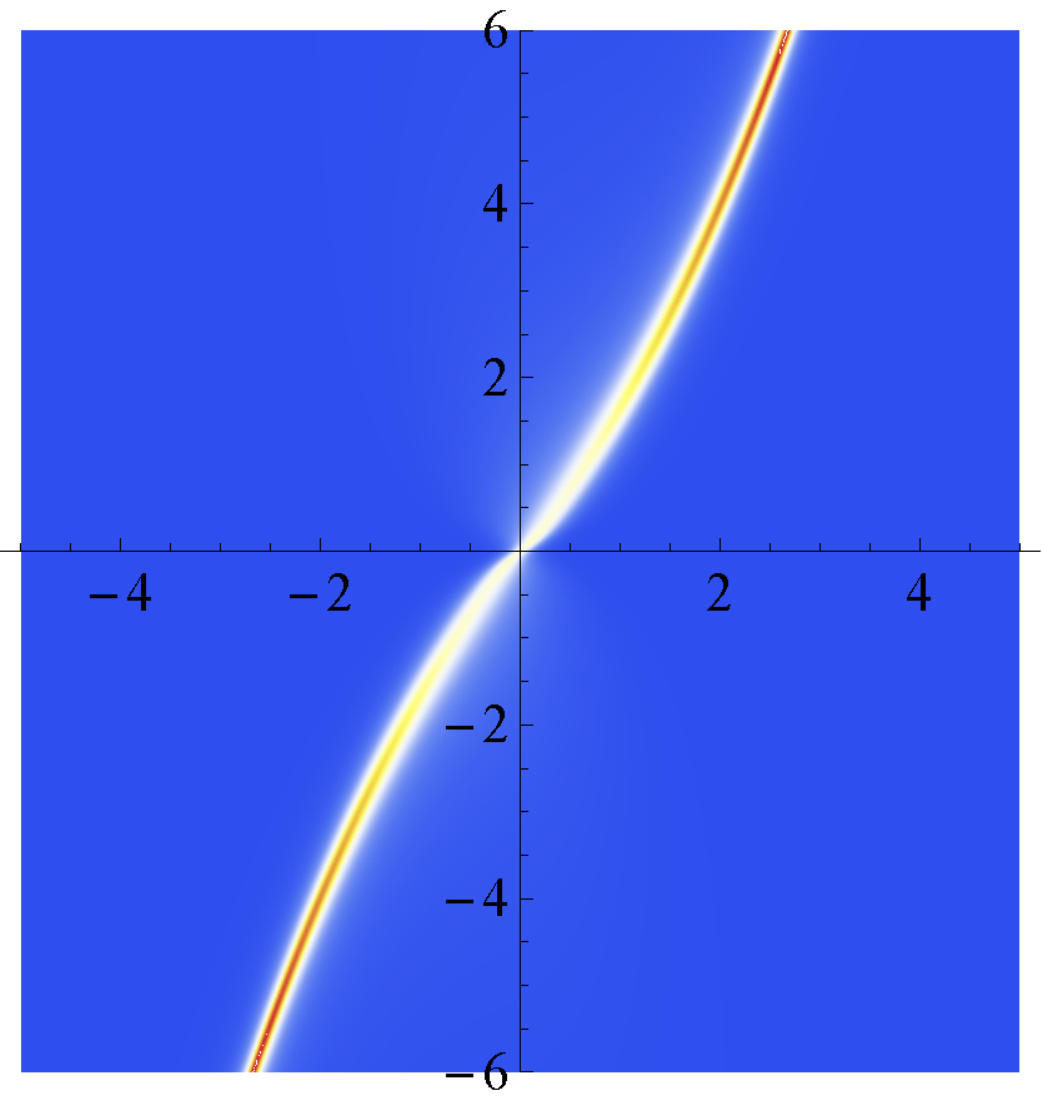}
\begin{picture}(0,0)
\put(-2,80){\scriptsize $k_3$}
\put(-82,165){\scriptsize $\omega$}
\end{picture}
}
\hspace{\stretch{1}}
\subfigure[Density plot of $\rho^+(\vec k,\omega)$ at $\lambda=0.5$]{
\includegraphics[width=0.38\textwidth]{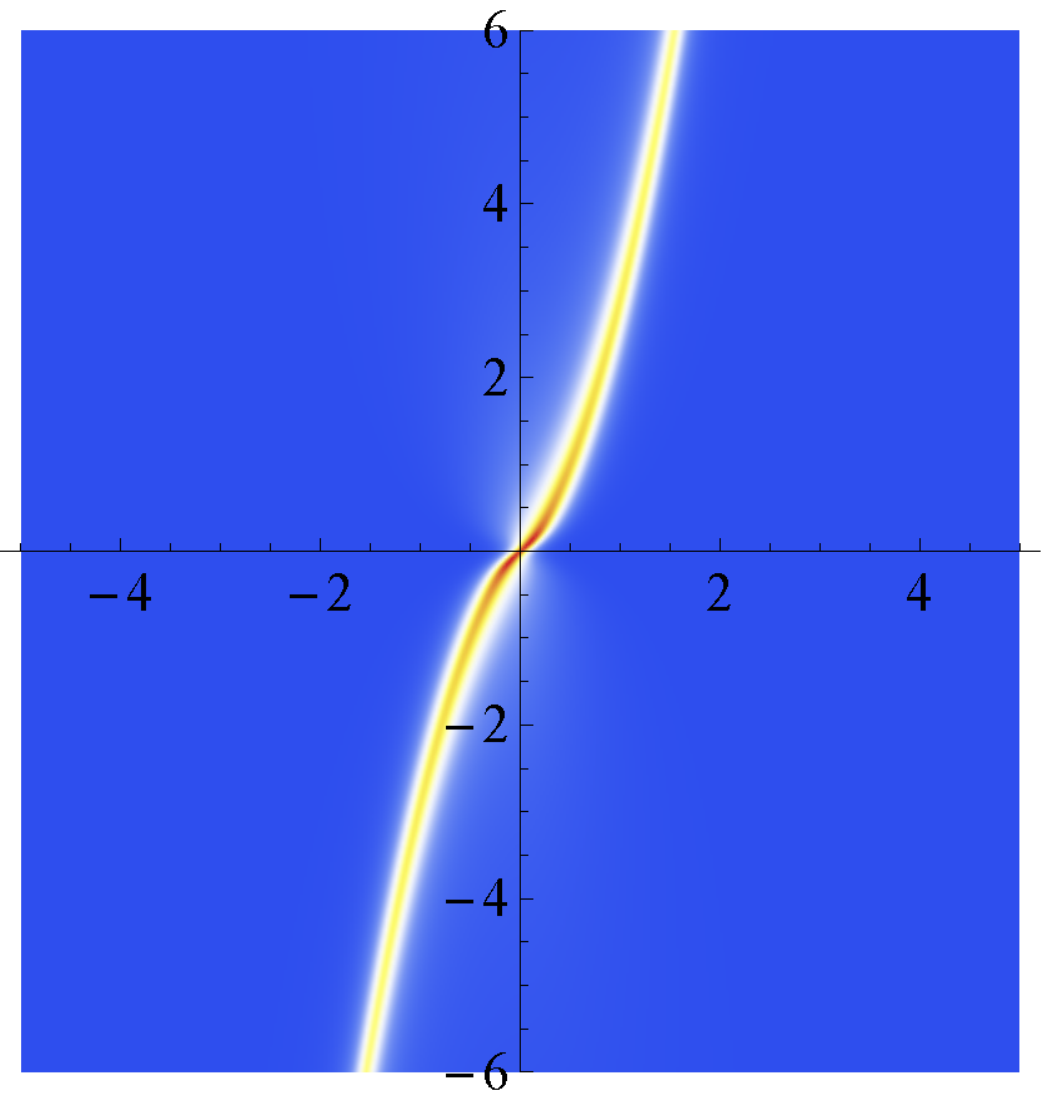}
\begin{picture}(0,0)
\put(-2,80){\scriptsize $k_3$}
\put(-82,165){\scriptsize $\omega$}
\end{picture}
}
\hspace{5pt} \\
\subfigure[Density plot of $\rho^+(\vec k,\omega)$ at $\lambda=0.1$]{
\includegraphics[width=0.38\textwidth]{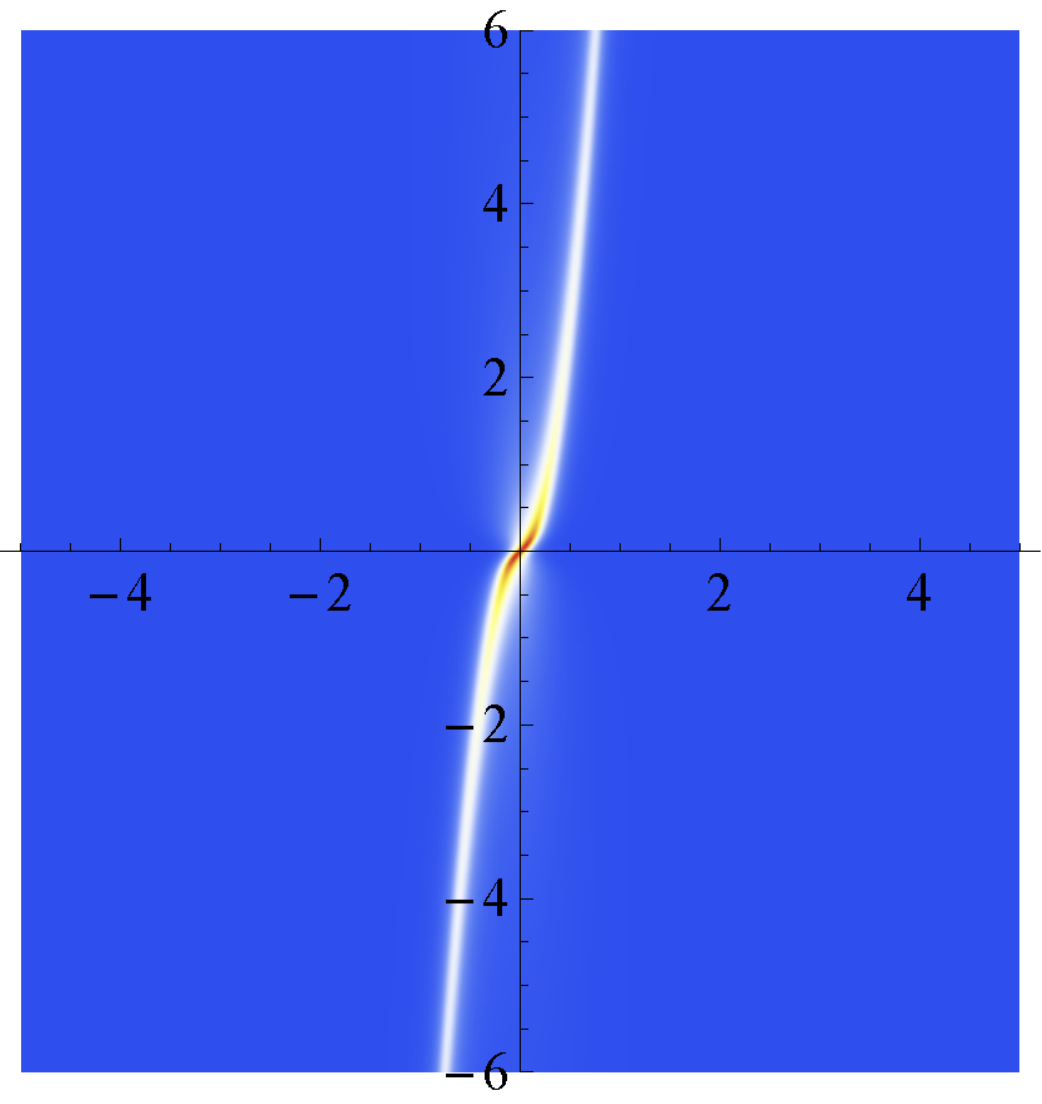}
\begin{picture}(0,0)
\put(-2,80){\scriptsize $k_3$}
\put(-82,165){\scriptsize $\omega$}
\end{picture}
}
\hspace{\stretch{1}}
\subfigure[Density plot of $\rho^-(\vec k,\omega)$ at $\lambda=-0.1$]{
\includegraphics[width=0.38\textwidth]{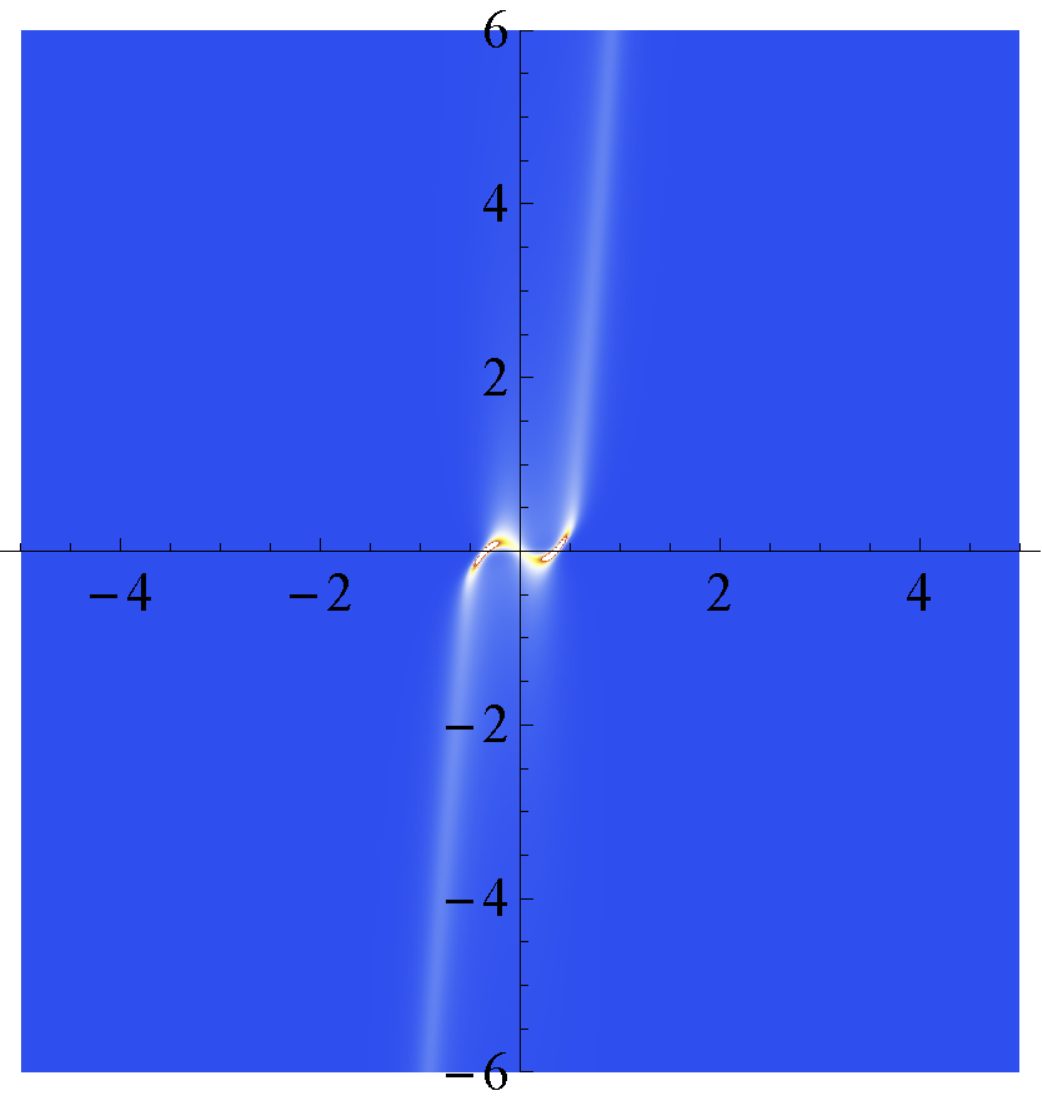}
\begin{picture}(0,0)
\put(-2,80){\scriptsize $k_3$}
\put(-82,165){\scriptsize $\omega$}
\end{picture}
}
\hspace{5pt} \\
\subfigure[Density plot of $\rho^-(\vec k,\omega)$ at $\lambda=-0.5$]{
\includegraphics[width=0.38\textwidth]{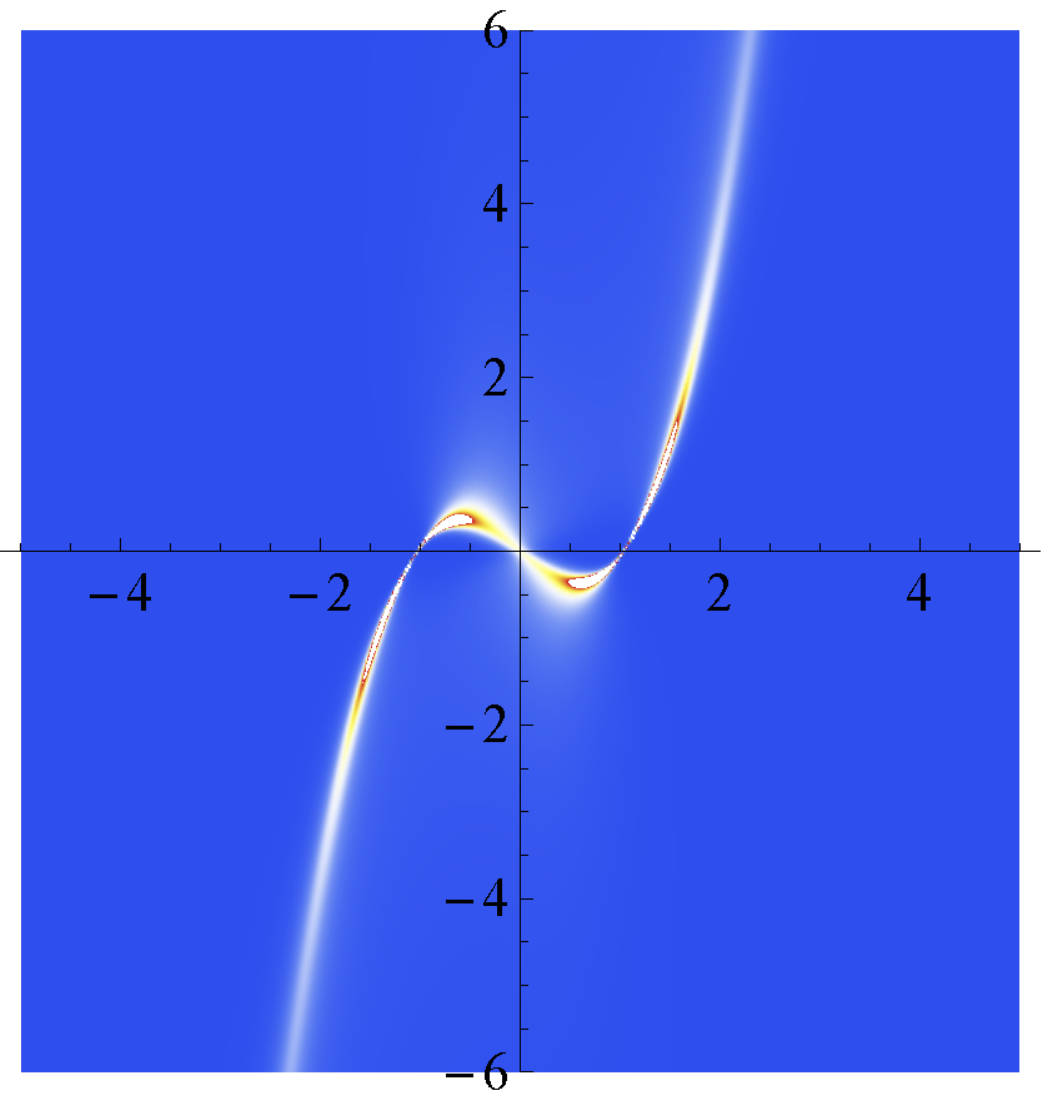}
\begin{picture}(0,0)
\put(-2,80){\scriptsize $k_3$}
\put(-82,165){\scriptsize $\omega$}
\end{picture}
}
\hspace{\stretch{1}}
\subfigure[Density plot of $\rho^-(\vec k,\omega)$ at $\lambda=-2$]{
\includegraphics[width=0.38\textwidth]{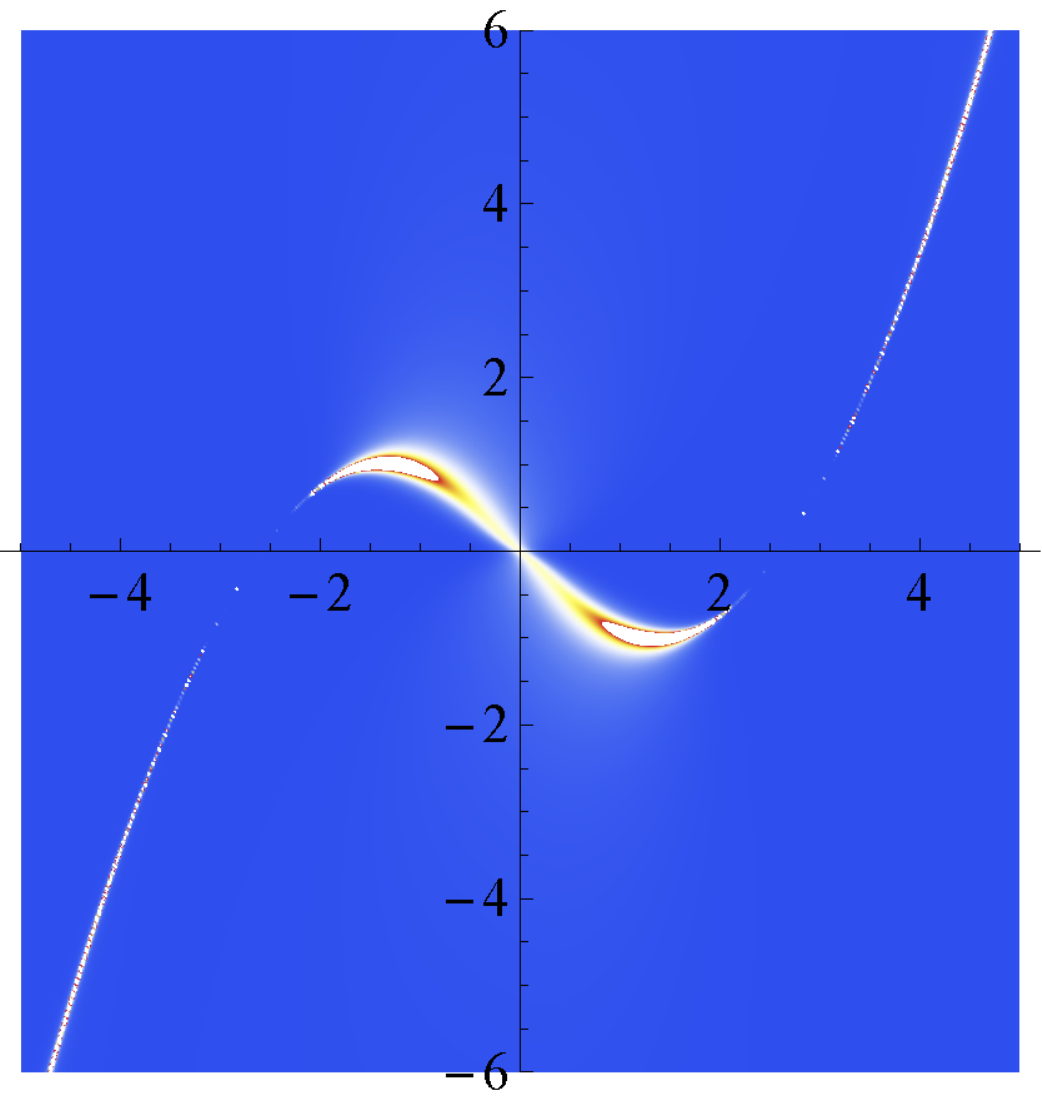}
\begin{picture}(0,0)
\put(-2,80){\scriptsize $k_3$}
\put(-82,165){\scriptsize $\omega$}
\end{picture}
}
\hspace{5pt}
\caption{Density plot of $\rho^{\pm}(\kv,\om)$ for $z=2$, $M=+1/4$, $T=1/30$, $g=1$.}
\label{figz2displambda}
\end{figure}
\begin{figure}[p]
\centering
\vskip-2pt
\subfigure[Density plot of $\rho^+(\vec k,\omega)$ at $\lambda=2$]{
\includegraphics[width=0.38\textwidth]{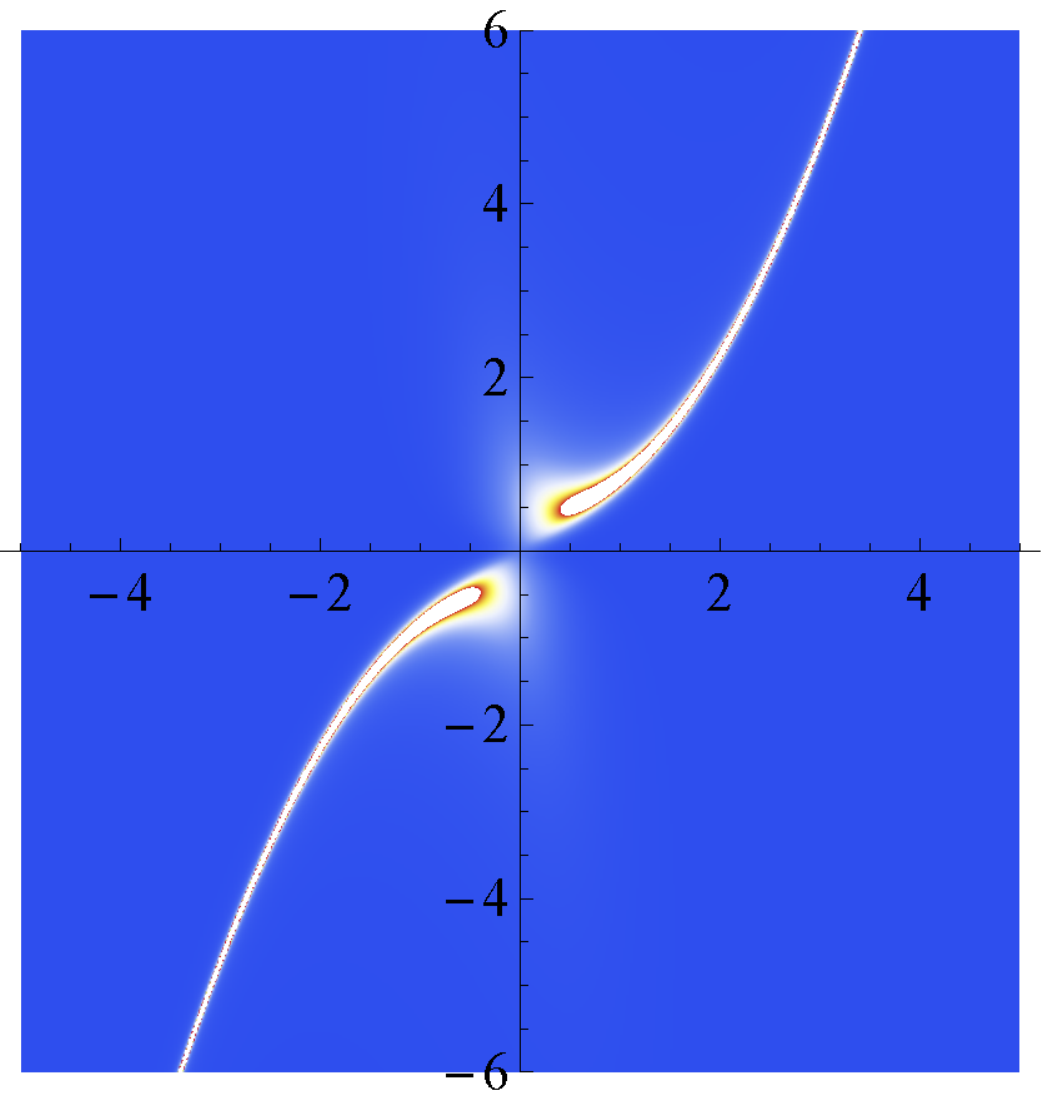}
\begin{picture}(0,0)
\put(-2,80){\scriptsize $k_3$}
\put(-82,165){\scriptsize $\omega$}
\end{picture}
}
\hspace{\stretch{1}}
\subfigure[Density plot of $\rho^+(\vec k,\omega)$ at $\lambda=0.5$]{
\includegraphics[width=0.38\textwidth]{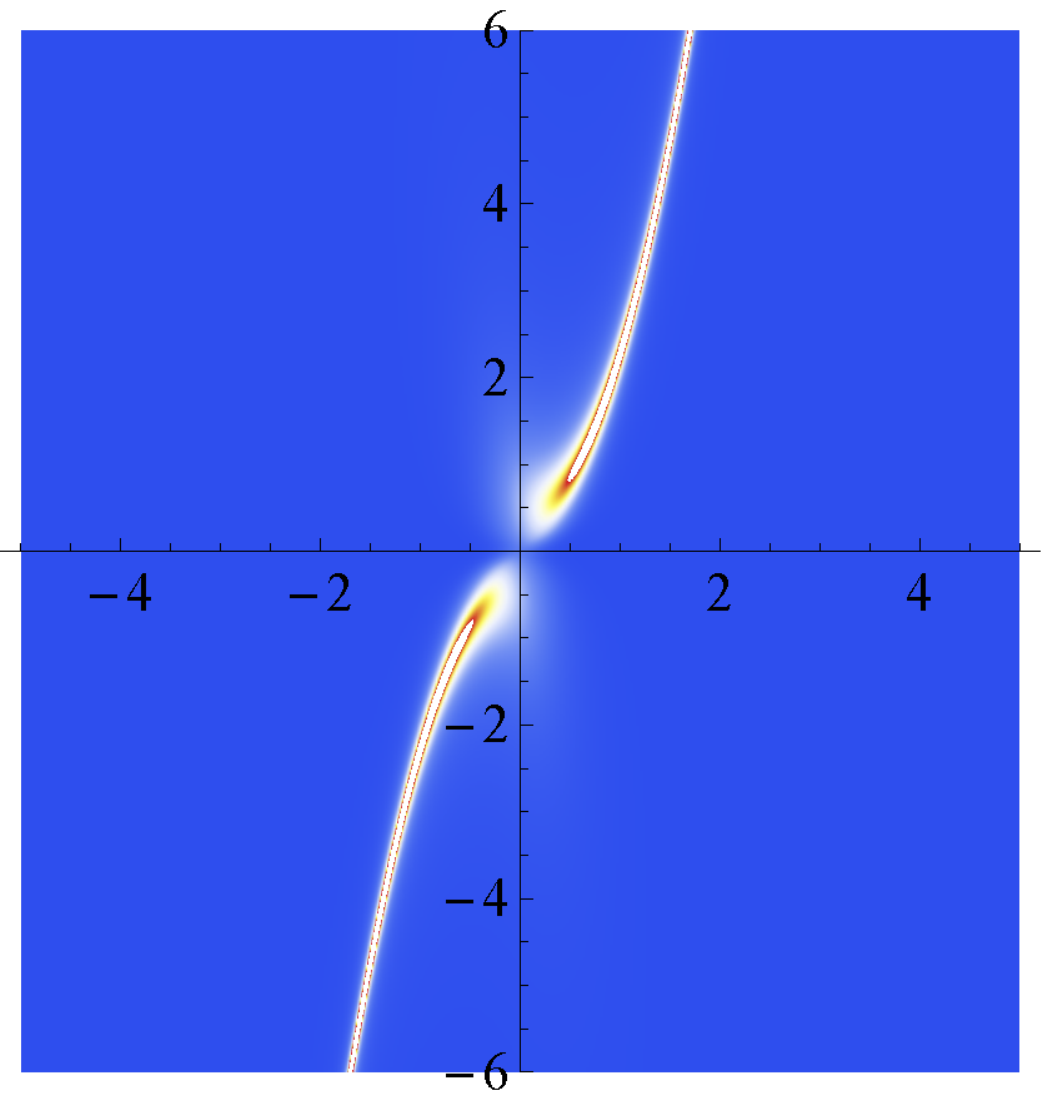}
\begin{picture}(0,0)
\put(-2,80){\scriptsize $k_3$}
\put(-82,165){\scriptsize $\omega$}
\end{picture}
}
\hspace{5pt} \\
\subfigure[Density plot of $\rho^+(\vec k,\omega)$ at $\lambda=0.1$]{
\includegraphics[width=0.38\textwidth]{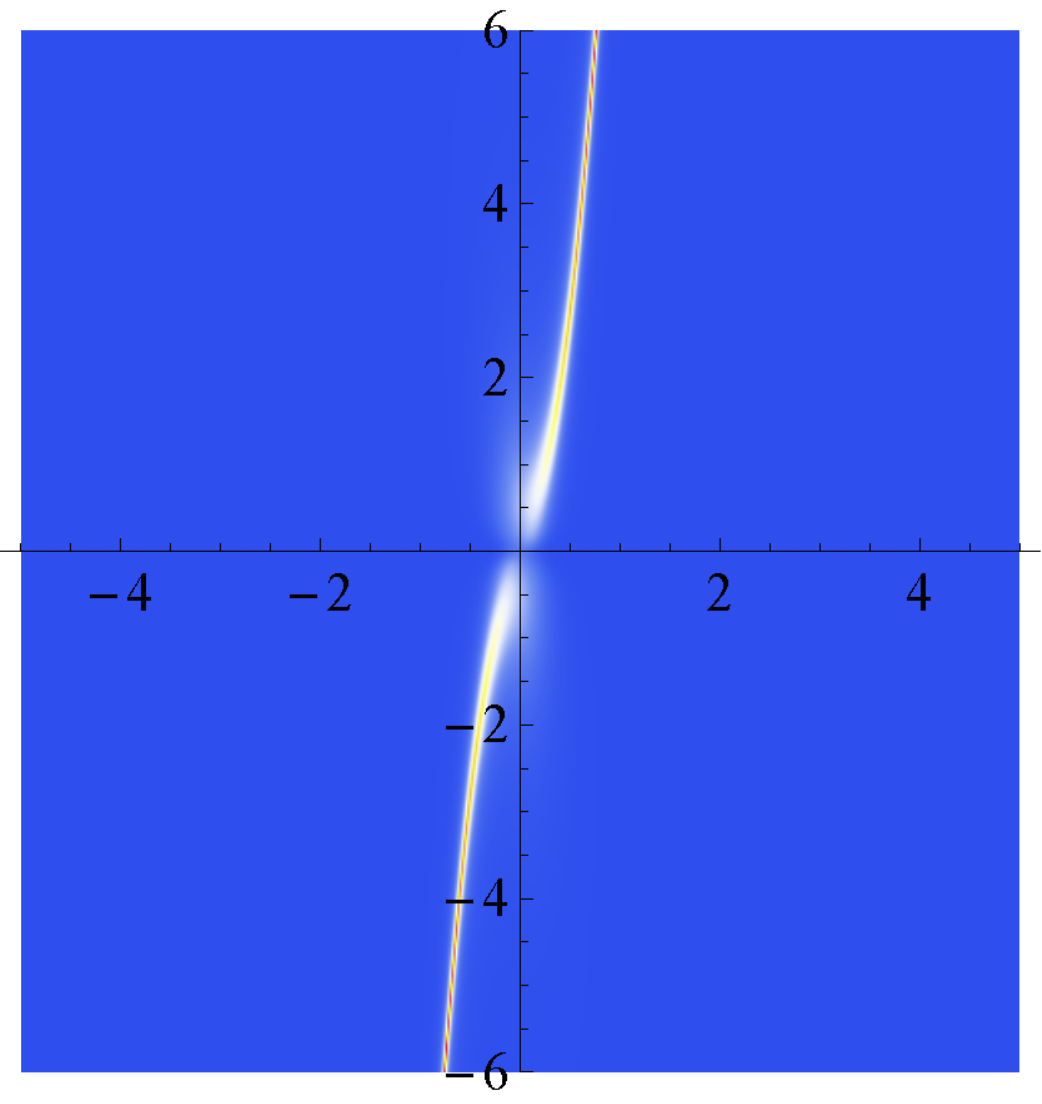}
\begin{picture}(0,0)
\put(-2,80){\scriptsize $k_3$}
\put(-82,165){\scriptsize $\omega$}
\end{picture}
}
\hspace{\stretch{1}}
\subfigure[Density plot of $\rho^-(\vec k,\omega)$ at $\lambda=-0.1$]{
\includegraphics[width=0.38\textwidth]{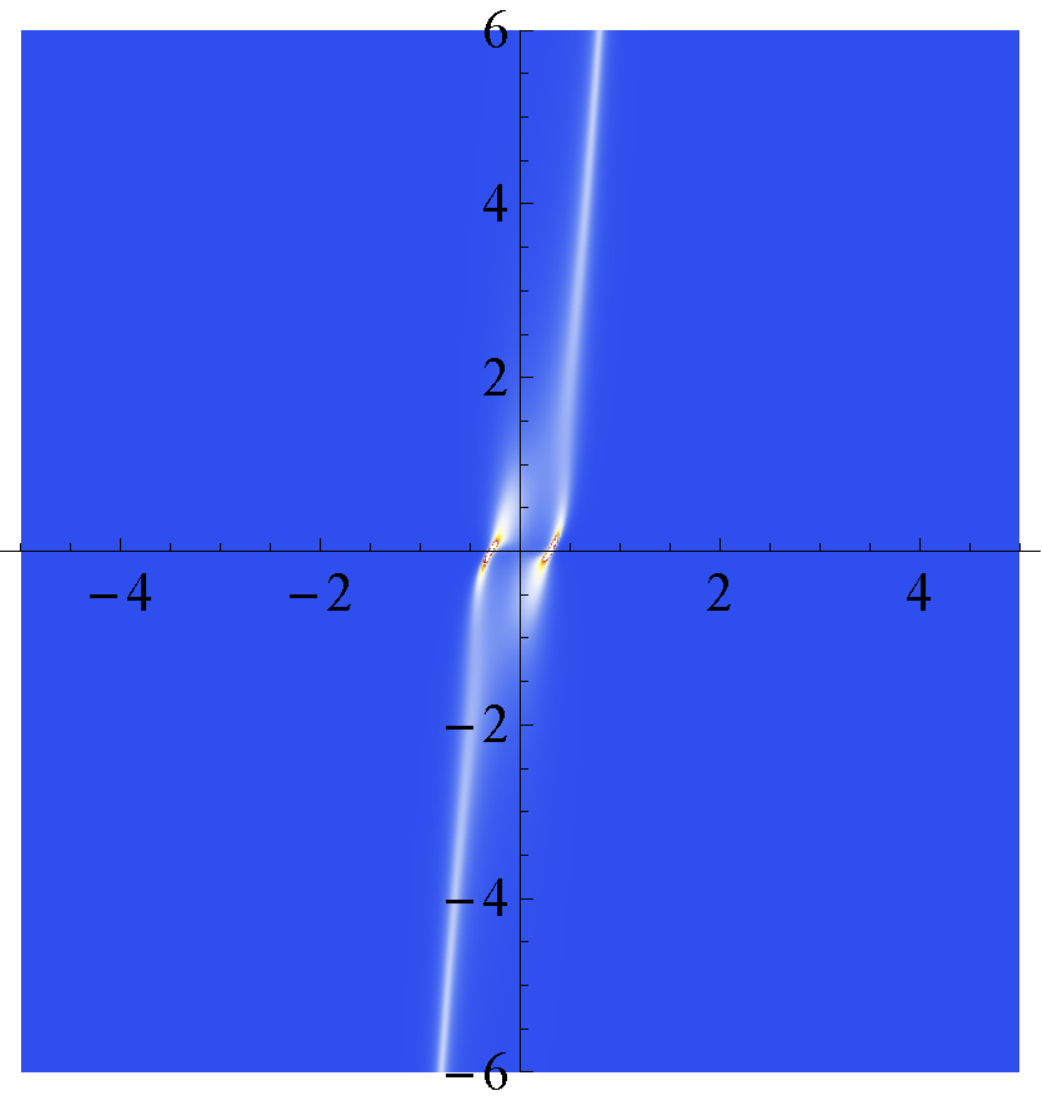}
\begin{picture}(0,0)
\put(-2,80){\scriptsize $k_3$}
\put(-82,165){\scriptsize $\omega$}
\end{picture}
}
\hspace{5pt} \\
\subfigure[Density plot of $\rho^-(\vec k,\omega)$ at $\lambda=-0.5$]{
\includegraphics[width=0.38\textwidth]{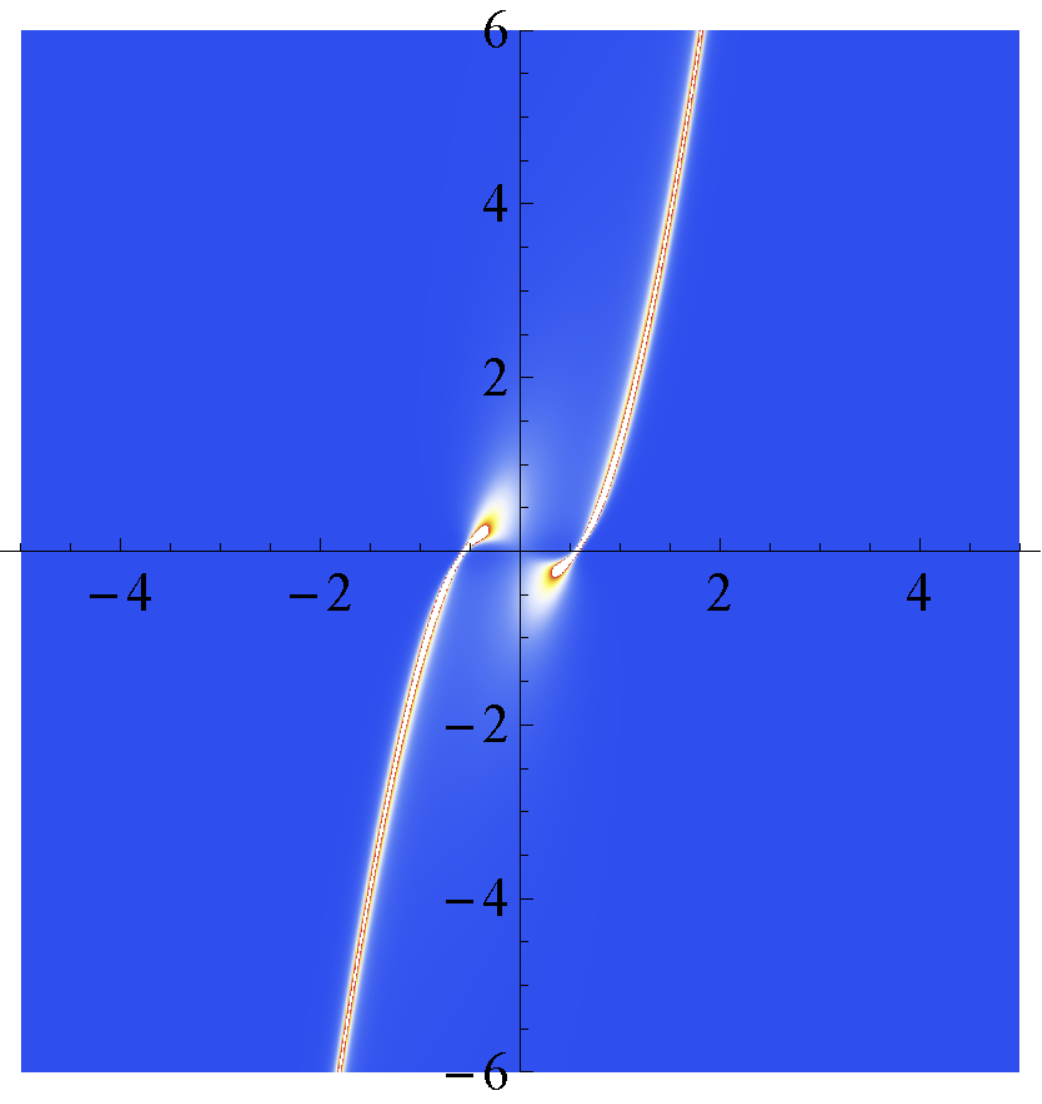}
\begin{picture}(0,0)
\put(-2,80){\scriptsize $k_3$}
\put(-82,165){\scriptsize $\omega$}
\end{picture}
}
\hspace{\stretch{1}}
\subfigure[Density plot of $\rho^-(\vec k,\omega)$ at $\lambda=-2$]{
\includegraphics[width=0.38\textwidth]{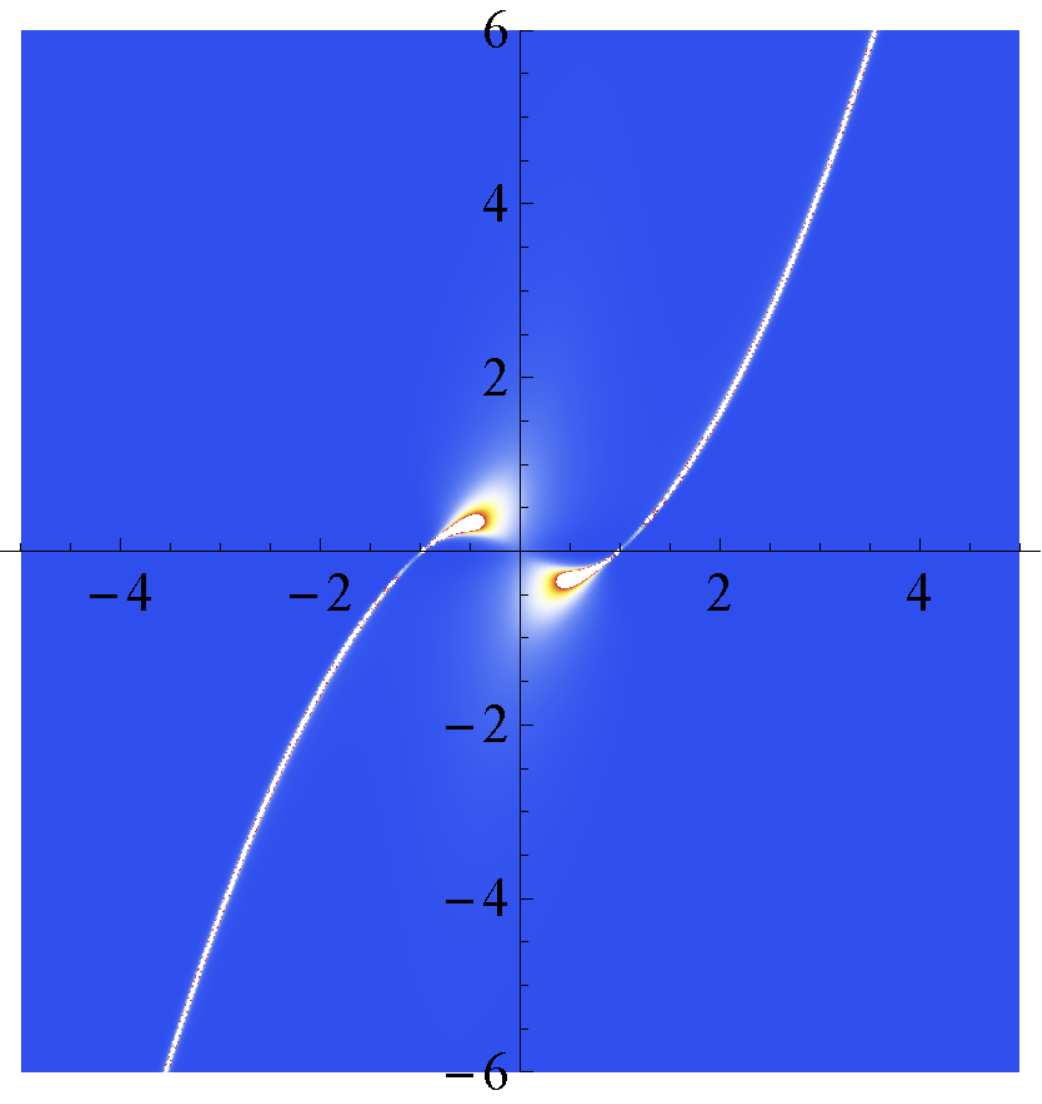}
\begin{picture}(0,0)
\put(-2,80){\scriptsize $k_3$}
\put(-82,165){\scriptsize $\omega$}
\end{picture}
}
\hspace{5pt} \caption{Density plot of $\rho^{\pm}(\kv,\om)$ for $z=2$, $M=-1/4$, $T=1/30$, $g=1$.}
\label{figz2displambdaMneg}
\end{figure}
\begin{figure}[t]
\centering
\vskip18pt
\includegraphics[width=0.7\textwidth]{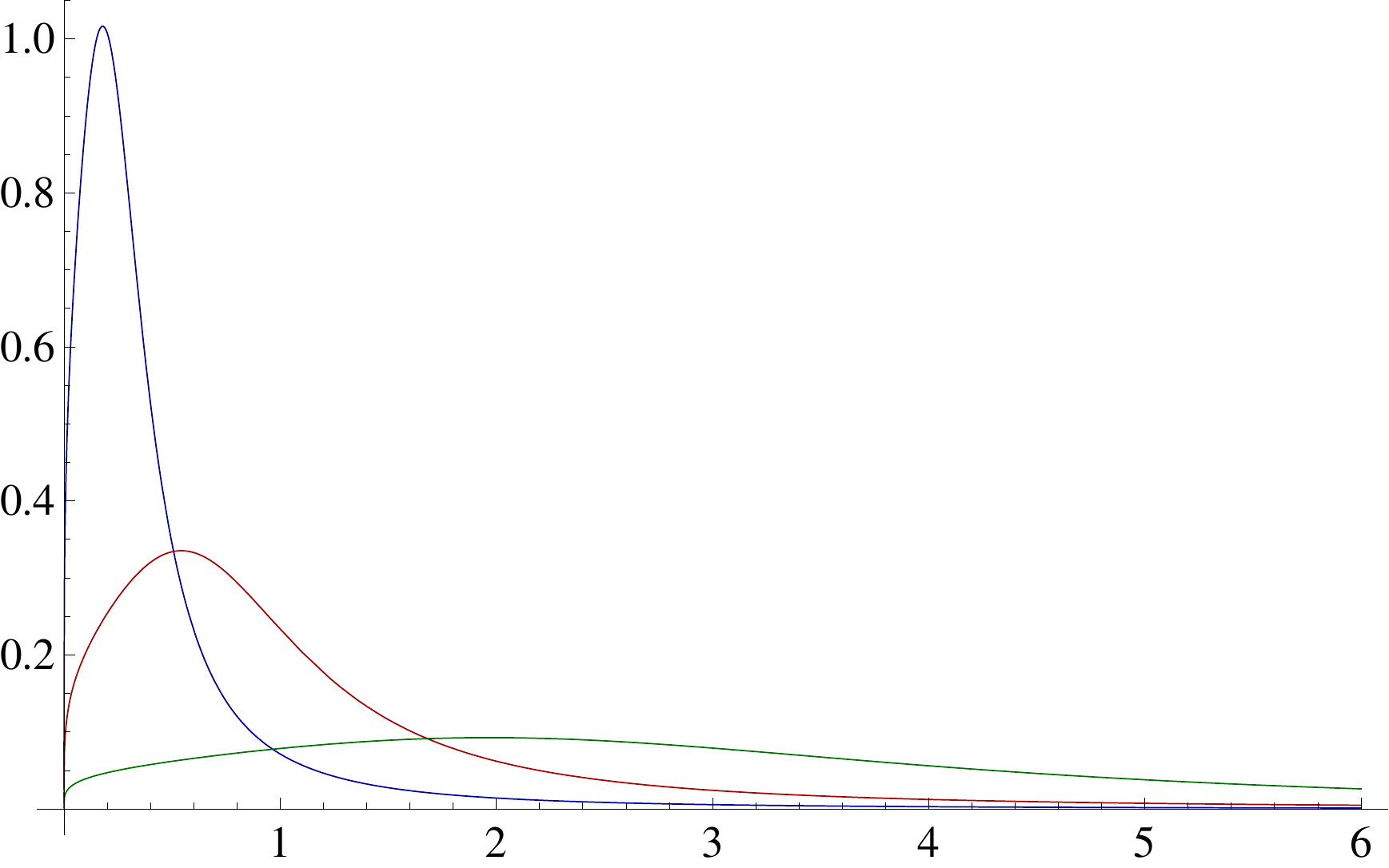}
\begin{picture}(0,0)
\put(0,8){\small $\omega$}
\put(-300,190){\small $\rho^{\pm}(0, \omega)$}
\put(-120,190){\footnotesize $k=0$, $T=0$, $M=-1/4$, $\mathcal{R}=0.4574$}
\put(-120,177){\footnotesize $g=0.25$ blue line}
\put(-120,164){\footnotesize $g=1$ red line}
\put(-120,151){\footnotesize $g=5$ green line}
\end{picture}
\caption{Spectral-weight function for vanishing momentum determined from \eqref{Gz}. Since the ratio of the location and the width of the peak remains constant ($\mathcal{R}=0.4574$ in the figure), there is no quasi-particle interpretation.
}
\label{figmassbump}
\end{figure}

To close our discussion about non-zero temperature effects in the case of $z=2$, let us consider the strict IR (or hydrodynamic) limit $\o\to 0$, $k\to 0$.
As explained in  appendix \ref{hydro},  in this limit the contribution from the free propagator  vanishes and the Green's function reduces to the inverse of the self-energy $\Sigma(p)$.
An analytical result for all values of the dynamical exponent $z$ and dimension $d$ and for non-zero temperature has been obtained for this case in equation
\eqref{GRhydro2}, which we recall here for convenience
\eq{\lab{GRhydro}
G_R(\vec 0,0) = \frac{i}{g}\:  2^{\frac{4M}{d+z-1}} \le( \frac{d+z-1}{4\pi }\ri)^{\frac{2M}{z}} T^{-\frac{2M}{z}}\;.
}


\subsubsection{Dispersion relation}
\lab{sec:z2disp}

We now consider the dispersion relation for $z\neq1$ at zero temperature, which can be derived from the general form of the Green's function given in
\eqref{Gzko}. In particular,  the dispersion relation is obtained by solving \eqref{Disp}, which in the present case reads
\eq{
\lab{disp5}
\omega -\frac{1}{\lambda}\; \sik\; k^{z-1} - g\;k^{2M}  \textrm{Re}\le( s_{1,M}\le(u\ri) +  \frac{ \sik}{k}\,s_{2,M}\le(u\ri) \ri)=0\;.
}
In the following we consider the upper-spin component for definiteness.
The latter means that  $\sik/k$ is replaced by $\textrm{sign}(k_3)$, where we again employed the rotational symmetry to align the momenta in the $z$-direction.
For positive (negative) momentum $k_3$ the dispersion relation then becomes
\eq{
\lab{disp6}
\omega \mp \frac{1}{\lambda}\: k^{z} - g\,k^{2M}  \R \bigl(
s_{1,M}\le(u\ri) \pm  s_{2,M}\le(u\ri) \bigr)=0 \;.
}
In order to determine the qualitative form of the dispersion relation, we  study various limits of the equation above.
\begin{itemize}

\item We first consider the UV limit $\o\to\infty$, $k\to\infty$ for which there are the three distinct possibilities: $\o/k^z\to \infty$, $k^z/\o \to \infty$ and $\o/k^z\to{\rm const.}$
The first two  do not allow for a solution of \eqref{disp6}, but the third possibility leads to
\eq{
\lab{disp7}
\o \approx  \pm\frac{1}{\lambda} \,k^z \hspace{50pt} \text{for}\,\, k\gg 1
}
in the allowed mass range $-z/2<M<z/2$.
This result confirms our general picture that the interaction of the elementary fermion with the CFT is irrelevant in the UV, hence the dispersion becomes that of a free fermion.

\item Next, we consider the limit $k_3 \to 0^+$, $\omega>0$.
In the mass range \mbox{$-z/2<M<z/2$} we determine the dispersion curve from \eqref{Gz} as
\eq{
\lab{disp8}
\o  + g\, c_z \left| \cos\Bigl[ \pi \bigl( \textstyle {\frac{M}{z}} + \tfrac{1}{2} \bigr) \Big]\right|
\, \textrm{sign}(M)\; \o^{\frac{2M}{z}} = 0 \;.
}
Noting that $g>0$ and $c_z>0$, we find that the only solution to this equation is $\o=0^+$ for $M>0$. On the other hand, for $M<0$ we obtain the only solution as
\eq{
\lab{node1}
\o =  \le(g\, c_z \left| \cos\Bigl[ \pi \bigl( \textstyle {\frac{M}{z}} + \tfrac{1}{2} \bigr) \Big]\right|\ri)^{\frac{z}{z-2M}}\equiv m_z \;, \hspace{40pt} M<0 \;.
}
These solutions correspond to the points where the dispersion curve touches the axis at $k=0$.

Studying the dispersion relation for the case of $k_3\to 0^+$ and $\o<0$, we find another solution at $\o = -m_z$, consistent with the particle-hole symmetry. This is the non-relativistic analog of the result we obtained  in equation \eqref{disp4}. As mentioned above, and similar to the case with $z=1$, the corresponding peaks in the spectral-weight function cannot be interpreted as a true gapped quasi-particle excitation. This is further illustrated by a plot of the analytic result in figure \ref{figmassbump}.

\item The most interesting case is when $0<M<1/2$ and $\lambda<0$. As one can see from figure \ref{figz2displambda}, this situation is quite different from $0<M<1/2$ and $\lambda>0$. The dispersion curve $\omega(\vec k)$ for $\lambda<0$ crosses the $\omega=0$ axis three times, instead of only once at $\omega=k=0$. It therefore seems that two Fermi surfaces appear at the momenta for which $\omega(\pm  k_F)=0$. This signals a phase transition as one crosses from positive to negative $\lambda$. The remainder of this paper is devoted to a detailed study of this phase transition.

\end{itemize}


\section{Quantum phase transition}\lab{sect:QPT}

In figures \ref{figz2displambda} and \ref{figz2displambdaMneg} we have already seen that for $z=2$ the parameter $\lambda$ controls features of the spectral density significantly. In particular, when switching the sign of $\lambda$ the number of zeros of the dispersion relation $\omega(\vec k)$ changes, as becomes clear when comparing for instance  figures \ref{figz2displambda}(a) and \ref{figz2displambda}(f). In the present section, we study this phenomenon in more detail. For simplicity we focus on the case $0<M<1/2$ since the quantum phase transition discussed here is also present for $-1/2<M<0$. However, we also briefly consider the case $|M|>1/2$ in section \ref{sec:FSMrange}.


\subsection{Momentum distribution}

\begin{figure}[p]
\centering
\vskip-2pt
\subfigure[$\lambda=10$]{
\includegraphics[width=0.43\textwidth]{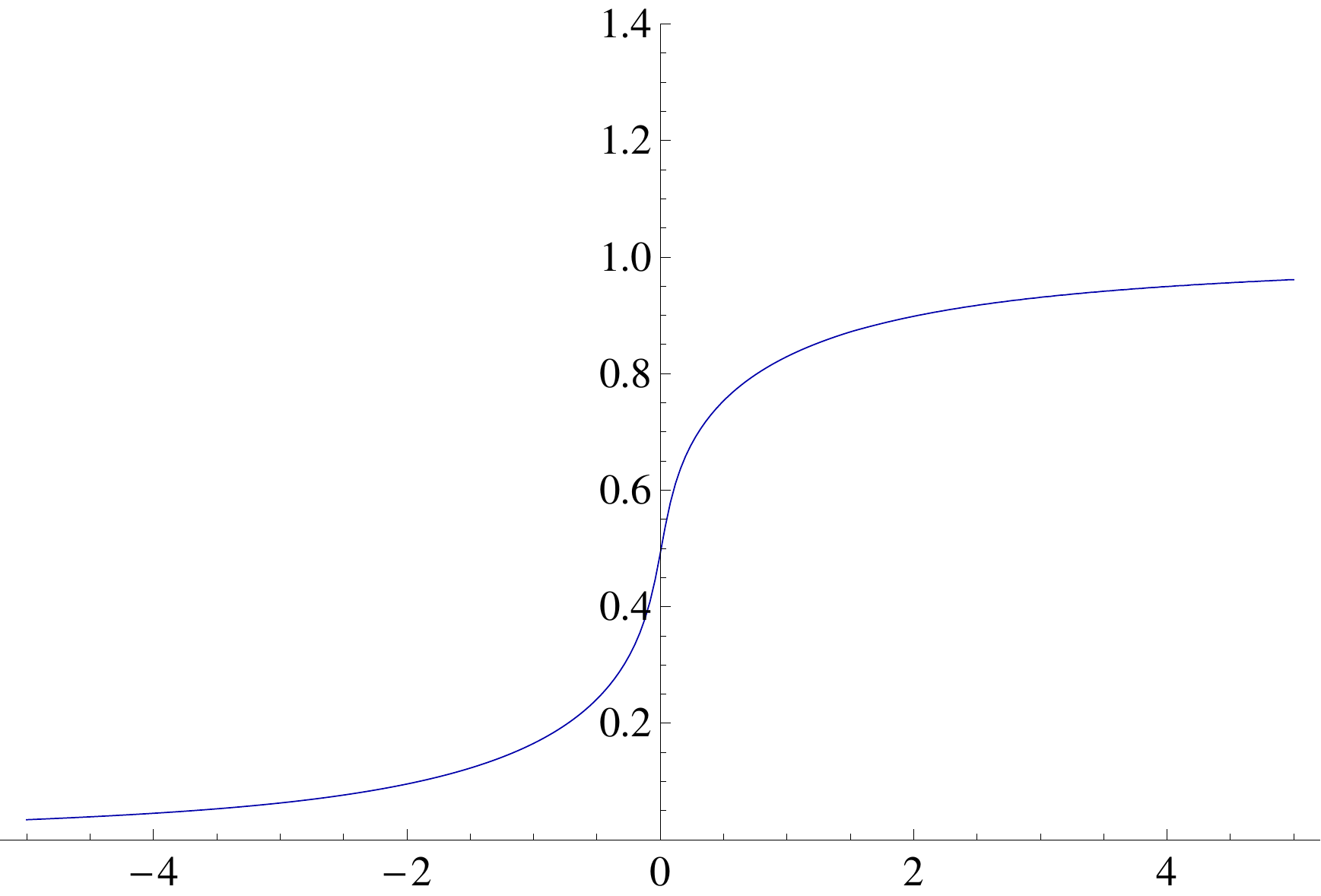}
\begin{picture}(0,0)
\put(-1,5){\scriptsize $k_3$}
\put(-100,128){\footnotesize $N^-_{\vec k}$}
\end{picture}
}
\hspace{\stretch{1}}
\subfigure[$\lambda=0.1$]{
\includegraphics[width=0.43\textwidth]{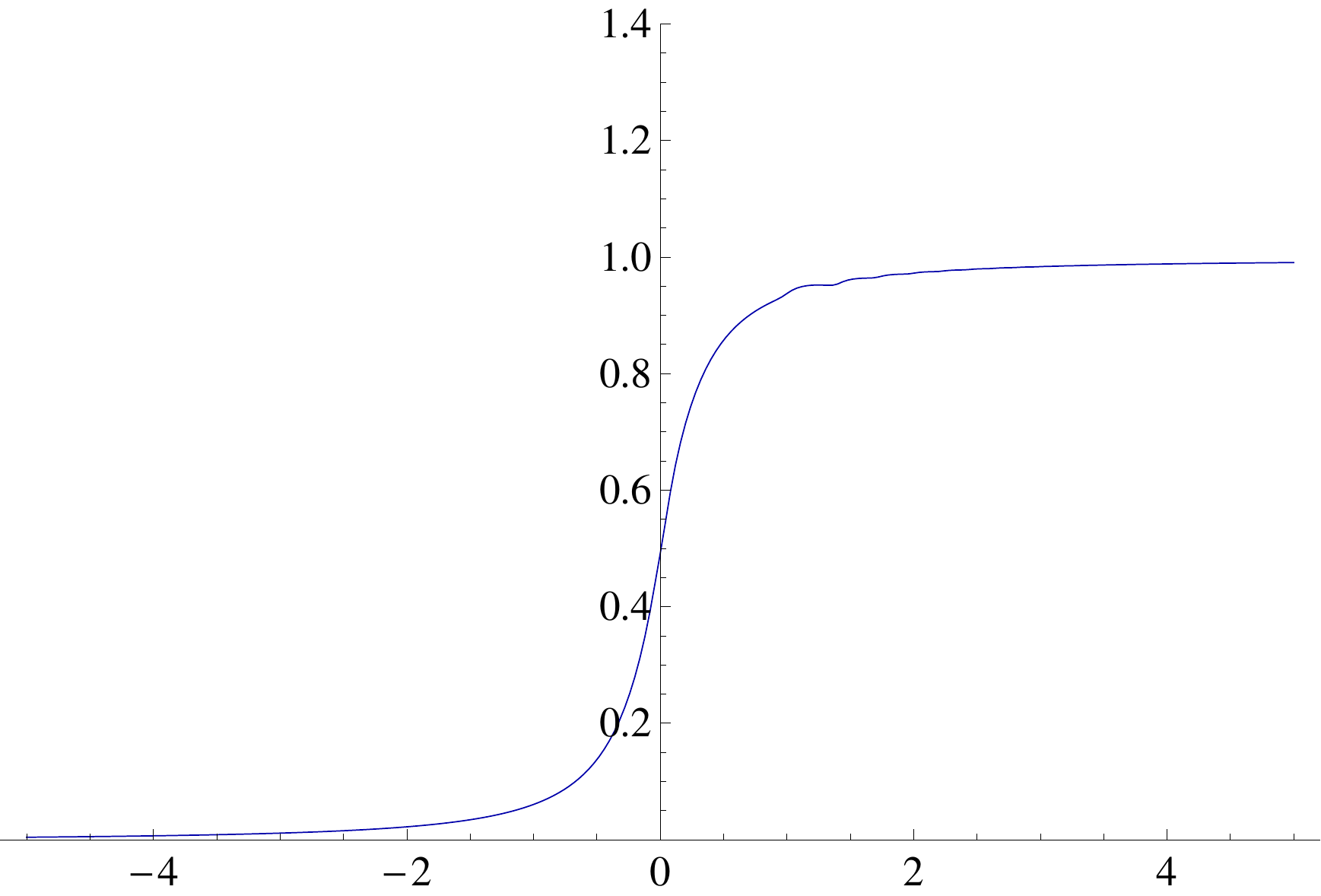}
\begin{picture}(0,0)
\put(-1,5){\scriptsize $k_3$}
\put(-100,128){\footnotesize $N^-_{\vec k}$}
\end{picture}
}
\hspace{5pt} \\
\vspace{10pt}
\subfigure[$\lambda=-0.1$, $k_F=0.35$]{
\includegraphics[width=0.43\textwidth]{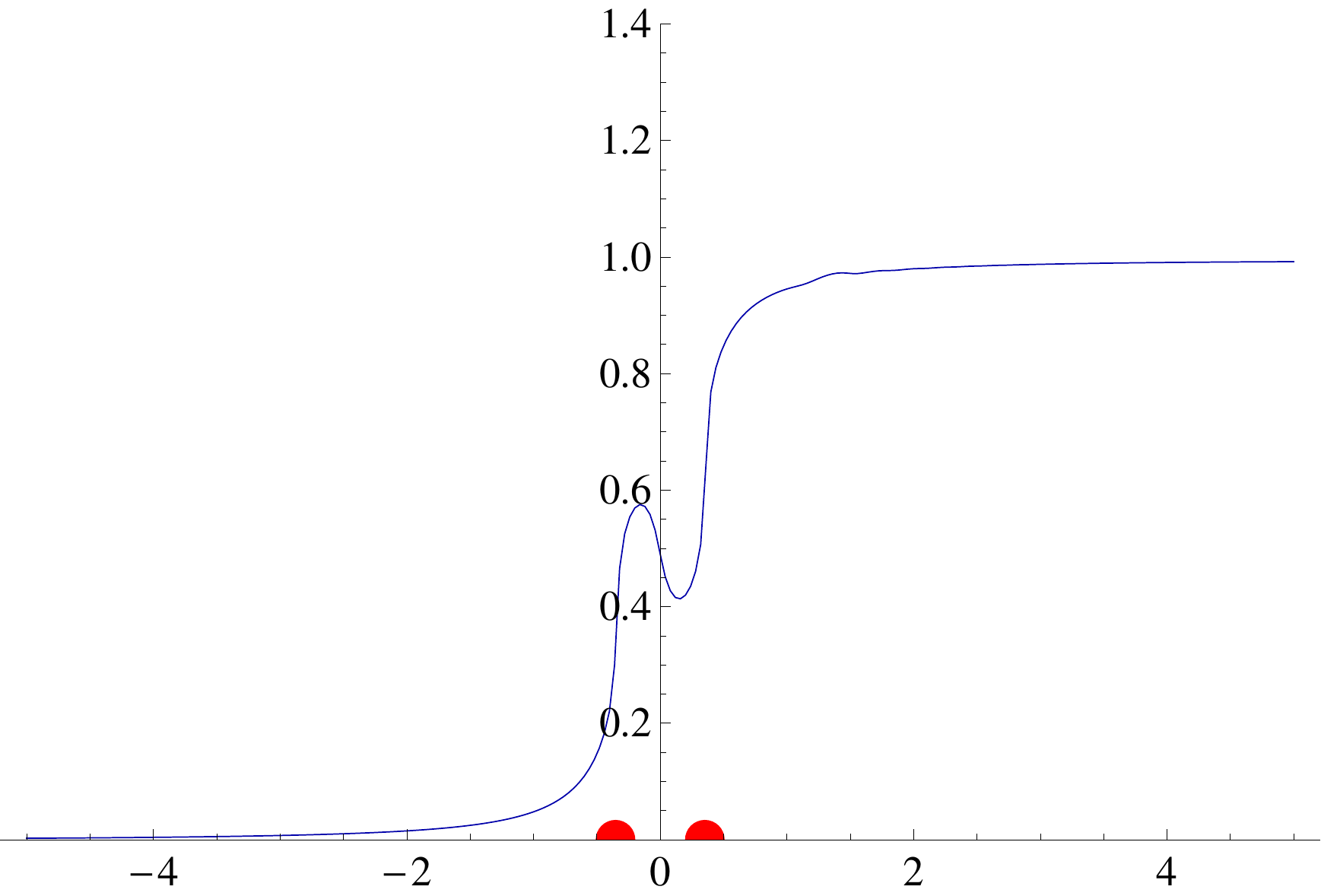}
\begin{picture}(0,0)
\put(-1,5){\scriptsize $k_3$}
\put(-100,128){\footnotesize $N^-_{\vec k}$}
\end{picture}
}
\hspace{\stretch{1}}
\subfigure[$\lambda=-0.5$, $k_F=1.03$]{
\includegraphics[width=0.43\textwidth]{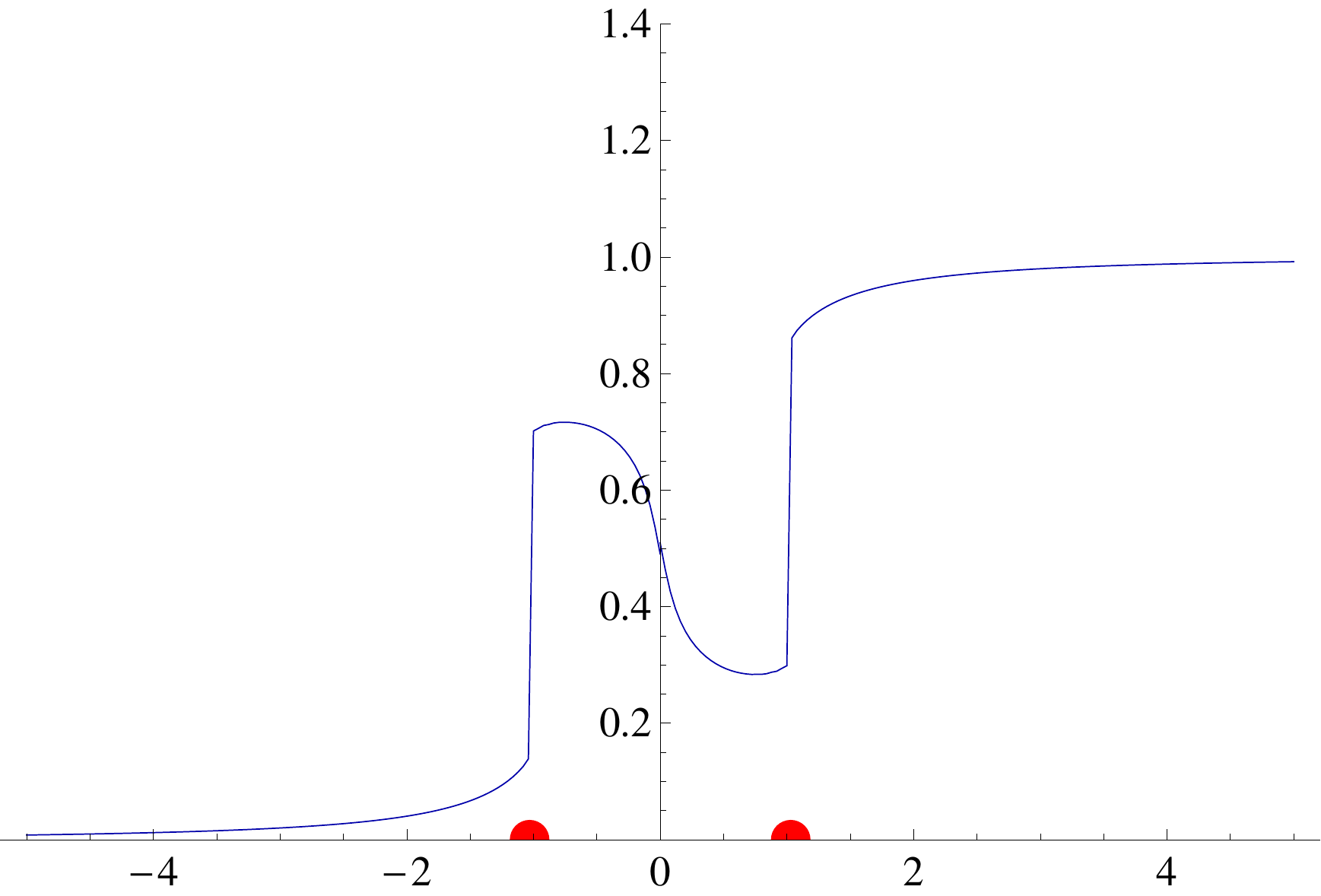}
\begin{picture}(0,0)
\put(-1,5){\scriptsize $k_3$}
\put(-100,128){\footnotesize $N^-_{\vec k}$}
\end{picture}
}
\hspace{5pt} \\
\vspace{10pt}
\subfigure[$\lambda=-1$, $k_F=1.63$]{
\includegraphics[width=0.43\textwidth]{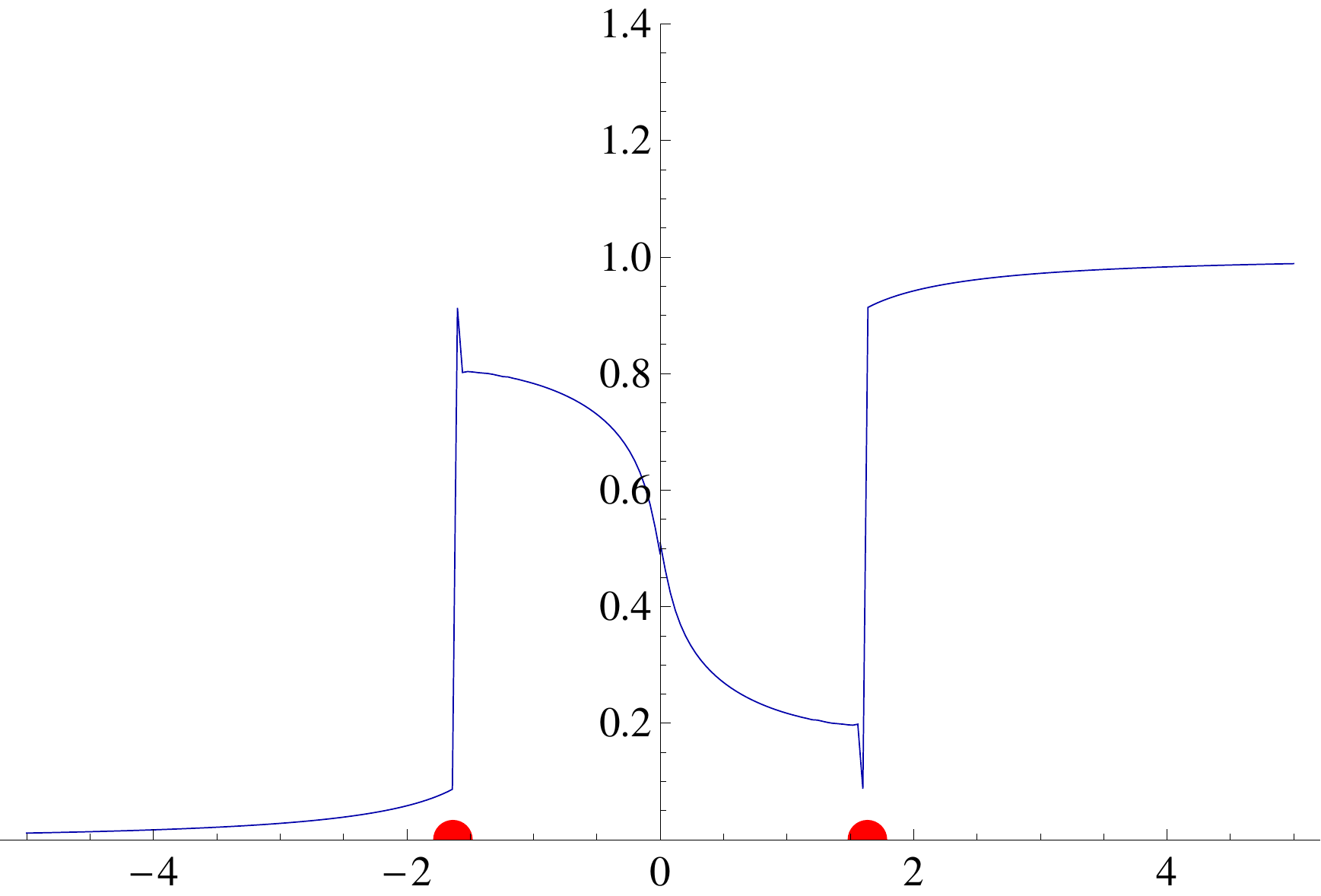}
\begin{picture}(0,0)
\put(-1,5){\scriptsize $k_3$}
\put(-100,128){\footnotesize $N^-_{\vec k}$}
\end{picture}
}
\hspace{\stretch{1}}
\subfigure[$\lambda=-100$, $k_F=35.2$]{
\includegraphics[width=0.43\textwidth]{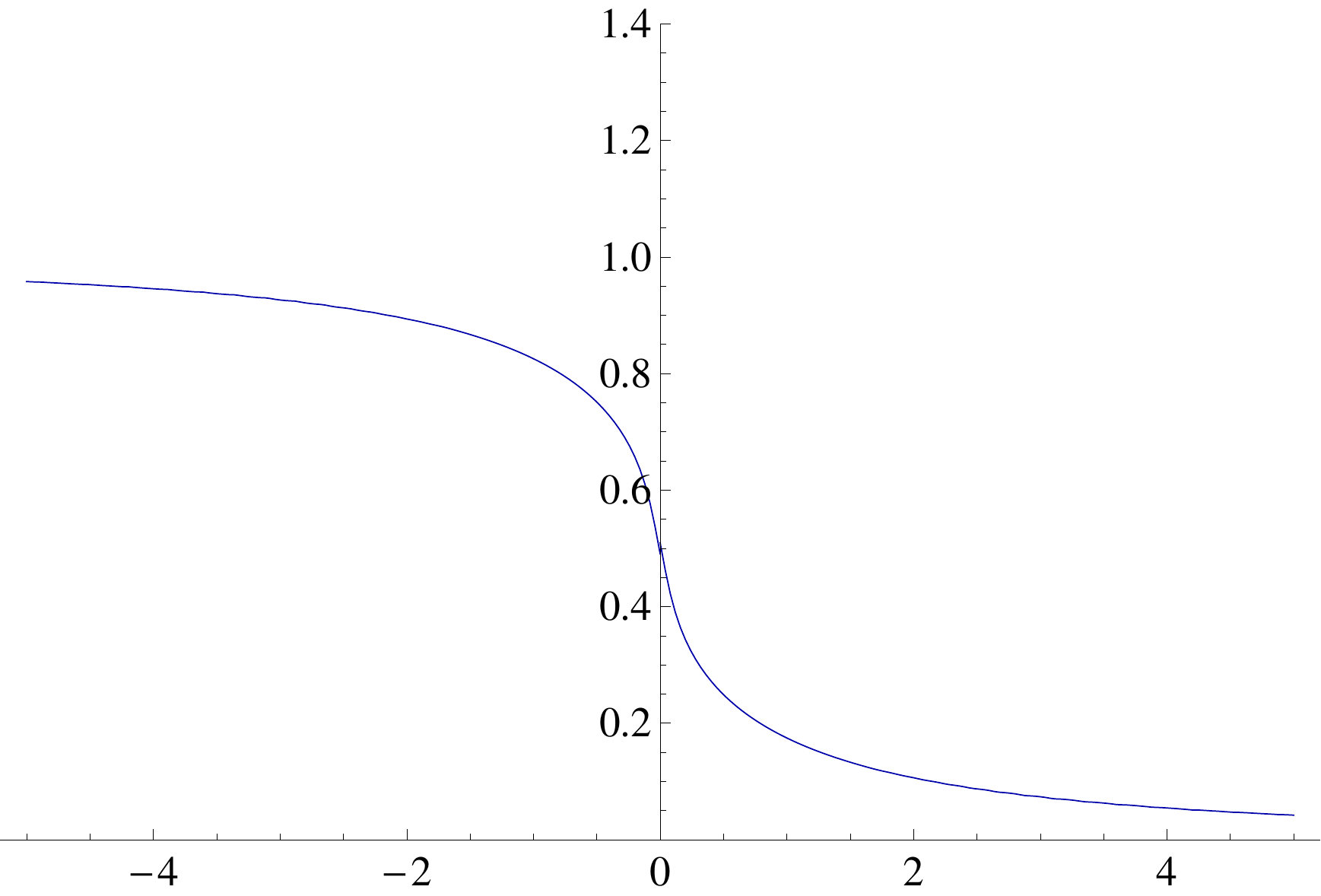}
\begin{picture}(0,0)
\put(-1,5){\scriptsize $k_3$}
\put(-100,128){\footnotesize $N^-_{\vec k}$}
\end{picture}
}
\hspace{5pt} \caption{Momentum distribution of the minus component for $M=+1/4$, $T=1/30$, $g=1$. For $\lambda >0$ it has a smooth kink-like behavior whereas for $\lambda <0$ two Fermi surfaces develop. The red dots give the analytic value \eqref{fermimom} of $k_F$. For small and negative $\lambda$, the Fermi surfaces are smeared out because of the non-zero value of the temperature.}
\label{figNkcombi}
\end{figure}
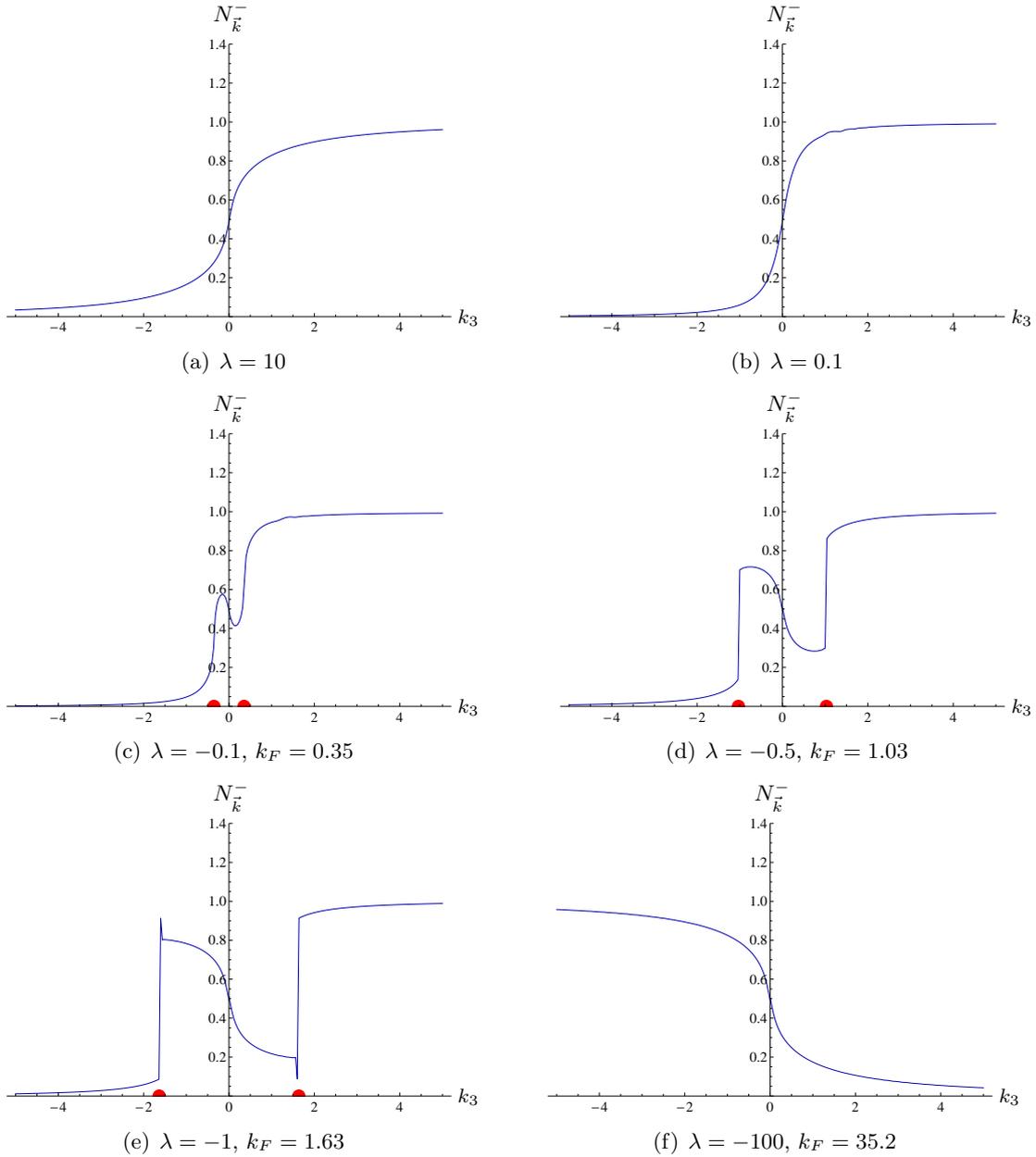

Let us start by defining the momentum distribution function. We choose again a convention such that the group velocity of a particle with one of the spin components always has the same sign at large momenta, irrespectively of the sign of $\lambda$. More concretely, with $n_F(\omega)$ the Fermi distribution we write
\eq{
\label{mom_dis_4}
N_{\kv}^{\pm} = \left\{\begin{aligned}
& \frac{1}{\pi} \int_{-\infty}^{\infty} d\om \;\I G_R^{\pm}(\om,\vec{k}) \;n_F(\om) \hspace{20pt}
&\text{for} \;\;\; \lambda >0\;,\\[2mm]
& \frac{1}{\pi} \int_{-\infty}^{\infty} d\om \;\I G_R^{\mp}(\om,\vec{k}) \;n_F(\om) &\text{for} \;\;\; \lambda <0\;.
\end{aligned}\right.
}
The behavior of \raisebox{0pt}[1mm]{$N^{\pm}_{\vec k}$} when  $\lambda$ changes sign is illustrated in figure \ref{figNkcombi} for non-zero temperatures, where we again employed the rotational symmetry to set \raisebox{0pt}[3mm]{$\vec k = (0,0,k_3)$}. In particular, for $\lambda<0$ the momentum distribution indicates a Fermi surface with a certain width, on which we comment later. When $\lambda$ switches sign, two extrema appear that develop into sharp discontinuities  as $\lambda$ increases. This is a clear signature of a Fermi surface. The locations of the jumps are determined analytically in the next section.

\subsection{Fermi momentum}\lab{sec:fs}

We now study analytically how the number of putative Fermi surfaces changes as we vary the parameter $\lambda$. To do so, we recall that Fermi surfaces are determined by the poles of the Green's function at vanishing frequencies $\o=0$. The latter can be computed by setting to zero the denominator in \eqref{GR2}.
Note also that the self-energy is given by \eqref{defsig} which satisfies the particle-hole symmetry derived from \eqref{symPH}. Employing the general form of the self-energy implied by  \eqref{Gzko}, we see that in the limit $\o\to0$ the imaginary part of $\Sigma(p)$ vanishes and the condition for the presence of a Fermi surface coincides with the presence of zeros of the dispersion relation \eqref{Disp}.
Thus, the loci of the Fermi surfaces are given precisely by the points where the dispersion curve $\o(\vec k)$ crosses the $\o=0$ axis.

We therefore consider equation \eqref{disp6} in the limit $\omega\to0$.
Using \eqref{flims1} and the expression \eqref{cz} for $c_1$,  we obtain the following formula for the loci of the Fermi surfaces
\eq{
\lab{fermi}
\frac{1}{\lambda}\; k^z + g\; k^{2M} c_1=0 \;.
}
Then, as we have mentioned in equation \eqref{posgot}, to avoid  violation of causality  we have to require $g >0$. Equation \eqref{fermi} can thus have non-trivial solutions in the range $-1/2 < M < 1/2$ only when $\lambda <0$, that is
\eq{
\lab{fermimom}
k =  k_F= \bigl(-g\hspace{1pt} \lambda \hspace{1pt}c_1\bigr)^{\frac{1}{z-2M}} \;.
}

Next, let us consider again the numerical results for the imaginary part of the Green's function shown in figure \ref{figz2displambda}. For $\lambda < 0$, there are zero-energy modes at non-vanishing momentum, suggesting that the system indeed has two Fermi surfaces at zero temperature.
However, due to the small but non-zero temperature which we have to employ in our  numerics, the spectrum shown in the plots has a finite width and the locations of the zero modes are approximately at the Fermi momentum \eqref{fermimom}. To investigate this point further and to confirm that we are indeed dealing with a genuine Fermi surface, in the following we compute  the quasi-particle weight $Z$ as a function of $\lambda$ and $g$, as well as the effective  mass and the lifetime at the Fermi surface as a function of $\om$, $T$, $g$ and $\lambda$.


\subsubsection{Quasi-particle residue}

In order to scrutinize the Fermi surfaces, in this subsection we determine the quasi-particle residue by linearizing the dispersion relation around the Fermi surface at $\om \approx0$, $ k\approx  k_F$.
For convenience, we choose the lower component of the Green's function which results in the spectral-weight function shown in figures \ref{figz2displambda}(d)-(f).
Furthermore, we employ the rotational symmetry to set $\kv=(0,0,k_3)$ with $k_3>0$, and thus $k_3=k$ in the following. We then compute up to first order in derivatives
\eq{
\label{lindisp}
 &- \R \le(G_R^{-}\ri)^{-1} \Big|_{k\approx k_F,\,\om \approx0} \\
 & \hspace{26pt}
 \approx \om\le (1-\partial_{\om} \R \Sigma^-(k_F,\om)\Big|_{\om=0}\ri)
 - \le(k-k_F\ri) \dau_{k}\le( \R \Sigma^-(k,0)-\frac{1}{\lambda}k^z\ri)\bigg|_{k=k_F}.\\
}
Using \eqref{lindisp} in the retarded Green's function, close to the Fermi surface we obtain the expression
\eq{
 \label{GRlin01}
G_R^-(k,\om) = \frac{-Z}{\om-Z (k-k_F)\, \dau_{k}\le( \R \Sigma^-(k,0)-\frac{1}{\lambda}k^z\ri)\big|_{k=k_F}-i \hspace{1pt}Z \I \Sigma^-(k,\o)} \;,
}
where the wavefunction renormalization factor, or quasi-particle residue $Z$, is given by
\eq{ \label{Z}
Z = \frac{1}{1 -  \dau_{\om} \R \Sigma^{\pm}(k_F,\om)\big|_{\om=0} } \;.
}
The residue is equal for both spin components because the $\om$ derivatives of $\R \Sigma^{\pm}$ are equal at $\om=0$.
Furthermore, note that $Z$ depends on the coupling $g$ explicitly due to the factor $g$ in $\Sigma(p)$, but also implicitly via the dependence of $k_F$ on $g$ shown in \eqref{fermimom}. The dependence  on $\lambda$ is only through $k_F$.
To be able to determine the quasi-particle residue, the real part of the self-energy has to be linear in $\o=0$, which is indeed the case, as shown in figure \ref{imsigma}. The first derivative of the real part of the self-energy is shown for $T=1/100$ in figure \ref{xideriv01}(a).

Let us illustrate the  calculation of $Z$ for a particular value of $\lambda$ and $g$, namely $\lambda=-0.5$ and $g=1$. For $T=1/30$, the first derivative of the real part of the self-energy has the non-zero value  $\partial_{\omega} \R \Sigma^+(k_F,0)\approx -0.8$ at $\omega=0$. This leads to a finite quasi-particle residue of about $Z \approx 0.56$, which is precisely the height of the step in the momentum distribution at $k_3=\pm k_F$, as shown in figure \ref{figNk}. Repeating this calculation for different values of $\lambda$, we have determined $Z(\lambda)$, which is shown in figure \ref{figZetaT}. Note that $0 \leq Z \leq 1$ as is required. Furthermore, the large deviation from 1 of the quasi-particle residue for small and negative $\lambda$, demonstrates that we are indeed describing a strongly interacting system.

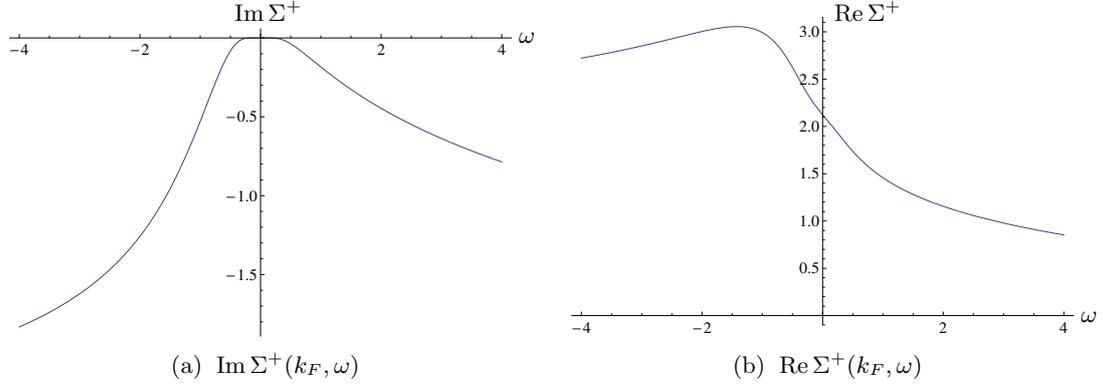
\begin{figure}[p]
\vskip10pt
\centering
\subfigure[$\I \Sigma^+(k_F,\omega)$]{
\includegraphics[width=0.45\textwidth]{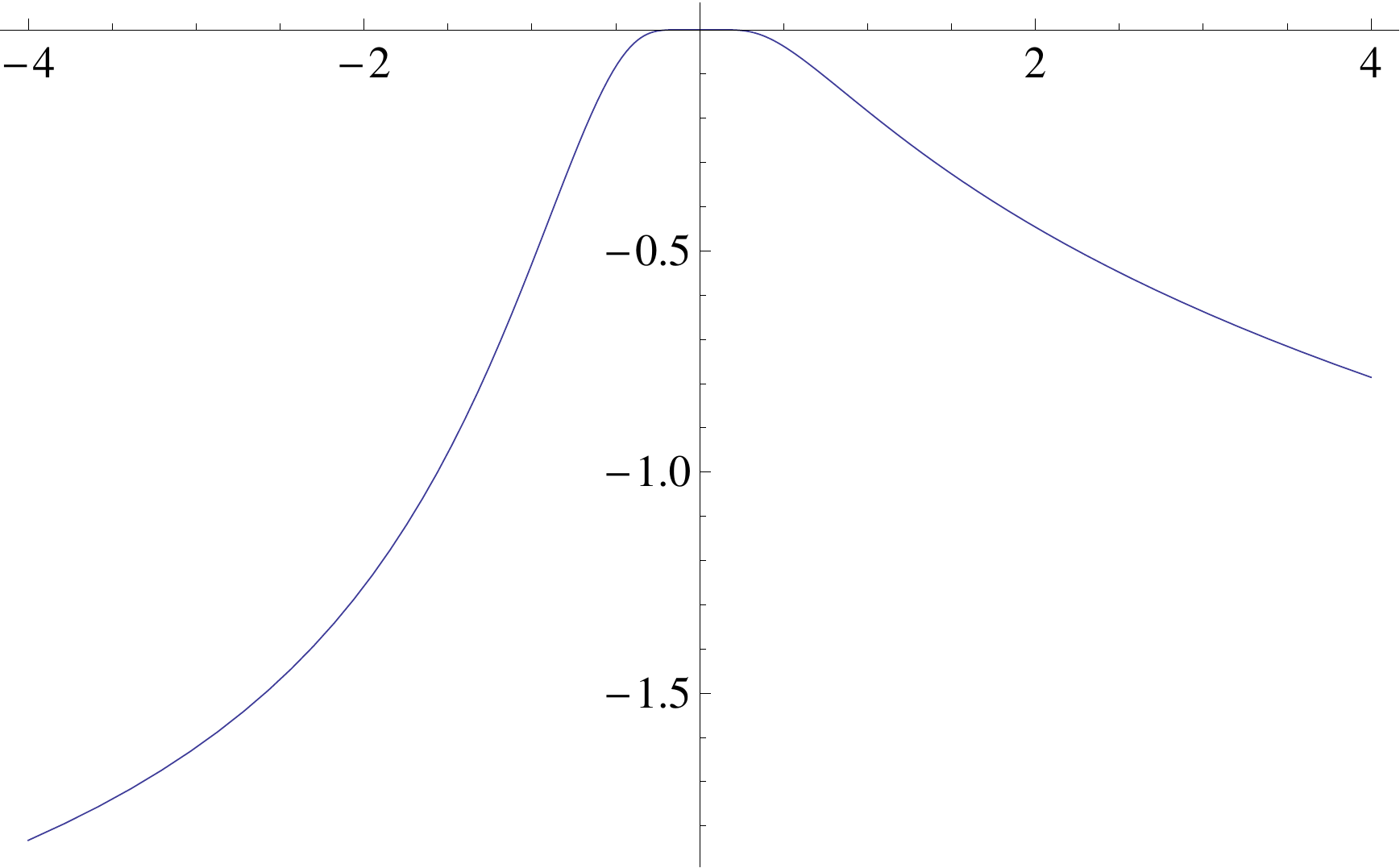}
\begin{picture}(0,0)
\put(-1,111){\footnotesize $\omega$}
\put(-109,119){\footnotesize $\I \Sigma^+$}
\end{picture}
}
\hspace{\stretch{1}}
\subfigure[$\R \Sigma^+(k_F,\omega)$]{
\includegraphics[width=0.45\textwidth]{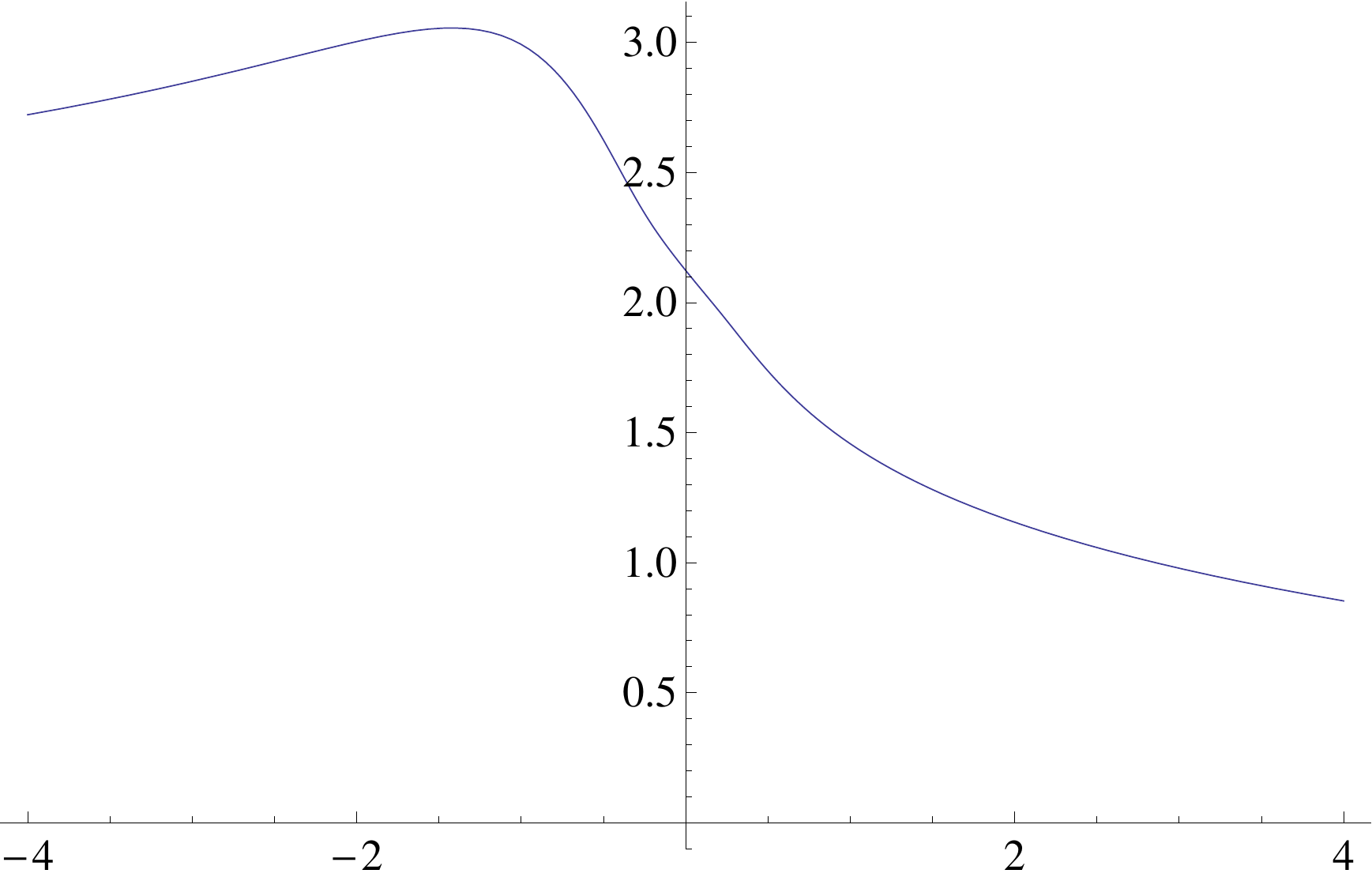}
\begin{picture}(0,0)
\put(-1,5.5){\footnotesize $\omega$}
\put(-95,119){\footnotesize $\R \Sigma^+$}
\end{picture}
}
\hspace{5pt}
\caption{Imaginary and real  part of the self-energy $\Sigma^+(p)$ evaluated at the Fermi momentum for $T=1/100$, $g=1$, $\lambda=-0.5$ and $M=1/4$. The imaginary part is zero around $\om=0$ and the real part is linear around $\om=0$, which are  both defining properties of a Fermi liquid.}
\label{imsigma}
\end{figure}
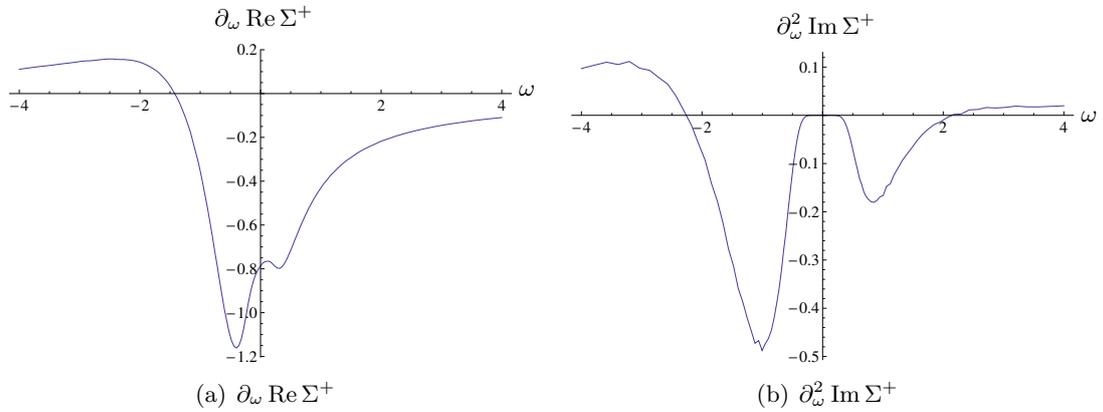
\begin{figure}[p]
\centering
\subfigure[$\partial_{\omega}\R \Sigma^+$]{
\includegraphics[width=0.45\textwidth]{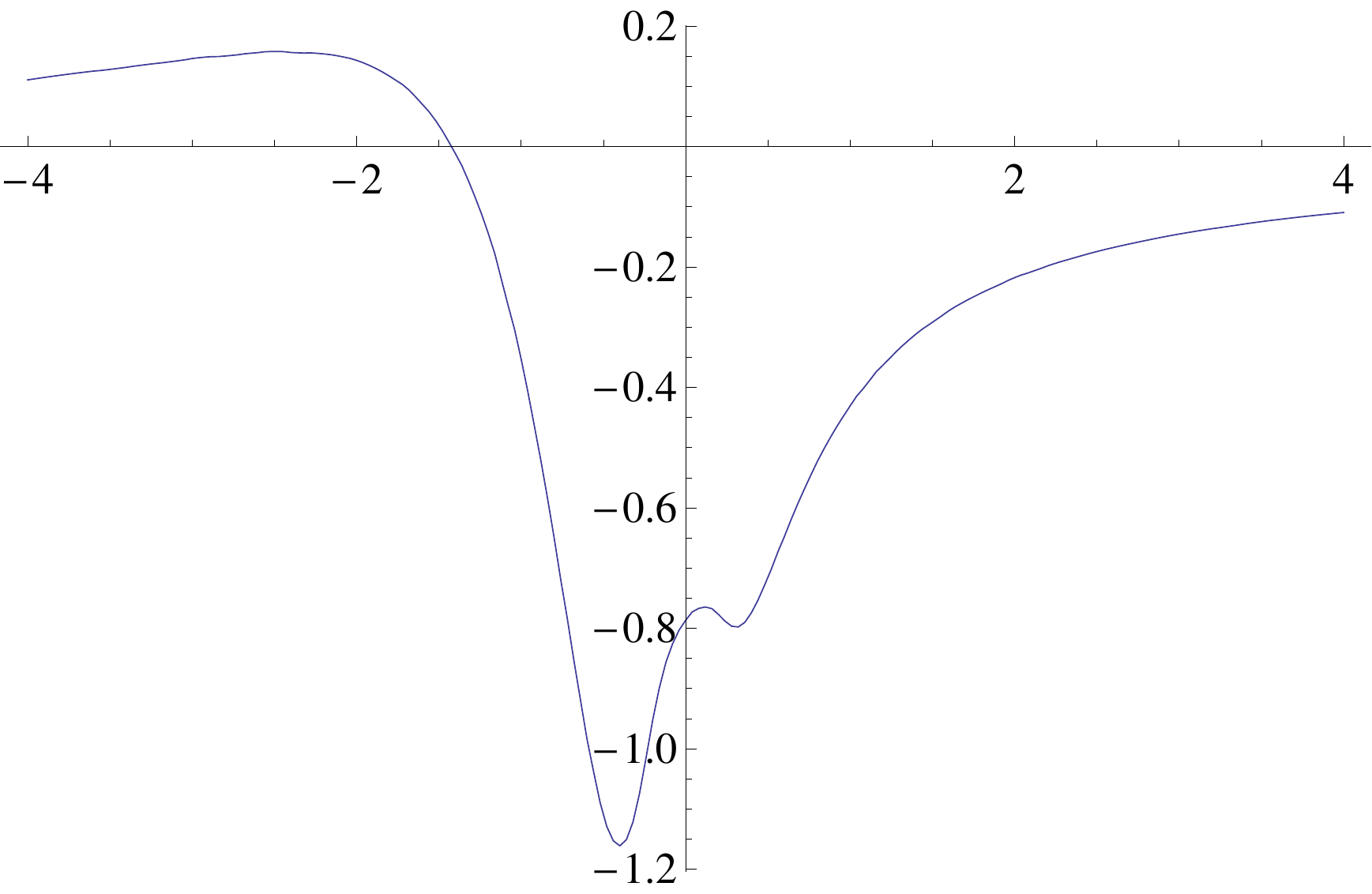}
\begin{picture}(0,0)
\put(-1,102){\footnotesize $\omega$}
\put(-115,128){\footnotesize $\partial_{\omega}\R \Sigma^+$}
\end{picture}
}
\hspace{\stretch{1}}
\subfigure[$\partial^2_{\omega}\I \Sigma^+$]{
\includegraphics[width=0.45\textwidth]{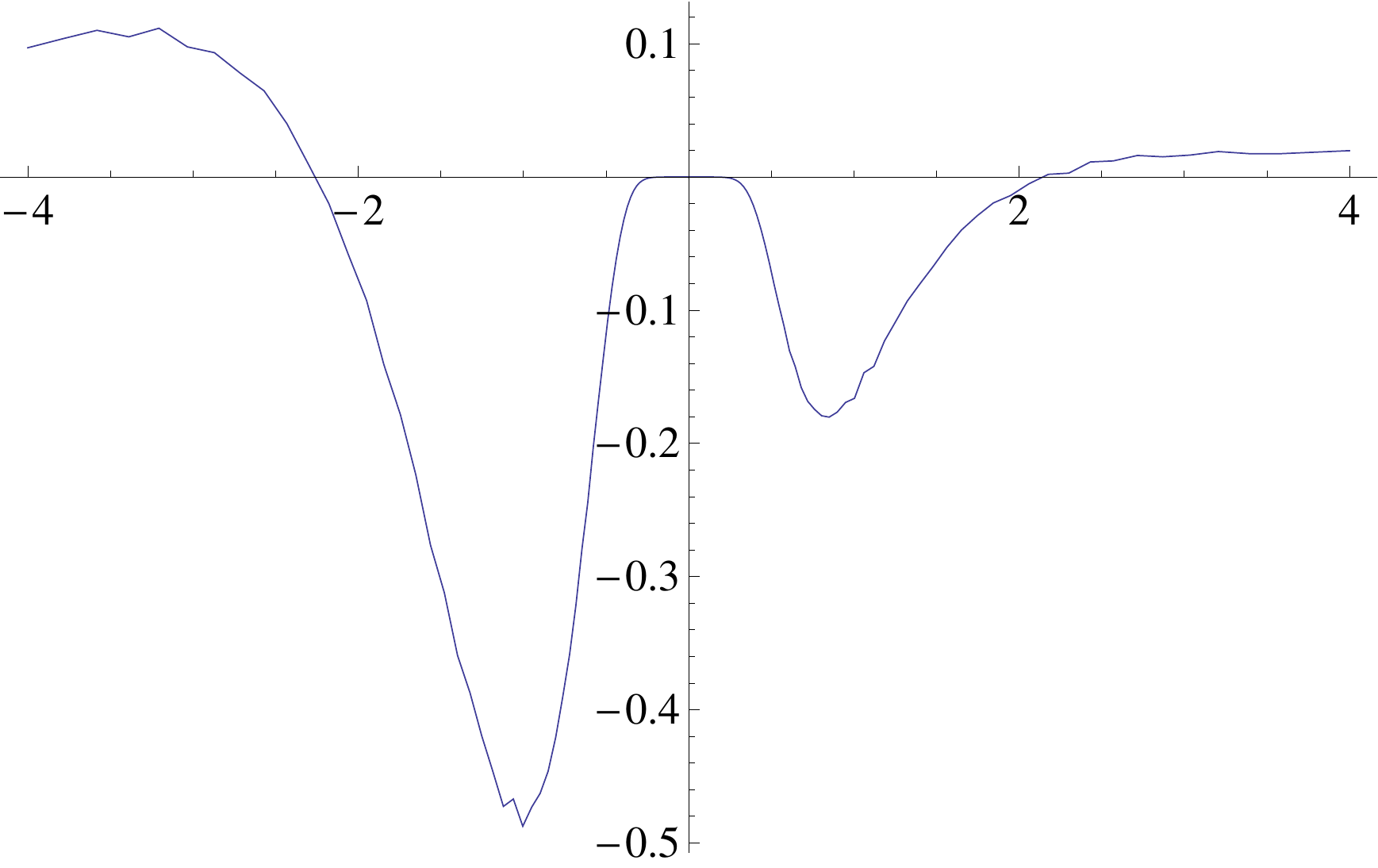}
\begin{picture}(0,0)
\put(-1,92){\footnotesize $\omega$}
\put(-115,125){\footnotesize $\partial^2_{\omega}\I \Sigma^+$}
\end{picture}
}
\hspace{5pt}
\caption{This figure shows the first derivative of $\R \Sigma^+(p)$ and the second derivative of $\I \Sigma^+(p)$, both at $\lambda=-0.5$, $T=1/100$, $M=1/4$ and $g=1$, evaluated at the Fermi momentum. }
\label{xideriv01}
\end{figure}

\begin{figure}[p]
\centering
\scalebox{0.8}{
\includegraphics[width=0.9\textwidth]{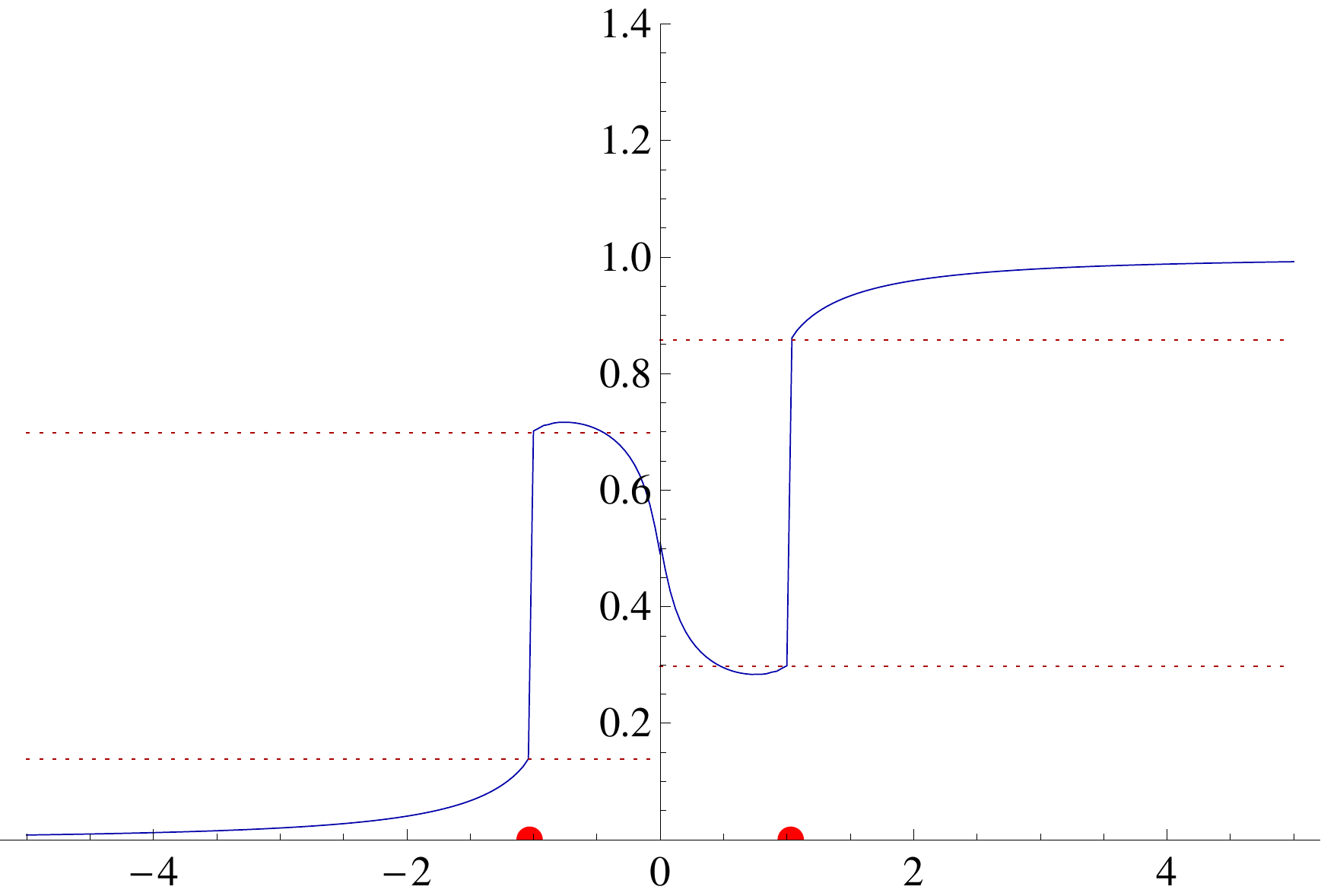}
\begin{picture}(0,0)
\put(-198,261){$N^-_{\vec k}$}
\put(0,12){$k_3$}
\put(-375,225){$k_F=1.0305$}
\put(-375,210){$Z=0.557$}
\put(-375,240){$\lambda=-0.5$}
\end{picture}
}
\caption{The momentum distribution for $M=1/4$, $T=1/30$ and $g=1$. The quasi-particle residue is given by the distance between the dotted lines: the difference in height is the numerical value of $Z$, which is 0.56 for $\lambda = -0.5$. The red dots give the analytic value of $\pm k_F$.}
\label{figNk}
\end{figure}
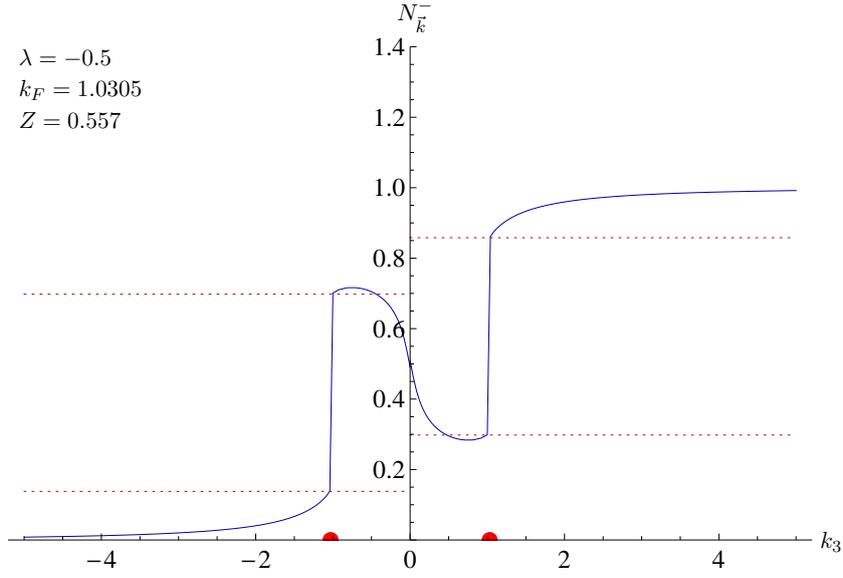

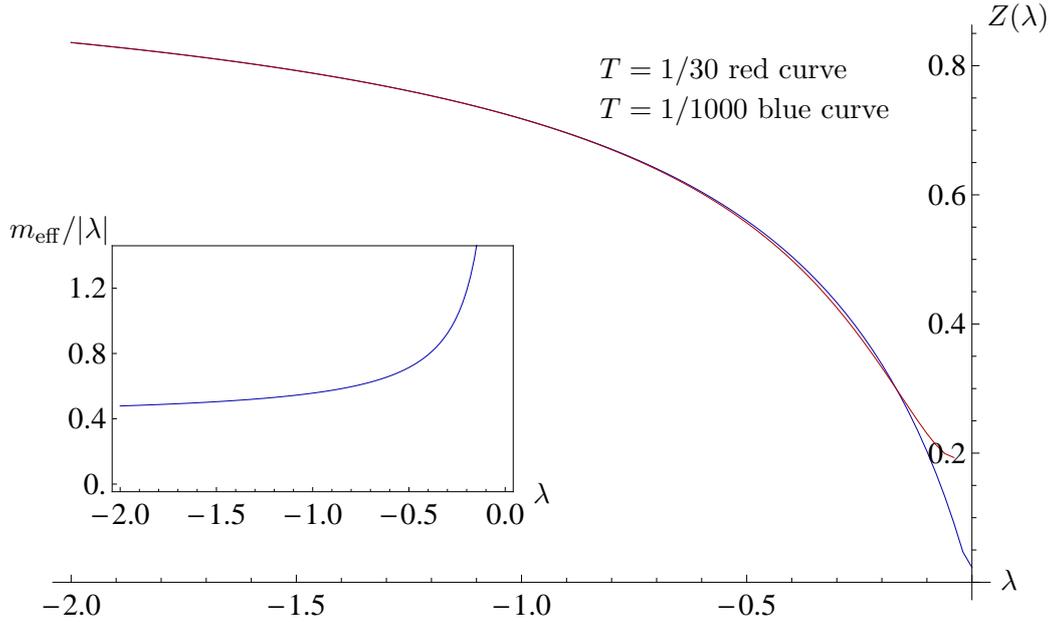
\begin{figure}[p]
\vskip10pt
\centering
\includegraphics[width=0.85\textwidth]{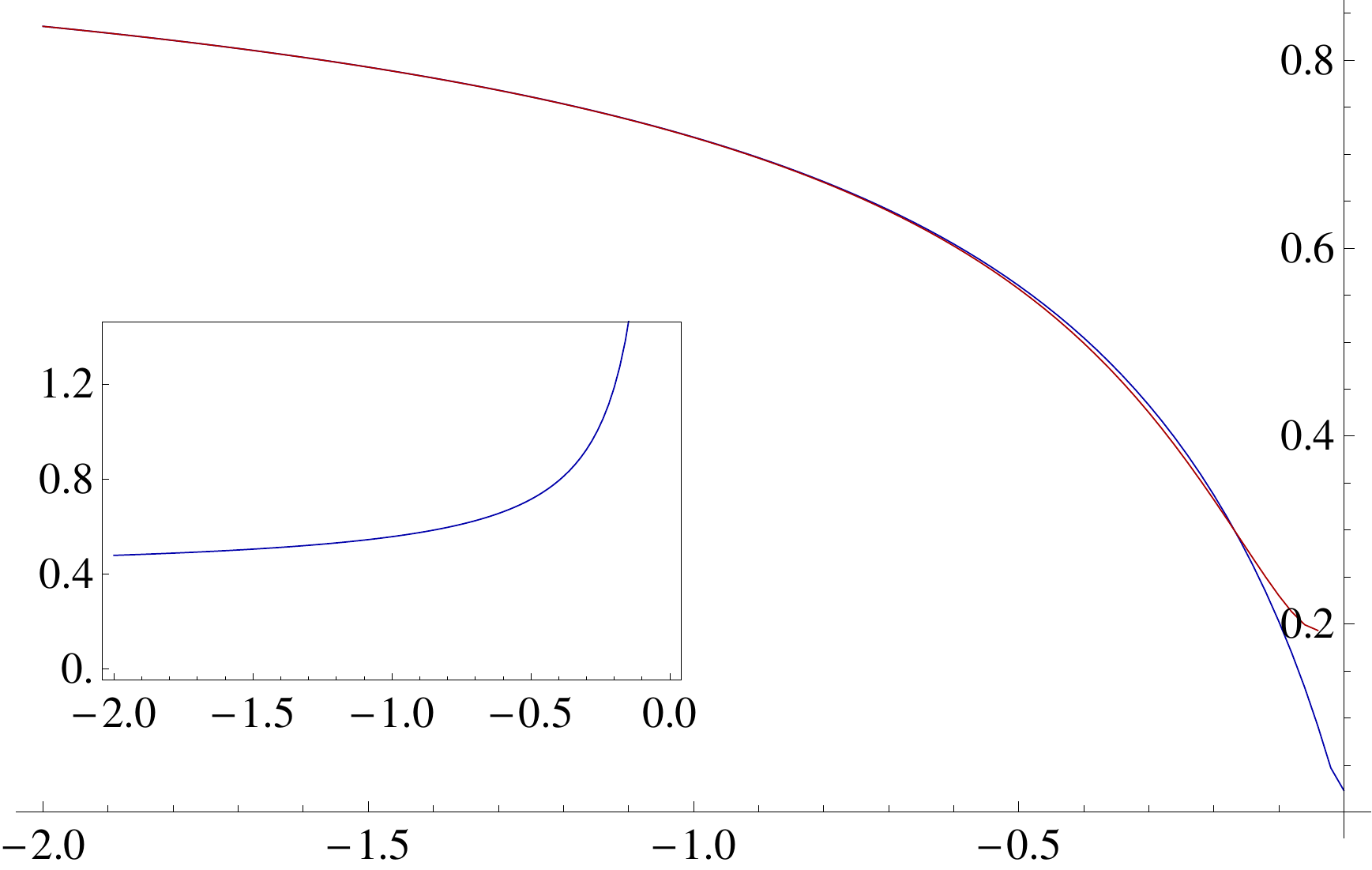}
\begin{picture}(0,0)
\put(0,12){$\lambda$}
\put(-5,225){$Z(\lambda)$}
\put(-150,205){$T=1/30$ red curve}
\put(-150,190){$T=1/1000$ blue curve}
\put(-175,45){$\lambda$}
\put(-371,145){$m_{\rm eff}/|\lambda |$}
\end{picture}
\caption{Wavefunction renormalization factor $Z$ as a function of $\lambda$ for $g=1$, $M=1/4$ and for different temperatures. $Z$ is only defined for negative $\lambda$. The quasi-particle weight lies between one and zero, as expected. It approaches unity for very large and negative values of $\lambda$, i.e., very far away from the phase transition. For $\lambda \to 0^-$ it vanishes only strictly at $T=0$. Due to non-zero temperature effects, the quasi-particle residue has a non-zero minimum at $\lambda=0$, which is  smaller for the lower temperature curve. The inset shows a plot of $m\eff/|\lambda|$ as a function of $\lambda$ for $M=1/4$.}
\label{figZetaT}
\end{figure}


\subsubsection{Effective mass}

In the last subsection we have seen that at the Fermi surfaces there are quasi-particles which are strongly renormalized. Now, to further characterize the Fermi liquid we should also compute their effective mass. The latter  is defined as the inverse of the slope of the quasi-particle dispersion around the Fermi surface, that is,
\eq{
\om(k) = \frac{1}{m\eff}\,(k-k_F)\,k_F  \hspace{40pt} \text{near the Fermi surface,}
}
where we are again employing $k_3>0$ and thus $k_3=k$.
Therefore, in our present situation the retarded Green's function near the Fermi momentum can be written as
\eq{
G^-_R(k,\om) = \frac{-Z}{\om- \frac{1}{m\eff}(k-k_F)k_F-i Z \I \Sigma^-(k,\o)} \;,
}
and by comparing with the explicit form given in  \eqref{GRlin01} we conclude that
\eq{
\frac{1}{m\eff} = \frac{Z}{k_F}\, \dau_k\le( \R \Sigma^-(k,0)-\frac{1}{\lambda}k^z\ri)\bigg|_{k=k_F}
\;.
}
At zero temperature, we can calculate the effective mass analytically. In particular, for $z=2$ we obtain from the exact result given in equation \eqref{Gzk} that
\eq{
  \R \Sigma^- (k,0) = g\hspace{1pt} c_1 k^{2M} \;, \hspace{40pt}
  \partial_k \R \Sigma^- (k,0) \bigl|_{k=k_F}  = -\frac{2M}{\lambda}\, k_F \;,
}
where we eliminated $g c_1$ in favor of $k_F$. Therefore, the effective mass for $z=2$ at zero temperature reads
\eq{
\label{meff}
m\eff =-\frac{\lambda}{2Z(\lambda)\le(M+1\ri)} \;.
}
The effective mass is shown in the inset in figure \ref{figZetaT}.

\subsubsection{Quasi-particle decay rate}\label{sec:lifetime}

\begin{figure}[t]
\centering
\vskip10pt
\subfigure[$T=1/20$]{
\includegraphics[width=0.45\textwidth]{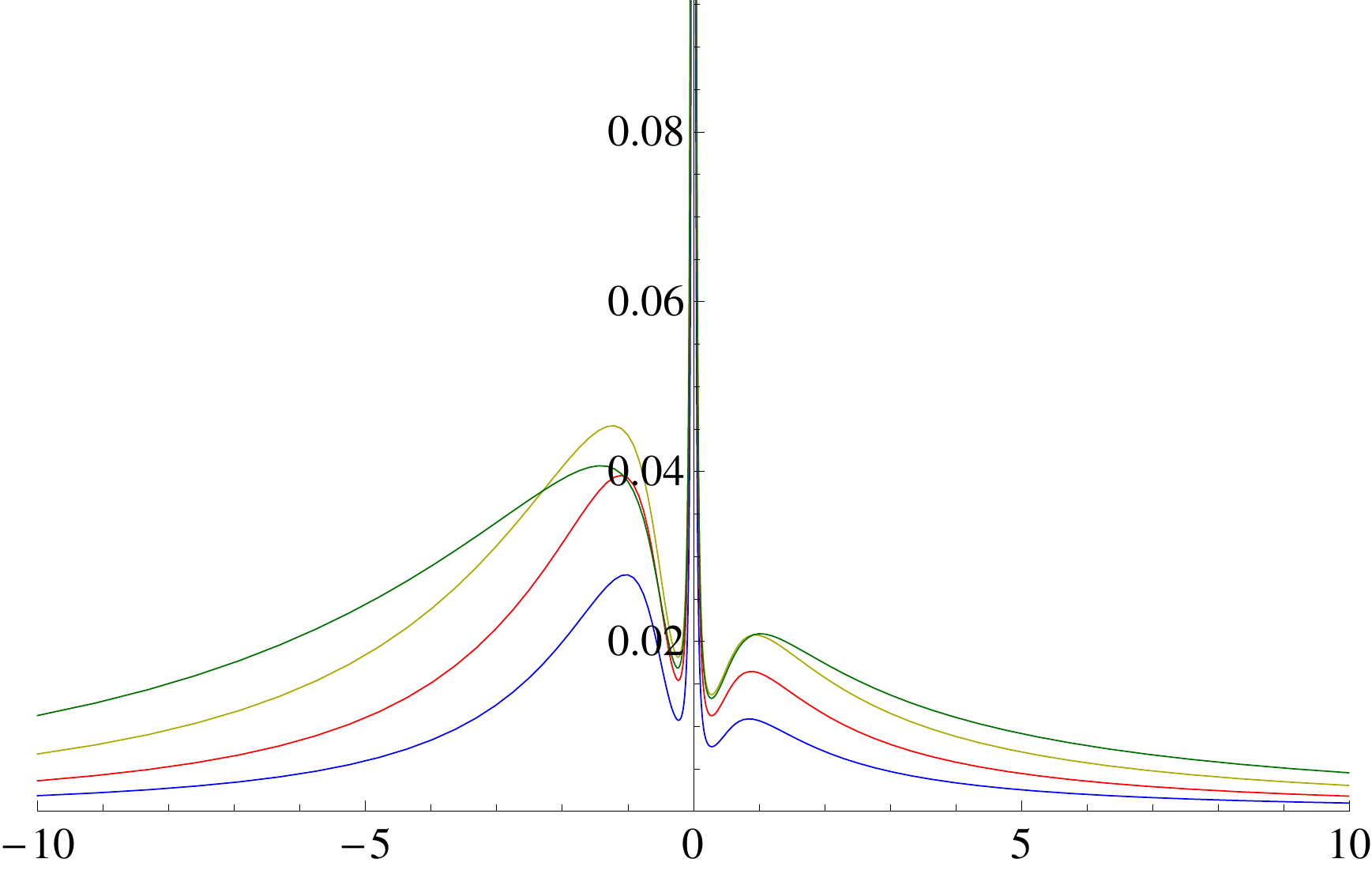}
\begin{picture}(0,0)
\put(-1,5){\scriptsize $\omega$}
\put(-100,124){\scriptsize $\rho^+$}
\end{picture}
}
\hspace{\stretch{1}}
\subfigure[$T=1/50$]{
\includegraphics[width=0.45\textwidth]{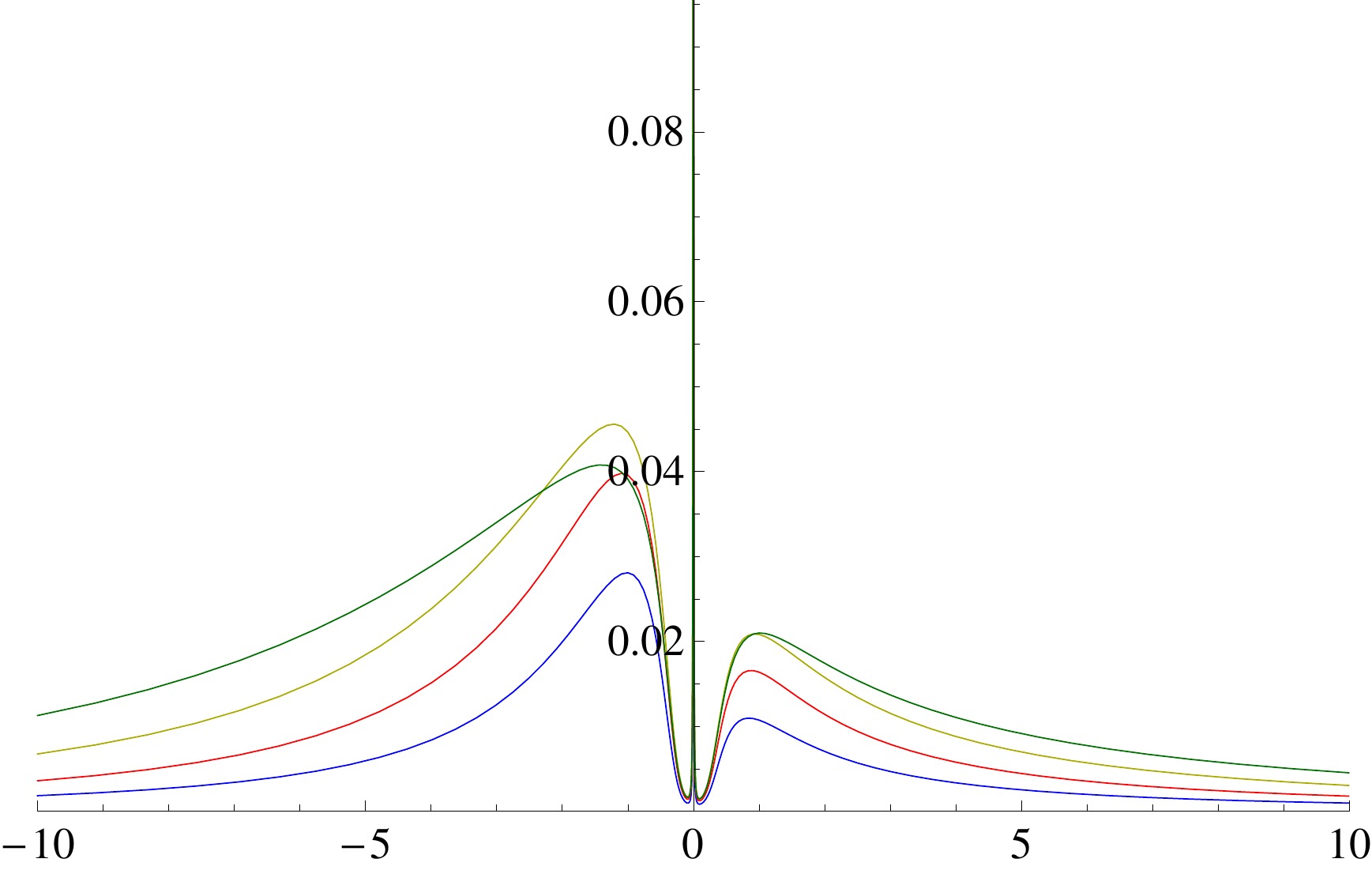}
\begin{picture}(0,0)
\put(-1,5){\scriptsize $\omega$}
\put(-100,124){\scriptsize $\rho^+$}
\put(-68,130){\scriptsize $g=0.25$ blue curve}
\put(-68,121){\scriptsize $g=0.5$ red curve}
\put(-68,112){\scriptsize $g=1$ yellow curve}
\put(-68,103){\scriptsize $g=2$ green curve}
\end{picture}
}
\hspace{4pt}
\caption{The spectral function for the spin-up component at the Fermi momentum as a function of $\om$. In both plots parameters are chosen as $g \lambda = -0.5$. This figure confirms numerically two important points that were made in the text. Firstly, for the higher temperature (a), a sharp peak is just visible at $\o=0$. For the lower temperature (b), this sharp peak is still there but is so narrow that it can barely be discerned. This indeed indicates that, at $T=0$, there is a $\delta$-function behavior at the Fermi surface, and that this $\delta$-function peak becomes broader as temperature increases. Secondly, close to the Fermi surface, that is, for small $\om$, the form of the spectrum depends only on the product $g \lambda $. Indeed, all four curves for different $g$ but equal $g \l$ overlap for small $\o$. This overlap is better for lower $T$. This shows that the  scale of the Fermi momentum is indeed set by the combination $g \l$ and that the spectra are close to the Fermi surface indeed governed by $k_F$ from (\ref{fermimom}) as we expect.}
\label{spectraldelta}
\end{figure}
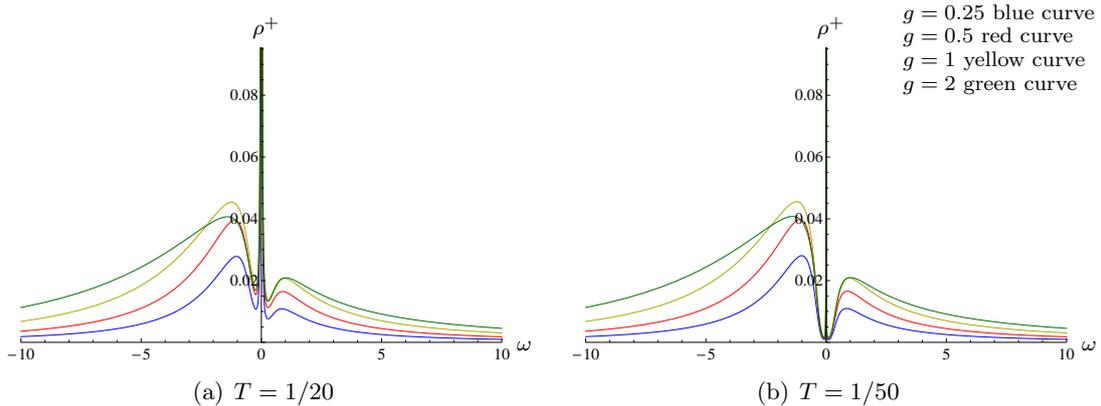

We now turn to the lifetime of the quasi-particles.
At zero temperature,  from the analytic result \eqref{Gzk} we infer that the imaginary part of the self-energy vanishes at $\om =0$, $k=k_F$. In figure \ref{imsigma}, which shows the imaginary part of the self-energy, we see that this behavior is confirmed numerically. This implies that the spectral function at the Fermi surface is a $\delta$-function at $T=0$, which is a well-known and defining property of a Fermi liquid. As a consequence, the momentum distribution shows a discontinuity at the Fermi surfaces.
Now, the inverse of the imaginary part of the self-energy  $\I \Sigma^{\pm}$ has SI units of time, and at $\o=0$, we associate it to a lifetime for the quasi-particles\footnote{Here, we ignore $\o$-dependent corrections to the lifetime, as the dominant contribution comes from the behavior at $\o=0$.}
\eq{
 \label{qplifetime}
 \tau_{\kv}  = \frac{1}{-2 Z \I \Sigma^{\pm}(\kv,0)} \;.
}
The lifetime $\tau_{k_F}$ is infinite precisely at the Fermi momentum. We can also define a frequency-dependent decay rate at the Fermi momentum as
\eq{
\Gamma(\om) = -2 Z \I \Sigma^{\pm}(k_F,\o) \;.
}
At the Fermi surface $\o=0$, the decay rate $\Gamma$ is the inverse of the quasi-particle lifetime, therefore it vanishes for $\o \rightarrow 0$. This behavior can also be observed in the spectra above. In particular, when $T$ and $\om$ are small but non-zero, the $\delta$-function is broadened, which is indeed visible numerically, as shown in figure \ref{spectraldelta}. For completeness, we note that the retarded Green's function given in \eqref{GRlin01} can now be written as
\eq{
G_R^-(k,\om) = \frac{-Z}{\om- \frac{1}{m\eff}(k-k_F)k_F+\frac{i}{2\tau_{k}}} \;.
}

A broadening of the $\delta$-function implies a non-zero decay rate for the quasi-particles. For a normal Fermi liquid, the imaginary part of the self-energy at the Fermi momentum vanishes at zero temperature and small frequency as $ \om^2$, which stems from Pauli blocking for scattering near the Fermi surface. For small non-zero temperatures, thermal averaging over  $\om^2$ yields a $T^2$ dependence of the decay rate.

Let us  compare this with our numerical results.
We observe that the decay rate vanishes much faster than in a conventional Fermi liquid.
In figure \ref{xideriv01}, we show the first derivative of $\R \Sigma^+(p)$ and the second derivative of $\I \Sigma^+(p)$  as a function of $\o$ at $k=k_F$. The former is non-zero at $\om =0$, which indicates a non-zero value of $Z$ as explained above. The latter is zero at $\om=0$, which shows that the quasi-particle decay rate vanishes faster than $\om^2$. The behavior shown in figure \ref{xideriv01} is in agreement with the existence of Fermi surfaces at $k=k_F$, however the decay rate is more suppressed than the typical $\om^2$ behavior of a conventional Fermi liquid. This indicates that, next to Pauli blocking, there is another mechanism at hand which strongly suppresses interactions around the Fermi surface for low temperatures $T$ and small frequencies $\om$. This appears to be a general feature of holographic Fermi liquids and can be understood better by performing a WKB analysis of the bulk Dirac equation.


\subsection{Second-order WKB analysis around the Fermi surface}
\lab{sec:WKBmain}

In section \ref{sec:lifetime} we have seen  that the fermionic quasi-particles are rather stable since their lifetime at the Fermi surface diverges faster than the power law expected from Fermi-liquid theory. To gain more insight into this property, we would like to obtain an analytic expression for the self-energy at the Fermi surface, that is, we want to know the form of the self-energy for both $\om \neq 0$ and $k = k_F \neq0$. As stated before, it is not possible to solve the differential equation for $\xi$ exactly for generic non-zero values of $\om$ and $k$,  however, we can deploy the WKB approximation to study the fermionic excitations around the Fermi surfaces analytically, as was done for instance in \cite{Iizuka,HHV}.

In principle, the WKB calculation gives the exact result for the retarded Green's function in the limit $k \to k_F $ and $\o\ll 1$. Indeed, as we will see, it yields very elegantly the functional form of the self-energy at the Fermi surface, from which we can derive the behavior of the quasi-particle lifetime at $k=k_F$. The behavior turns out to be exponential, i.e.,
\eq{
\label{expalpha}
\I \Sigma \sim e^{- \alpha\hspace{1pt} \frac{k^2}{\om}}\;,
}
with $\alpha$ a constant  to be determined. This result is consistent with the exponential behavior found in \cite{Iizuka,HHV}. Furthermore, we also find, to first approximation, an analytical result for the wavefunction renormalization $Z$, which is consistent with the strong quasi-particle renormalization found numerically. Notably, to obtain the latter result, we have to use the second-order WKB approximation.

However, before presenting the explicit result for the self-energy of our second-order WKB calculation, we have to mention the following. Obtaining the exact prefactors of the various terms in the final result, e.g. the constant $\alpha$ in \eqref{expalpha}, turns out to be non-trivial. The reason is that several integrals in the computation cannot be performed analytically (see appendix \ref{app:WKB} for details) but have to be evaluated  numerically.
Alternatively, we could make an approximation to the integrals which, however, turns out to be similarly difficult. Our approach here  is to deploy the WKB approximation  to gain an understanding of the functional form of the quasi-particle lifetime at the Fermi surfaces. Therefore, the result that we present in equation \eqref{WKBresSigma} is obtained using the latter method, and so the prefactors, including $\alpha$, are not exact. At the end of appendix \ref{app:WKB}, we show some numerical evidence for the  conjectured value of the constant $\alpha$.

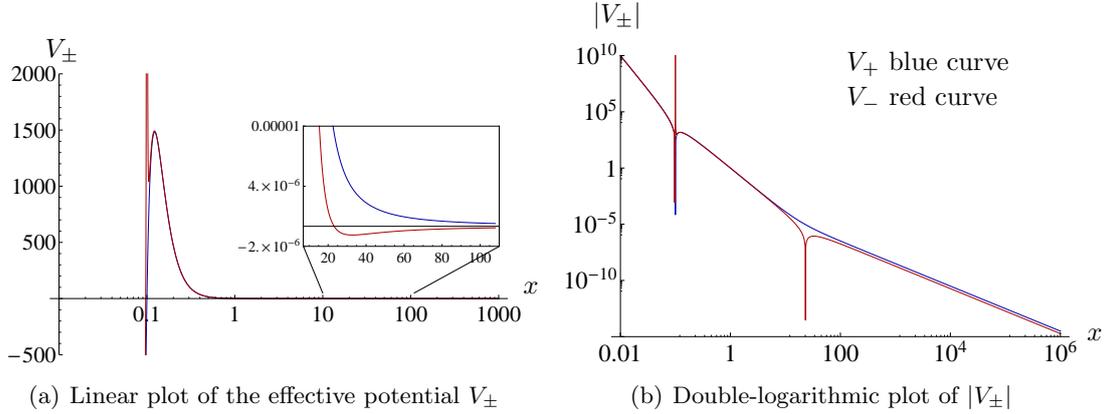
\begin{figure}[t]
\centering
\vskip20pt
\subfigure[Linear plot of the effective potential $V_{\pm}$]{
\includegraphics[width=0.46\textwidth]{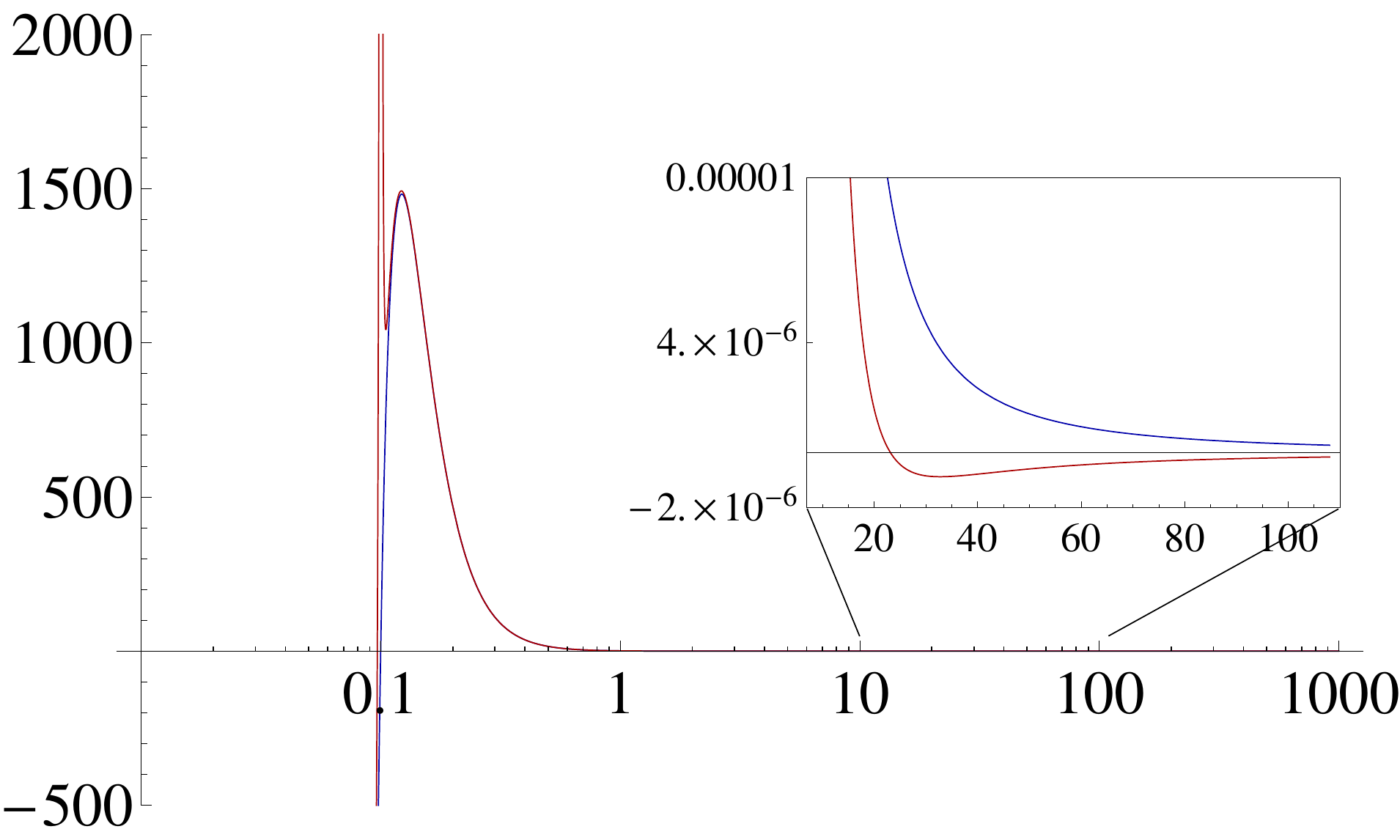}
}
\hspace{\stretch{1}}
\subfigure[Double-logarithmic plot of $|V_{\pm}|$ ]{
\includegraphics[width=0.46\textwidth]{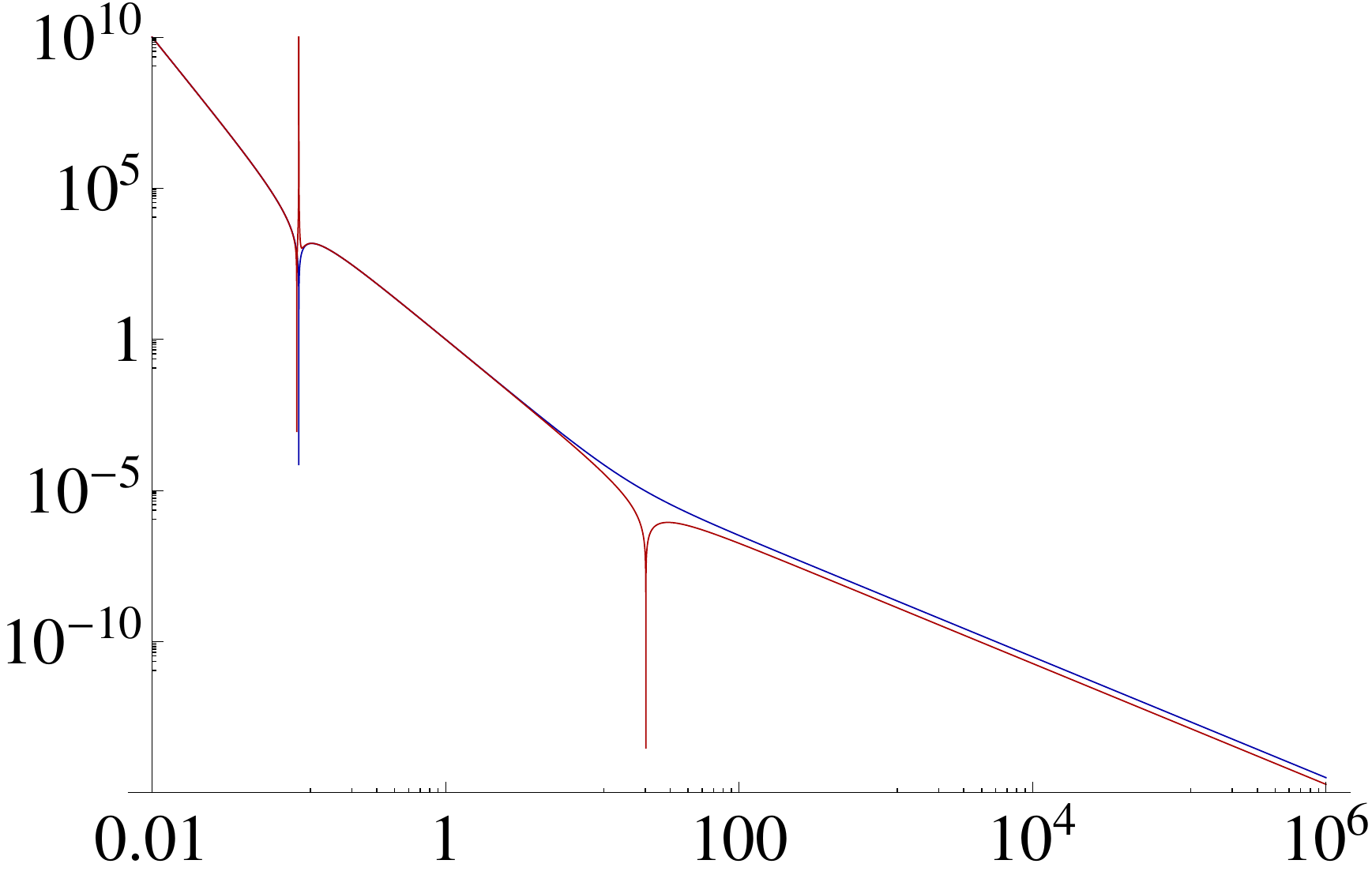}
\begin{picture}(0,0)
\put(-211,28){\small $x$}
\put(0,10){\small $x$}
\put(-390,117){\small $V_{\pm}$}
\put(-185,130){\small $|V_{\pm}|$}
\put(-90,112){\small $V_+$ blue curve}
\put(-90,99){\small $V_-$ red curve}
\end{picture}
}
\hspace{3pt}
\caption{Plots of the effective potentials $V_+$ and $V_-$ (blue and red line) as function of $x=r/\sqrt{\om}$ at $T=0$ for $\bk=10$ and $M=1/4$.
In the near-horizon region $0 < x < \xtp$, both effective potentials are negative. The potential $V_+$ has a turning point at $\xtp$, which can be seen as the first dip from the left in (b). $V_-$ has a pole located approximately at the position of the classical turning point of $V_+$. $V_+$ is always positive for $x>\xtp$, while $V_-$ has a different turning point at an intermediate value of $x$. This point can be seen in the inset in (a) and  as the second dip from the left in (b).
}
\label{Schrofig}
\end{figure}

\medskip
Let us outline the basic steps of our WKB calculation  and refer to appendix \ref{app:WKB} for further details.
We determine the Green's function at {\em zero temperature}.
The expansion parameter that plays the role of $1/\hbar$ in the conventional WKB approximation, and that we take to be large in order to satisfy the WKB conditions, is the rescaled momentum  (see appendix   \ref{app:sca})\footnote{We will specialize to the case of interest $z=2$ in this section, although the calculation can easily be generalized to arbitrary $z>1$.}
\eq{
\lab{resk}
\bk = \frac{k}{\sqrt{\o}} \gg 1\;.
}
To simplify the calculation we also assume that both $\o$ and $\bk$ are positive, and we introduce the rescaled variable $x= r/\sqrt{\o}$. Choosing the up and down components $u_\pm$  and $d_\pm$, respectively, of $\psi_\pm$ introduced in \eqref{FTspinor} as
\eq{
 \label{defupm}
\psi_\pm =  \frac{1}{\sqrt{r^5 V(r)}} \, \binom{ u_\pm}{ d_\pm  } \;,
}
we define new fluctuation fields $y_\pm$
in terms of the original fields $u_\pm$ as
\eq{\lab{schro1}
y_{\pm} =  \frac{\sqrt{x \bk  \pm 1 }}{x^{3/2}} \,u_\pm \;.
}
Then, the once-iterated Dirac equation can be written as a Schr\"odinger-like equation
\eq{
\frac{1}{\bk^2} \frac{d^2}{dx^2}\,y_\pm(x) - \bV_\pm(x) \,y_\pm(x) = 0 \;,
}
with the effective Schr\"odinger potential at $T=0$ of the form
\eq{\lab{schro2}
\bV_\pm(x)=\frac{1}{x^4}-\frac{1}{\bk^2 x^6}
+\frac{\left(\half\pm M\right)\left(\frac32\pm M\right)}{\bk^2 x^2}
+\frac34 \frac{1}{\left(x\bk \pm 1\right)^2}-\frac{\frac32\pm M }{x\bk\left(x\bk \pm 1\right)}\ .
}
The characteristic behavior of the Schr\"odinger potentials for the chiral components is illustrated in figure \ref{Schrofig}. In particular, $V_+$ has a single classical turning point at $\xtp$ and $V_-$ has a pole and a turning point. For $\bk \gg 1$, both effective potentials have a very high potential barrier close to $x\rightarrow0$.
For simplicity we present here the calculation  only for the plus-component, the result for the minus-component is easily obtained  using the symmetry explained in appendix \ref{app:symm}. Also, we only perform the calculation for the case $M>0$ so that the single turning point of $V_+$ at $\xtp$ divides the entire range into two regions, a classically allowed region where $V<0$ and a classically disallowed region where $V>0$.

The strategy to determine the WKB solution is as follows. We first incorporate
the in-falling near-horizon boundary condition
into the WKB wavefunction in the classically allowed region $x<\xtp$. Then, we apply the usual linearized connection formulae at $x=\xtp$  to continue the solution into the classically disallowed region $x>\xtp$. Next, we match this solution onto the exact near-boundary solution to fix the remaining integration constants. Finally, from the asymptotic expansion of the solution we read off the coefficients $A_\pm$ (see appendix \ref{app_asymps}) that determine $G_R$ via $\xi=i A_-/A_+$. The details of this calculation are explained in appendix \ref{app:WKB}, and the final result for the WKB self-energy reads
\eq{\label{WKBresSigma}
\Sigma^+( \vec k,\om) = g\, c_1\, k^{2M} \, \frac{\g_+}{\epsilon_+} \, \frac{1-\frac{i}{2} \frac{\g_-}{\g_+}e^{-i\pi M} e^{-2\frac{k^2}{\o}}}{1+\frac{i}{2}  \frac{\epsilon_-}{\epsilon_+}e^{i\pi M} e^{-2\frac{k^2}{\o}}} \;,
}
where we employed again $\vec k=(0,0,k_3)$ with $k_3>0$,
so that the corresponding Green's function in the limit $\bk \gg 1$ reads
\eq{\lab{WKBres}
G_R^+( \vec k,\om) = -\le(\o-\frac{1}{\l} k^z - g\, c_1\, k^{2M} \, \frac{\g_+}{\epsilon_+}\,\frac{1-\frac{i}{2} \frac{\g_-}{\g_+}e^{-i\pi M} e^{-2\frac{k^2}{\o}}}{1+\frac{i}{2}  \frac{\epsilon_-}{\epsilon_+}e^{i\pi M} e^{-2\frac{k^2}{\o}}} \ri)^{-1}\;.
}
The constant $c_1$ was given in (\ref{cz}), and we defined
\eq{\lab{gep}
\epsilon_\pm = 1\pm \frac{M}{2\bk^2}\log\bk \mp \frac{1}{4\bk^2}\;,
\hspace{40pt}
\g_\pm  = 1\pm \frac{M}{2\bk^2}\log\bk \mp \frac{3}{4\bk^2} \;.
}
As mentioned before, the result \eqref{WKBresSigma} comes from performing an expansion \textit{within} the framework of the WKB calculation.  In this case, we find $\alpha=2$ for the aforementioned coefficient in \eqref{expalpha}, which we can improve  by taking into account more terms in the expansion. In particular, as explained in more detail in
appendix \ref{app:WKB}, the correct prefactor in the exponent seems to be approximately $\alpha \approx \pi/2$.

\medskip
After having obtained our result, let us now study the Green's function \eqref{WKBres} in more detail.
As a first consistency check, we see that in the limit $\o\to 0$, which corresponds to $\bk \equiv k/\sqrt{\o}\to \infty$, we find the same locus for the Fermi surface at $k=k_F$ as in  equation (\ref{fermimom}).
Furthermore, the decay rate of the quasi-particle excitations around the Fermi surface
is calculated using \eqref{qplifetime} and employing the imaginary part of the self-energy at $k=k_F$, which reads
\eq{\label{decayrate}
\I \Sigma^+( \vec k_F,\om) =  \frac{1}{2}\, g\,  c_1\,  k_F^{2M} \sin\left(\pi\left(M-\tfrac12\right)\right)
 \frac{\g_+}{\epsilon_+}\le(\frac{\g_-}{\g_+} + \frac{\epsilon_-}{\epsilon_+}\ri) \,e^{-2\frac{k_F^2}{\o}}\ ,
}
where $k_F$ is given by (\ref{fermimom}). We note that $\g_\pm$ and $\epsilon_\pm$ become unity in the strict Fermi-surface limit, and the result thus shows an exponentially suppressed decay rate with additional  $\o/k_F^2$ corrections.
This result demonstrates the extremely stable nature of the quasi-particle excitations on the Fermi surface and is consistent with previous WKB calculations that similarly find exponentially narrow decay rates \cite{Iizuka, HHV}. A comparison with the numerical results is made in figure \ref{figImSWKB}.

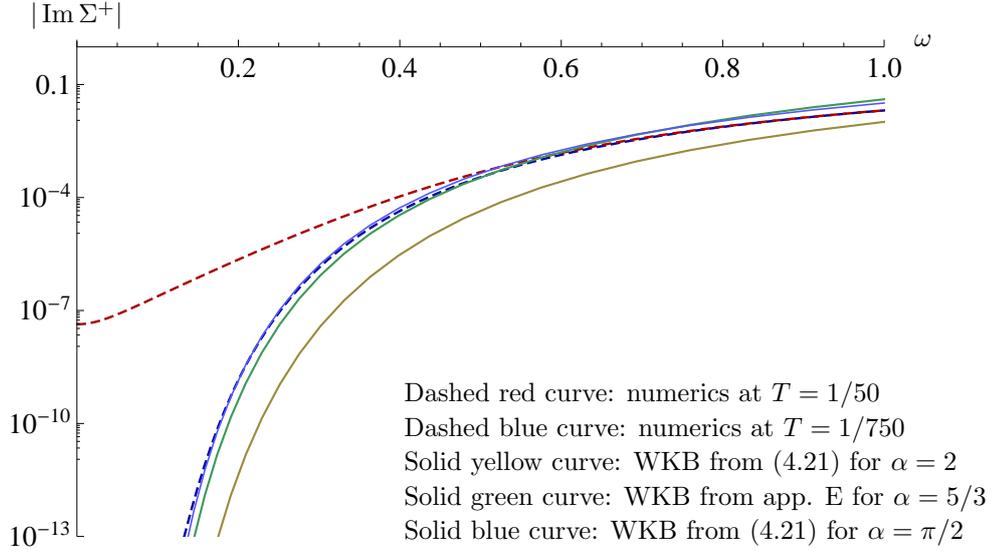
\begin{figure}[p]
\centering
\includegraphics[width=.8\textwidth]{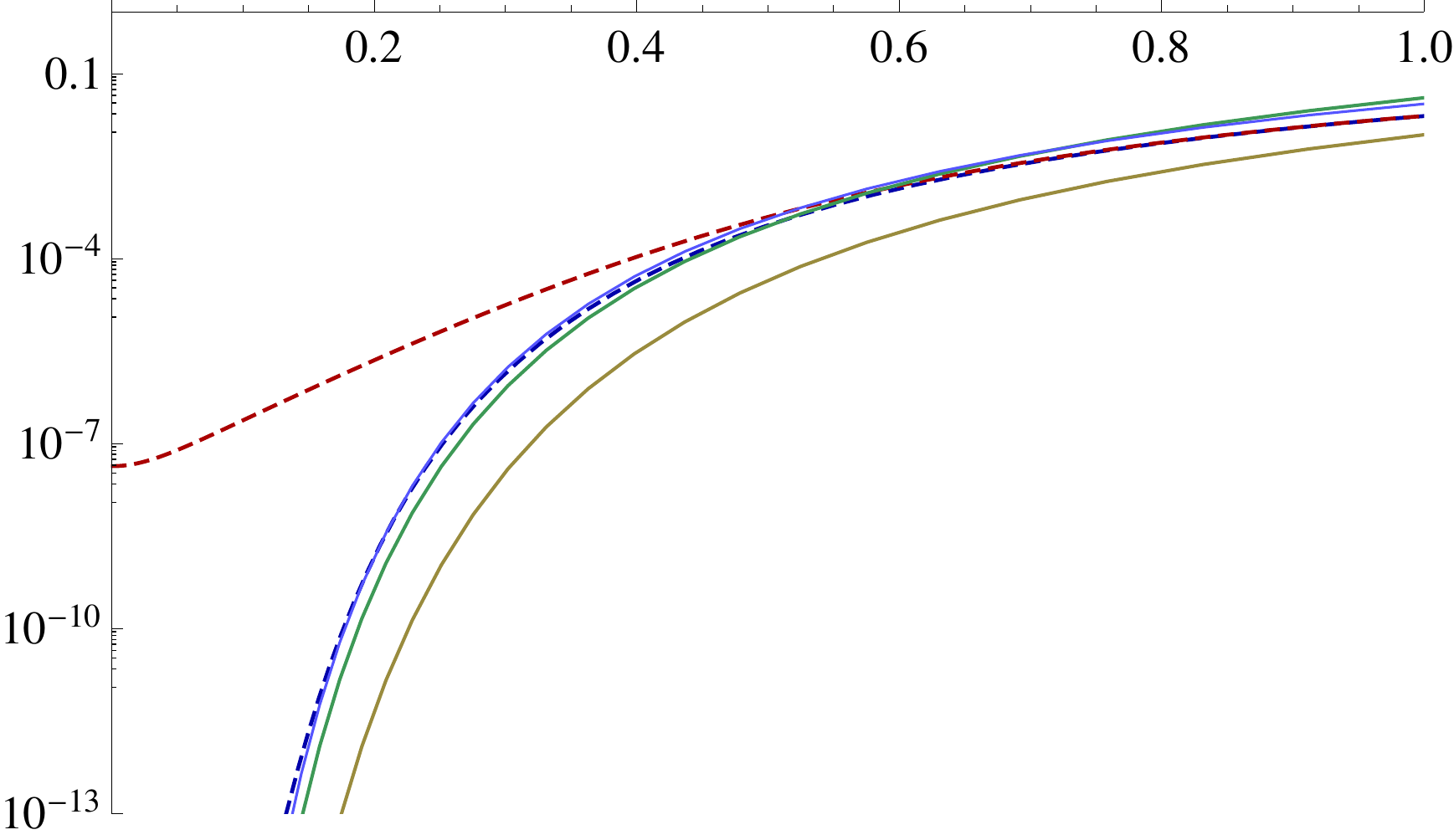}
\begin{picture}(0,0)
\put(1,191){\small $\o$}
\put(-330,200){\small $|\I \Sigma^+|$}
\put(-190,57){\small Dashed red curve: numerics at $T=1/50$}
\put(-190,44){\small Dashed blue curve: numerics at $T=1/750$}
\put(-190,31){\small Solid yellow curve: WKB from \eqref{WKBresSigma} for $\alpha=2$}
\put(-190,18){\small Solid green curve: WKB from app. \ref{app:WKB} for $\alpha=5/3$}
\put(-190,5){\small Solid blue curve: WKB from \eqref{WKBresSigma} for $\alpha=\pi/2$}
\end{picture}
\caption{Comparison of WKB and numerical results for $|\I \Sigma^+|$ at $k=k_F$ with
$\lambda =-1$ yielding $k_F \approx 1.64$, and $M=1/4$ as a function of $\o$.
The dashed curves are numerical results, the solid curves are second-order WKB results at $T=0$.
Note that the WKB result  \eqref{WKBresSigma} for $\alpha=\pi/2$ agrees very well with the low temperature numerical result.
}
\label{figImSWKB}
\end{figure}
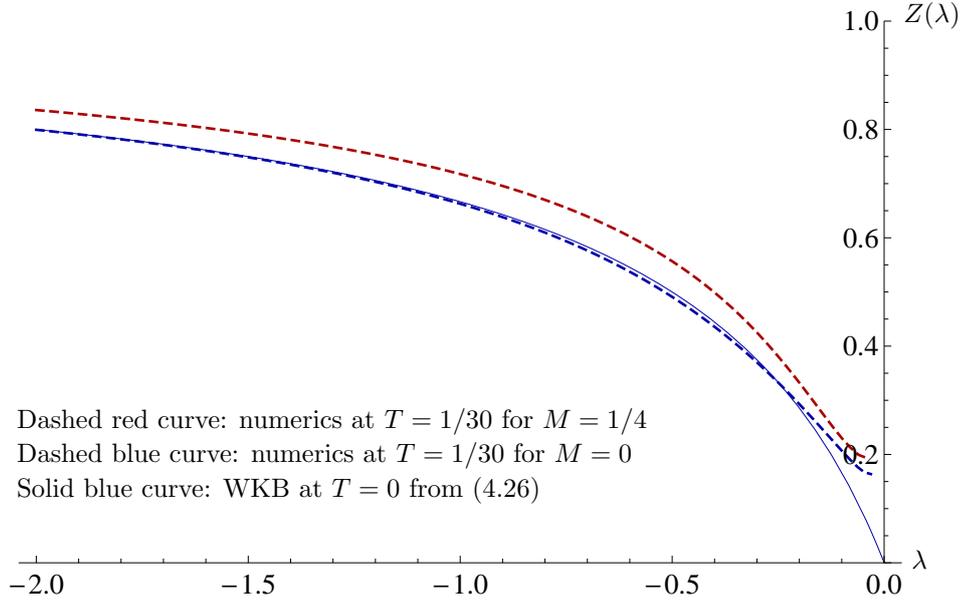
\begin{figure}[p]
\centering
\includegraphics[width=.8\textwidth]{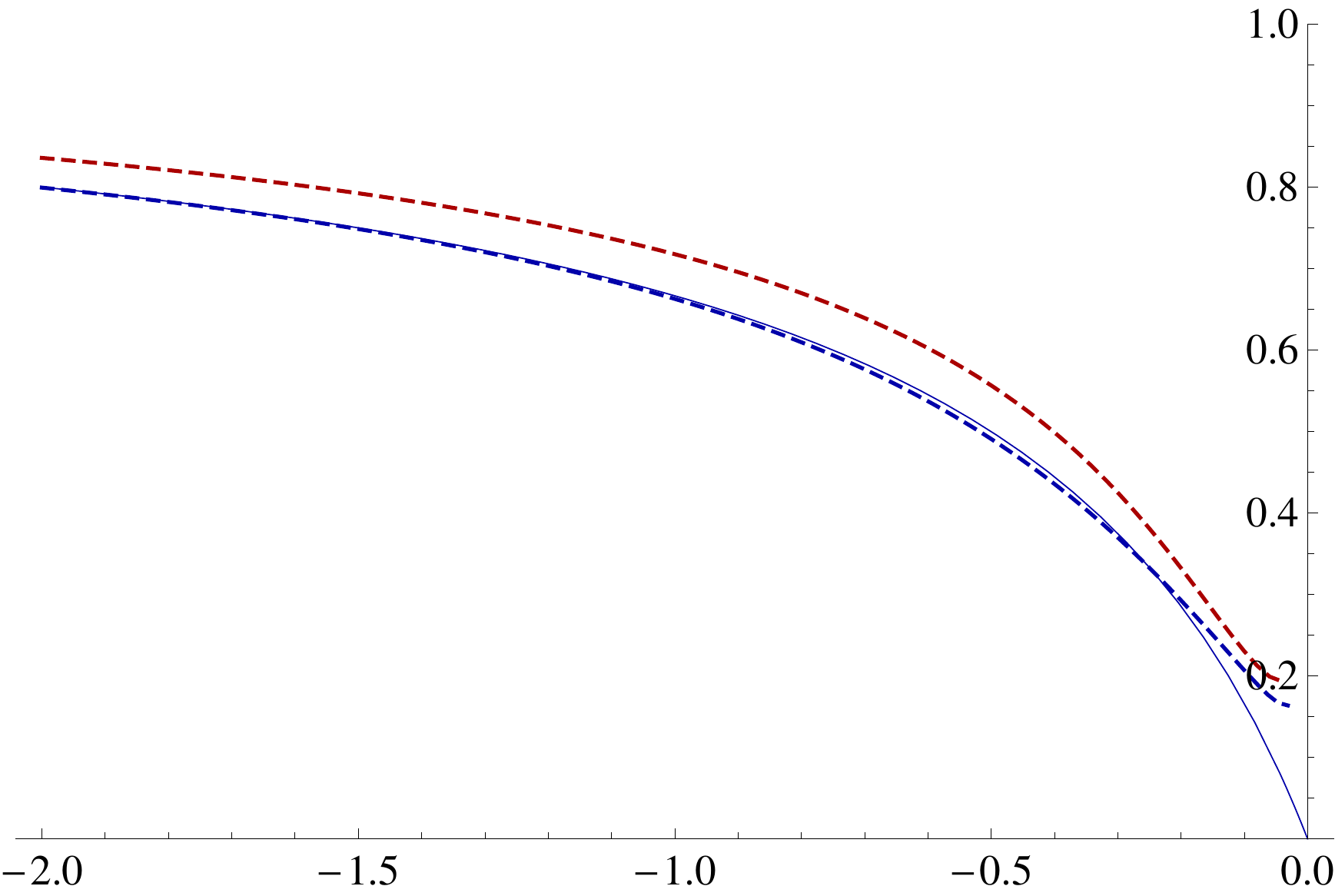}
\begin{picture}(0,0)
\put(0,12){\small $\lambda$}
\put(-3,218){\small $Z(\lambda)$}
\put(-335,65){\small Dashed red curve: numerics at $T=1/30$ for $M=1/4$}
\put(-335,52){\small Dashed blue curve: numerics at $T=1/30$ for $M=0$}
\put(-335,39){\small Solid blue curve: WKB at $T=0$ from \eqref{ZWKB}}
\end{picture}
\caption{Comparison of WKB and numerical results for the quasi-particle residue $Z$ from (\ref{Z}) as a function of $\lambda$. The dashed curves are numerical results,
the solid curve is the second-order WKB result at $T=0$.
This figure shows that the WKB result for $Z$ resembles the $M=0$ result, indicating that the analytic expression may be the $M=0$ part of a more general result.
}
\label{figZWKB}
\end{figure}

Another analytic result that we extract from (\ref{WKBresSigma}) is the wavefunction renormalization (\ref{Z}) on the Fermi surface. This follows from the real part of the self-energy  which reads
\eq{\lab{ReSigma}
\R \Sigma^+(\vec k,\om) =  g \,c_1 \,k^{2M}\, \frac{\g_+}{\epsilon_+} \;,
}
up to exponentially small terms multiplied by $\exp(-\alpha k^2 / \om)$ that will drop out in the limit $\om\rightarrow 0$  after differentiation.
Combining \eqref{ReSigma}  with \eqref{fermimom} and \eqref{Z}, we arrive at an analytic result for the wavefunction renormalization factor of the form
\eq{ \lab{ZWKB}
Z = \frac{2\l}{2\l-1}\ ,
}
which for $\lambda<0$ satisfies $0\leq Z \leq 1$ as required.
We note that, to obtain this non-trivial result different from one, we have to go beyond the leading order in the WKB approximation, since otherwise the derivative of the self-energy in (\ref{WKBresSigma}) would  vanish due to the exponential suppression.
Interestingly, because all factors $M$ in (\ref{gep}) are multiplied by $\om$, these drop out in the limit $\om \rightarrow 0$ so that the analytic result (\ref{ZWKB}) is independent of $M$. This is in contrast to the numerical results which do depend on the value of $M$. The analytic result (\ref{ZWKB}) is therefore expected to be the $M=0$ part of a larger expression, that one may obtain using a more sophisticated WKB calculation. This presumption is supported by figure \ref{figZWKB}, which shows that both results are consistent in the case $M=0$. However, the aim of the WKB calculation was merely to support our numerical results, and finding perfect agreement is beyond the scope of this work.


\subsection{Phase diagram}

The results obtained in the sections above, in particular the plots in figure \ref{figNkcombi}, can be explained by the existence of a quantum phase transition  at $\lambda=0$ and $g\ge0$. Firstly, we can understand why there must be different phases. Consider again the Green's function \eqref{GR2} and recall that $\lambda$ can change sign whereas $g$ is required to be positive (in the range $-1/2<M<1/2$) in order to satisfy the Kramers-Kronig relation.
Furthermore, the kinetic term scales as $k^z$, but the self-energy is proportional to $k^{2M}$ which is a lower power of $k$. As a consequence, there will always be a small but non-zero $\vec{k}$ interval at low values of the momentum in which the self-energy term dominates. When there is a difference in the relative sign, there can be a cancellation between the kinetic and the interaction term in the Green's function at low momenta. This is what brings the Fermi surfaces into existence.
When there is no relative sign difference, the self-energy contribution adds up to the kinetic term, which results in a different behavior of $N^{\pm}_{\vec k}$ and therefore in a different phase. Thus, the crucial property is the possibility of a relative sign change between the kinetic and interaction term which is dominant at low momenta.

We sketch the phase diagram of our system in figure \ref{phasediagram}, and relate it to our results for the momentum distribution at various values of $\lambda$ and $T$.
\begin{itemize}

\item For $\lambda<0$ and  temperatures $T$ small compared to the Fermi energy $\epsilon_F = k_F^z/|\lambda|$, we have a phase with two Fermi surfaces: one for the particles and one for the holes, which can be seen as two sharp discontinuities in the momentum distribution at $k_3=\pm |k_F|$. The system is formally only a Fermi liquid at exactly zero temperature. Numerically we can obtain only non-zero temperature results, and the effect of temperature is to smear out the discontinuities over an interval around $\pm |k_F|$. However, when $T/\epsilon_F$ is small (outside the critical region) we still see the Fermi-liquid-like or degenerate behavior. Indeed in figure \ref{figNkcombi} for small $\lambda <0$ the jumps are not infinitely sharp but still very much located around a single value of $\kv$.

\begin{figure}[t]
\vskip10pt
\centering
\includegraphics[width=.7\textwidth]{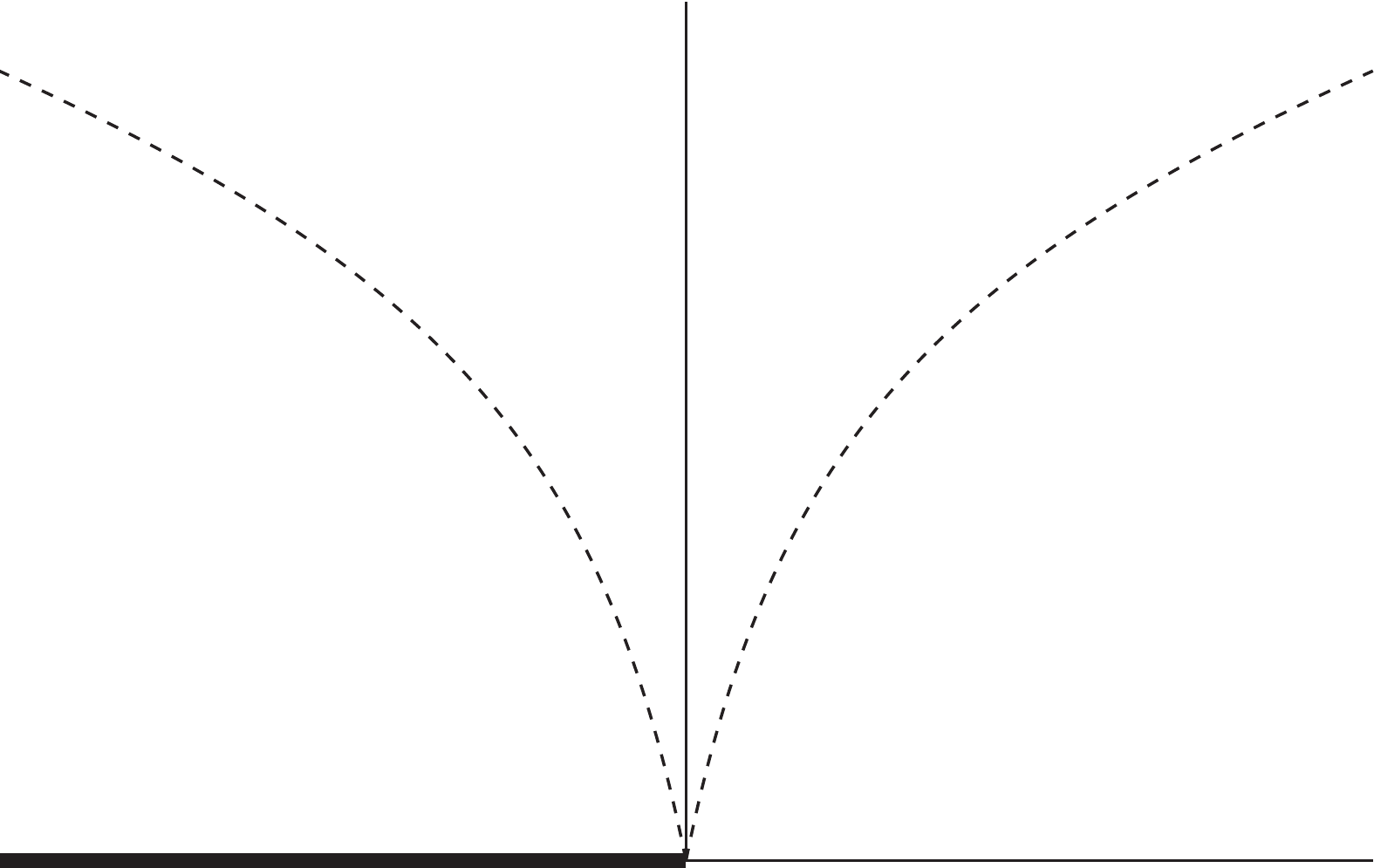}
\begin{picture}(0,0)
\put(0,-1){\small $\lambda$}
\put(-153,191){\small $T$}
\put(-110,60){\small non-Fermi-liquid phase}
\put(-263,5){\small Fermi-liquid phase}
\put(-232,160){\small quantum-critical}
\put(-232,148){\small phase}
\put(-12,-1){\small $\rightarrow$}
\put(-152.51,179){\small $\uparrow$}
\end{picture}
\caption{Sketch of the phase diagram in the case $-1/2<M < 1/2$ and $g=1$, showing the non-Fermi-liquid phase for low $T$ and positive $\l$, the Fermi-liquid phase for low $T$ and negative $\l$, and two qualitatively different quantum-critical phases for high temperatures and $\l$ positive and negative. The system is strictly only a Fermi liquid at $T=0$ (thick line), but has Fermi-liquid-like behavior for low temperatures. The dashed curves correspond to the crossovers to the quantum-critical regimes. There is a quantum phase transition at $\l=0$, $T=0$.}
\label{phasediagram}
\end{figure}
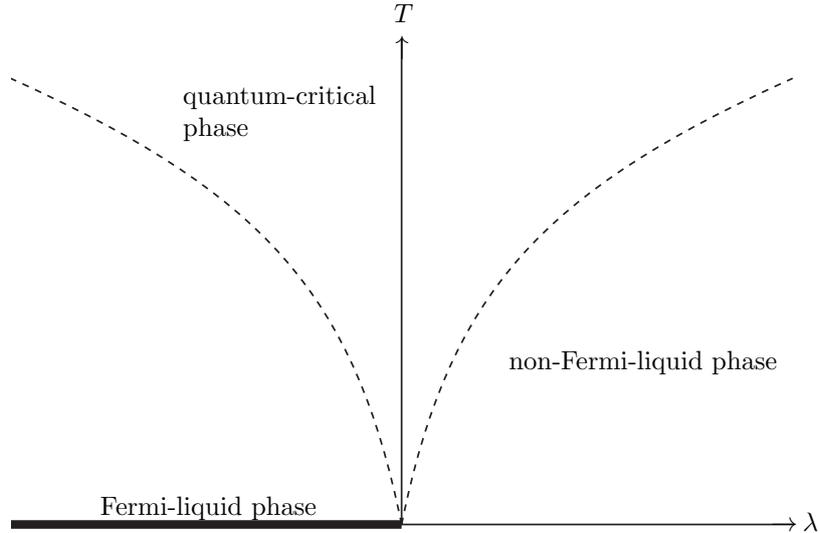

\begin{figure}[p]
\centering
\vskip10pt
\subfigure[$\lambda=-1$]{
\includegraphics[width=0.45\textwidth]{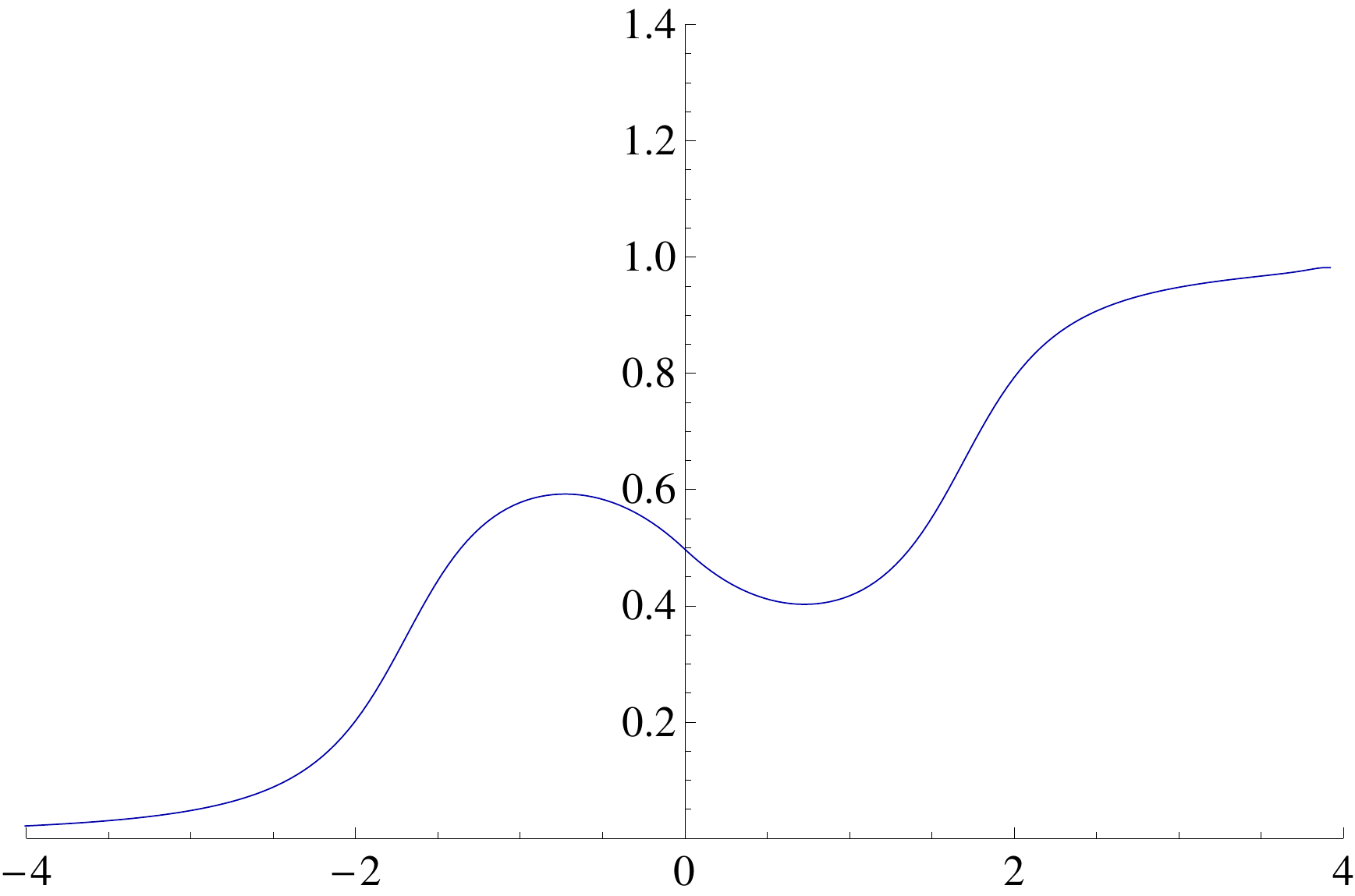}
\begin{picture}(0,0)
\put(0,6){\scriptsize $k_3$}
\put(-103,130){\scriptsize $N^-_{\vec k}$}
\end{picture}
}
\hspace{\stretch{1}}
\subfigure[$\lambda=-0.002$]{
\includegraphics[width=0.45\textwidth]{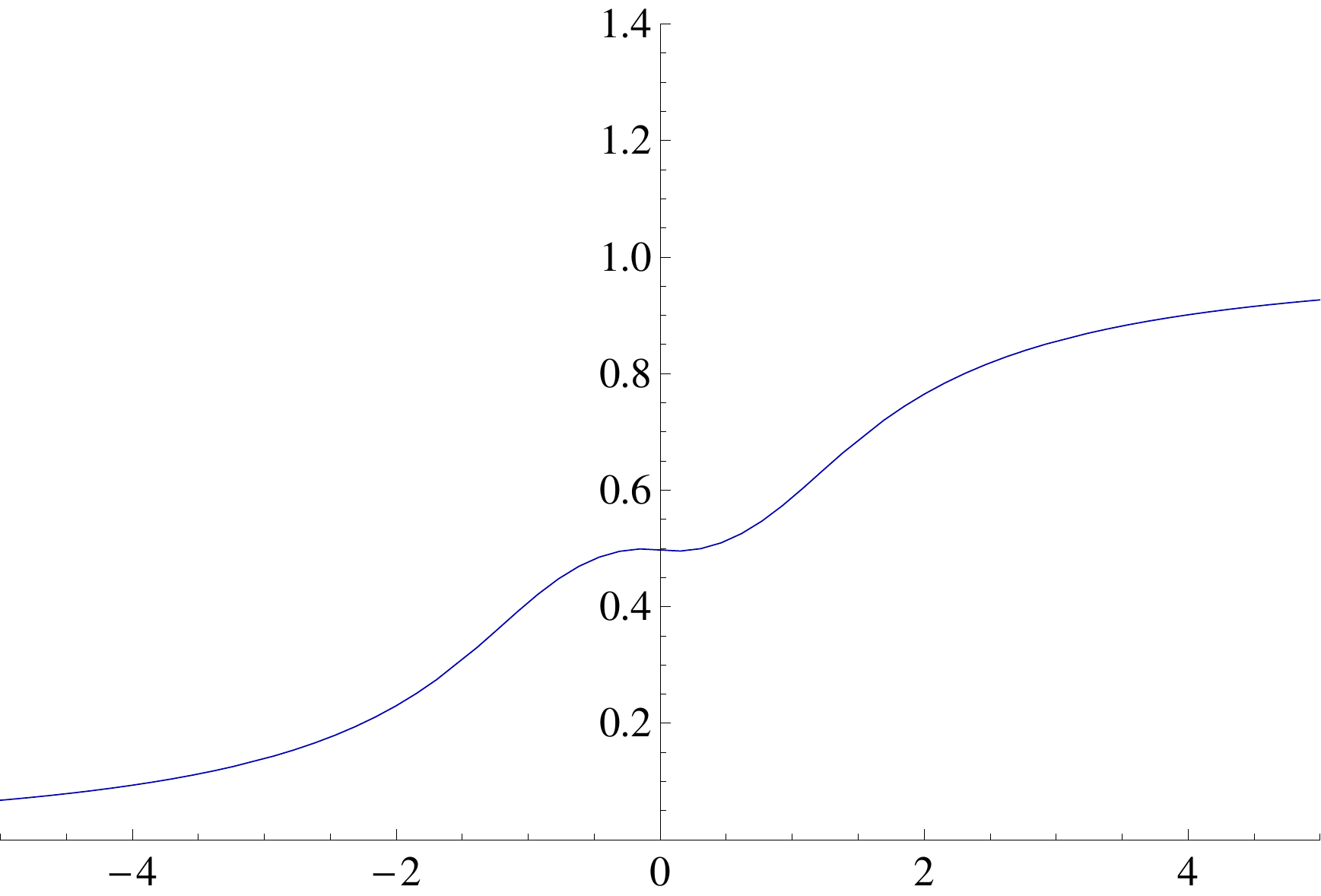}
\put(1,6){\scriptsize $\frac{k_3}{k_F}$}
\put(-103,135){\scriptsize $N^-_{\vec k}$}
}
\hspace*{10pt}
\\
\subfigure[$\lambda=+1$]{
\includegraphics[width=0.45\textwidth]{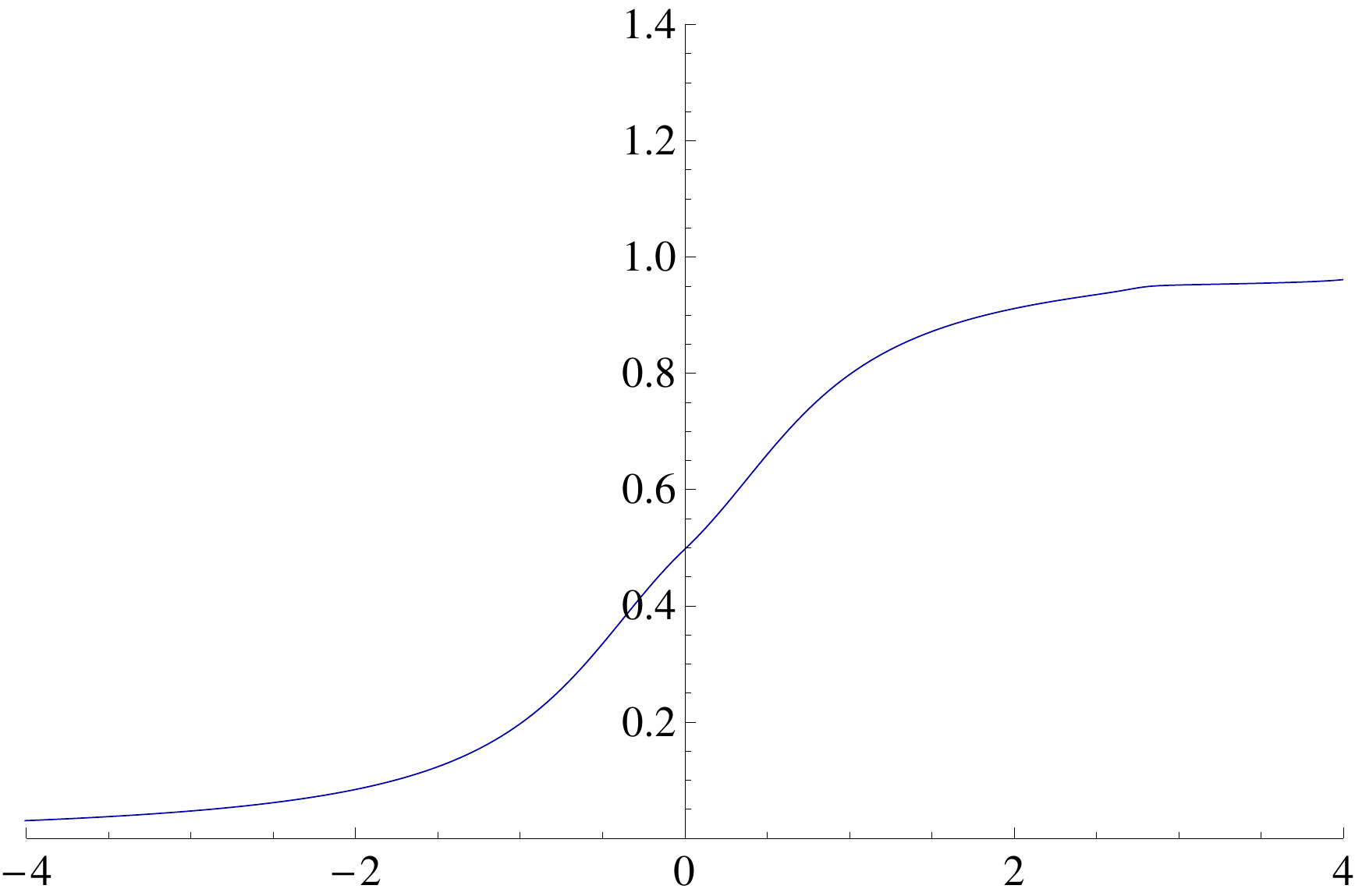}
\put(1,6){\scriptsize $k_3$}
\put(-101,130){\scriptsize $N^-_{\vec k}$}
}
\caption{Momentum distribution in the quantum critical region for $T=5$, $g=1$, $M=1/4$ and  various values of $\lambda$.
For $\lambda<0$, the momentum distribution always shows two extrema at non-zero $k_3$, which develop into Fermi surfaces when the temperature is lowered. This is shown in (a).
The behavior for small negative $\lambda$ is illustrated in figure (b) with $k_F \approx 0.026$. Note that because $k_F$ is relatively small, we plotted $N^-_{\vec k}$ here as a function of $k_3/k_F$ for clarity. For $\l>0$ the momentum distribution is very broad as can be seen in (c).
}
\label{NkT5}
\end{figure}

\item When increasing temperature, in the region $T \sim \epsilon_F$ there is a crossover between the Fermi-liquid phase and the quantum critical phase, indicated by the dashed lines in figure \ref{phasediagram}. To calculate the crossover temperature as a function of $\lambda$ we realize that the crossover occurs when $T \approx \epsilon_F$, thus at
\eq{
T \approx \epsilon_F = \frac{1}{|\lambda|} \,k_F^{z}
= |\lambda|^{\frac{2M}{z-2M}} = g \hspace{1pt}c_1 k_F^{2M} \;.
}
In the case $M=1/4$, $z=2$, the crossover is at $T \sim |\lambda|^{1/3}$.

\item When $T \gg\epsilon_F$ the Fermi surfaces will be smeared out to an extend that they are difficult to identify, as it can be seen in figure \ref{NkT5}(a). Then the system is in the quantum critical phase. As $\lambda \to 0^-$ , $\epsilon_F$ decreases, and as a consequence the Fermi surfaces develop into smooth bumps for very small $\lambda$. Numerically, we will always observe temperature effects for small values of $|\lambda|$.

\item At $\lambda =0$ and $T=0$, there is a quantum phase transition from a phase with two Fermi surfaces to a phase with no Fermi surfaces. At $\lambda = 0$ and $T>0$, the system is in the quantum critical phase. Interestingly, there is a qualitative difference between the quantum critical $\lambda<0$ and $\lambda>0$ phases, see figure \ref{NkT5}(b) for an impression of the behavior of $N^{\pm}_{\vec k}$ for very small $\lambda$.

\item In the case $\lambda>0$, the momentum distribution has a kink-like behavior and is centered around $k=0$. For $T\ll\epsilon_F$, the width of the kink is determined by $k_F \equiv \left( g \l c_1 \right)^{1/(z-2M)}$, whereas for $T\gg\epsilon_F $ it is determined by $T$, see figure \ref{NkT5}(c).

\end{itemize}


\subsection{Fermi surfaces and the quantum phase transition for $\half<|M|<\frac{z}{2}$}\lab{sec:FSMrange}

In the previous sections, we have studied the system in the particular range $0<M<1/2$ and we have demonstrated the appearance of Fermi surfaces for $\lambda<0$. However, we obtain a similar behavior also for positive values of $\lambda$ if we consider $1/2<M<z/2$ and $-z/2<M<-1/2$.

Let us demonstrate this for $1/2<M<z/2$. In this case, the relation between the self-energy $\Sigma(p)$ and the bulk quantity $\xi(r,p)$ is given by equation \eqref{defsigspec}. However, the  expression \eqref{Gzko2} for the Green's function is  unchanged because this form is determined solely by the symmetry properties of $\xi(r,p)$,  which are also satisfied by the counter-term in \eqref{defsigspec}. In particular, the $\o\to 0$ limit of the dispersion relation is again given by  equation \eqref{fermi}
which we recall for convenience,
\eq{
\lab{fermi1}
\frac{1}{\lambda}\; k^z + g\, k^{2M} c_1=0\; .
}
Now, we can find a non-trivial, real solution to this equation in the range $1/2>M>z/2$ {\em for positive $\lambda$}
because the constant $c_1$ given by \eqref{cz} is negative in this range.
Note that the same conclusion applies for $-z/2<M<-1/2$.
Thus, we now obtain Fermi surfaces for positive $\lambda$ in these ranges for $M$.
Furthermore, we also find a quantum phase transition  at $\lambda =0$ below which the aforementioned Fermi surfaces  disappear.


\section{Conclusion and discussion}\lab{sect:concl}

In our recent work \cite{Gursoy:2011gz} we proposed the construction of a holographic fermionic retarded Green's function modified to describe single-particle correlations, using the AdS/CMT correspondence. Two aspects of this prescription that make it particularly interesting for condensed-matter applications, are the following. Firstly, the retarded Green's function satisfies the zeroth frequency sum rule and the Kramers-Kronig relation both at zero and also at non-zero temperatures. We have checked that this is the case analytically where possible, see appendix \ref{app:sum_rule}, and otherwise numerically. This shows that the correlation function is a genuine single-particle correlation function that can be compared to experimental ARPES data. Secondly, it can describe quantum critical points with relativistic as well as non-relativistic dynamical exponent $z >1$, as a consequence of the (approximate) Lifshitz isometry of the used bulk spacetime.

In this paper, we have investigated the physics arising from this modified holographic prescription. The resulting spectral-weight function and dispersion relations that we presented in section \ref{sect:results}, show that we are describing a particle-hole symmetric, gapless, chiral boundary system that behaves as a strongly interacting Weyl semimetal in the sense that it satisfies the Weyl equations in the non-interacting and low-energy limit. To be more precise, our results show that the system has strong interactions in the infra-red, where the self-energy is dominant over the kinetic energy, while in the far ultra-violet the system becomes free.

An important point mentioned in the introduction is repeated here. We use holography here in a bottom-up approach. This means that the model is an effective low-energy theory which has the properties of a Weyl semimetal as described above. As with all holographic AdS/CMT models, the microscopic structure of the actual boundary system is unknown. We do not claim to describe the emergence of a Weyl fermion from a chirality-invariant microscopic structure. What we present is a specific model that may or may not be able to capture the IR physics of generic realistic Weyl semimetals. Our working assumption is that the Weyl semimetal under consideration gives rise to a strongly-coupled, chiral, scale-invariant theory in the IR. The main problem here is that in principle, we cannot be sure that our holographically obtained self-energy corresponds sufficiently accurately to a self-energy that stems from Coulomb interactions between the (chiral) electrons. This should be scrutinized for instance using feedback from future experiments on realistic materials.

On the other hand, bottom-up model building has the advantage of being able to capture a wide range of possible situations, and it provides an exploratory study of possible strongly-coupled dynamics. The model that we consider is specified by the following features:
\begin{enumerate}
\item The IR theory is Lifshitz-invariant with dynamical scaling exponent $z$.
\item Single-particle excitations are coupled to the CFT only through a single channel, specified by the scaling dimension of the chiral CFT operator that is determined by the corresponding bulk fermion mass $M$.
\item The chemical potential, corresponding to the number of particles, vanishes. In condensed-matter language this means that there is particle-hole symmetry.
\item Considering the single-particle dispersion relation, the single-particle cones of definite chirality are separated in momentum space in the IR and we are considering only the physics of one of these cones.
\end{enumerate}
Assumption 1.\ is fairly general and covers a wide range of interesting IR physics. Assumption 2.\ is merely a simplifying working assumption. The single fermion in this model is described by a dynamical source field that is coupled to gauge-invariant composite operators in the dual CFT, and we do not claim that this generally happens in nature. However, many CFTs exhibit a discrete spectrum of scaling dimensions, making it plausible that one particular channel will contribute dominantly to the self-energy of a single fermion in the IR. Assumptions 3.\ and 4.\ are merely a convenient restriction of this paper. What is presented in this work corresponds to a subspace in the phase diagrams of these types of models specified by zero chemical potential and neglecting (large-momentum) Umklapp processes that couple the two cones. However, this subspace is not unrealistic. The chemical potential couples to the number of particles minus the number of holes in the spectrum of the theory. In the case of zero chemical potential there is still non-trivial IR dynamics in the form of particle-hole excitations. Indeed, at zero doping the various phases and the corresponding quantum phase transition described in section \ref{sect:QPT} are realized. The scale that governs the quantum phase transition is an appropriate combination of the parameters $g$ and $\l$ appearing in our prescription. As a generalization of our holographic model for a Weyl semimetal we will extend the prescription to non-zero doping in future work. This is interesting for condensed-matter purposes as it corresponds for instance to turning on a gate voltage in the material. Holographically this can be done by adding another U(1) gauge field to the bulk spacetime under which the fermions are charged, and by considering charged Lifshitz black branes \cite{Tarrio:2011de}. It will be interesting to investigate the consequences for the phase diagram described in section \ref{sect:QPT}.
Finally, the simplified description obtained by focusing on a single chiral sector is reasonable, as we are interested in energy scales lower than the scale separating the two chiral cones in momentum space. Indeed we are currently working on generalizing this picture to two (opposite) chiral cones.

Weyl semimetals have recently received considerable interest in the condensed-matter community because of their fascinating and unusual properties that result from the topological nature of the band structure \cite{Weyl1,Weyl2,Meng,Burkov,Spivak,Zhang,Hosur,Voskresensky}. In order to investigate the band structure, the Weyl semimetals are usually treated as free or weakly interacting systems. Using our holographic prescription for models of Weyl semimetals, we are now also able to explore these systems when they are strongly interacting or critical. In particular, the inclusion of a holographic self-energy leads to the existence of several different phases including a Fermi-liquid phase and a non-Fermi-liquid phase, separated by a quantum phase transition, as we have shown in section \ref{sect:QPT}.

There are various interesting directions in which our prescription can be extended. Firstly, our current prescription is intended to describe single-particle correlation functions, which appear often in condensed matter. However, in experiments one also encounters two-particle correlation functions. Important examples include the current-current correlation function that determines the electrical conductivity, $\sigma$, and the heat current-heat current
 correlation function which is related to the thermal conductivity $\kappa$. The electrical conductivity has been investigated in a holographic context before, starting with \cite{Hartnoll:2007ih}, but it would be interesting to also consider this particular function using our holographic prescription. In particular, since our present discussion is at zero doping, that is, it contains an equal amount of particles and holes, we expect the corresponding conductivity in the boundary system to remain finite in the $\o \rightarrow 0$ limit even without impurities. Additionally, in Fermi-liquid theory the thermal conductivity $\kappa$ scales as $1/T$, whereas the electrical conductivity $\sigma$ scales as $1/T^2$.
Their ratio is therefore proportional to temperature, i.e., $\kappa/\sigma \sim T$, which is the famous Wiedemann-Franz law. It would be interesting to investigate whether this law also holds here using the holographic results for electrical and thermal conductivity.

A topic often discussed in the AdS/CMT literature is the well-known Luttinger theorem from Fermi-liquid theory, which states that the total particle density in a system is proportional to the total volume in momentum space enclosed by its Fermi surfaces. In our case, this can be checked as follows. The system we describe is at zero doping, so the total number density of the particles and holes together is zero. The Fermi surfaces for particles and holes separately enclose a non-zero volume, but the difference in volume enclosed by the Fermi surfaces is zero. Therefore, the Luttinger theorem is trivially satisfied in our case.

Finally, whereas our current prescription takes into account only the leading-order contributions in the $1/N$ expansion, it would be interesting to also consider quantum or $\mathcal{O}(1/N)$ corrections. In particular, it is sometimes proposed \cite{ssprivate} that the typical feature of holographic Fermi liquids, an exponentially suppressed quasi-particle decay rate at the Fermi surface, may be a large $N$ remnant, i.e., a consequence of the fact that the leading order does not take into account all possible quasi-particle decay processes. Therefore, considering also $1/N$ contributions may resolve this problem and restore the conventional $\o^2$ power-law behavior, which is rather robust in Fermi liquids.
An example of $1/N$ corrections that are ignored here, are the back-reaction of the massive fermions on the gravitational background in the bulk, but one may also think of other $1/N$ corrections.

\section*{Acknowledgments}

It is a pleasure to thank R. Duine, C. Herzog, S. Hartnoll and S. Sachdev for discussions. This work was partially supported by the Netherlands Organization for Scientific Research (NWO) under the VICI grant 680-47-603.


\clearpage
\appendix \lab{sect:app}


\section{Conventions}


\subsection{A note on dimensions and units}
\label{app:units}

In this paper, we work with natural units in which $\hbar = c =k_B=1$, where $k_B$ is Boltzmann's constant, and we use dimensionless coordinates $r, t, \vec{x}$. As a consequence,
Newton's  constant $G_{d+1}$ as well as the metric are dimensionless.
However, to convert the Lifshitz metric \eqref{metric_02}
to standard SI units, we introduce a length scale $l$ and define the dimensionful expression
\begin{equation}
{\rm d}{\tilde s}^2\equiv l^2 {\rm d}s^2=\frac{l^2}{{\tilde r}^2}\frac{{\rm d}{\tilde r}^2}{V^2({\tilde r})} - V^2({\tilde r}) \, \frac{{\tilde r}^{\,2z}}{l^{2z}} c^2{\rm d}{\tilde t}^{\,2}
  + \frac{\tilde{r}^{\,2}}{l^2} {\rm d} {\vec {\tilde x}}^{\:2} \;,
  \end{equation}
where
\begin{equation}
{\tilde r}=l\,r\ ,\qquad {\vec {\tilde x}}=l\,\vec{x}\ ,\qquad {\tilde t}=\frac{l}{c}\,t\ ,
\end{equation}
which has the correct units of meter and second, respectively. Note that $l$ can be interpreted as the scale characterizing the size of the Lifshitz spacetime. The temperature of the black brane, obtained in \eqref{T}, can be converted to have dimension of Kelvin by the rescaling
\begin{equation}
{\tilde T}=\frac{\hbar c}{k_B}\frac{T}{l}\ ,
\end{equation}
and the mass of the fermions in SI-units takes the form
\begin{equation}
{\tilde M}^2=\frac{\hbar c}{{\tilde G}_{d+1}}\,\,M^2l^{d-3}\ .
\end{equation}
Here, ${\tilde G}_{d+1}$ is Newton's constant which has dimension ${\rm m}^d\cdot {\rm s}^{-2}\cdot {\rm kg}^{-1}$. To convert the dimensionless frequencies $\omega$ and momenta $\vec k$ to SI units, one should rescale
\begin{equation}
\tilde \omega =\frac{c}{l}\,\omega \ ,\qquad {\vec {{\tilde k}}}=\frac{\vec k}{l}\ .
\end{equation}
Employing the results of this section, the reader can at any time convert to physical units. However, for ease of notation, we present our analysis and results in dimensionless coordinates.


\subsection{Vielbeins and Dirac matrices}
\label{app:spinors}

Here, we collect some formulae related to the spin connection and vielbeins, and state our choice of Dirac matrices.
The vielbeins are denoted by $e^{\ul a}{}_{\mu}$ with $\mu=0,\ldots,d$ and $\ul a=0,\ldots,d$, and satisfy
\eq{
  g_{\mu\nu} = e^{\ul a}{}_{\mu} \, \eta_{\ul a\ul b} \,e^{\ul b}{}_{\nu} \;,\hspace{50pt} \eta_{\ul a\ul b} =
  {\rm diag}\, (-1,+1,\ldots, +1)
  \;.
}
We will furthermore employ the following notation and relations
\eq{
  \label{vielbein_01}
  e^{\ul a} = e^{\ul a}{}_{\mu} {\rm d}x^{\mu} \;,\hspace{50pt}
  \Omega^{\ul a}{}_{\ul b} = (\Omega_{\mu})^{\ul a}{}_{\ul b} \, {\rm d}x^{\mu} \;,\hspace{50pt}
  0 = {\rm d} e^{\ul a} + \Omega^{\ul a}{}_{\ul b} \wedge e^{\ul b} \;,
}
where $\Omega$ is the torsion-free spin connection. Given the metric \eqref{metric_02}, we can  read off
\eq{
  e^{\ul r} = \frac{{\rm d}r}{rV} \;,\hspace{60pt}
  e^{\ul t} = r^{z} V\, {\rm d}t \;,\hspace{60pt}
  e^{\ul i} = r\,{\rm d} x^i \;,
}
where $\ul a=\{\ul t,\ul r, \ul i\}$ with
$i=2,\ldots, d$ labeling the spatial directions. Using the relation in \eqref{vielbein_01} as well as the anti-symmetry of $\Omega$, we determine its non-vanishing components as
\eq{
  \Omega^{\ul t}{}_{\ul r} = r V \,\partial_r \bigl( r^{z} V \bigr) \, {\rm d}t \;,\hspace{60pt}
  \Omega^{\ul i}{}_{\ul r} = r V {\rm d}x^i \;.
}
For the spinors, we use the following notation
\eq{
  \label{def_01}
  \ov\Psi = \Psi^{\dagger} \Gamma^{\ul t} \;,\hspace{40pt}
  \mathcal{D}_{\mu} = \partial_{\mu} + \frac{1}{4}\, (\Omega_{\mu})_{\ul a\ul b} \Gamma^{\ul a\ul b}
  \;,\hspace{40pt}
  \slashed{\mathcal{D}} = \Gamma^{\ul a} e_{\ul a}{}^{\mu} \mathcal{D}_{\mu} \;.
}
Here, $e_{\ul a}{}^{\mu}$ is the inverse transpose of $e^{\ul a}{}_{\mu}$ and the symbol $\Gamma^{\ul a\ul b}$ is defined in terms of the Gamma matrices $\Gamma^{\ul a}$ as
\eq{
  \Gamma^{\ul a\ul b} = \frac{1}{2} \: \bigl[ \Gamma^{\ul a} , \Gamma^{\ul b} \bigr] \;.
}
We also perform chiral projections using
\eq{\label{chir-proj}
  \Psi_{\pm} \equiv \half\,\bigl(1\pm \Gamma^{\ul r}\bigr)\,\Psi\;,
  \hspace{40pt}
  \Gamma^{\ul r}\,\Psi_\pm=\pm \,\Psi_\pm\ .
  }
A convenient choice of Gamma matrices reads as follows  \cite{Liu2}:
\begin{itemize}

\item For odd dimensions $d$, we can choose
\eq{\lab{odd}
  \Gamma^{\ul r} = \left(\begin{array}{cc} \mathds{1} & 0 \\ 0 & -\mathds{1} \end{array}\right) \;,\qquad
  \Gamma^{\ul t} = \left(\begin{array}{cc} 0 & \gamma^{\ul 0} \\ \gamma^{\ul 0} & 0 \end{array}\right)\;,
  \qquad
  \Gamma^{\ul i} = \left(\begin{array}{cc} 0 & \gamma^{\ul i} \\ \gamma^{\ul i} & 0 \end{array}\right)\;,
}
where the  $\gamma$-matrices are the $d$-dimensional
Dirac matrices of the boundary theory.
We then decompose $\Psi$ into chiral components of the bulk spinor
\eq{
  \Psi = \binom{\psi_+}{\psi_-} \;,
}
where the components $\psi_\pm$ are chiral in the bulk, but they are Dirac spinors on the boundary.

\item In the case when $d$ is even, a natural choice is given by
\eq{
\lab{even}
\Gamma^{\ul r} = \gamma^{d+1}\;,\hspace{60pt}
\Gamma^{\ul t} = \gamma^{0}\;,\hspace{60pt}
\Gamma^{\ul i} = \gamma^{\ul i} \;,
}
where $\gamma^{d+1}$ is the analog of $\gamma^5$ in four
dimensions and $\gamma$ are the gamma matrices of the
boundary theory. In this case $\Psi_{\pm}$ correspond to operators
with {\em definite chirality} on the boundary.

\item Most of our interest lies in the case $d=4$. Using \eqref{chir-proj}, and with a slight abuse of notation, we write the four-component  Dirac bulk spinor $\Psi$ in terms of two-component spinors $\Psi_+$ and $\Psi_-$ as
\begin{equation}
\lab{defcomp}
\Psi=\begin{pmatrix}\Psi_+ \\ \Psi_-\end{pmatrix} \;.
\end{equation}
The four-dimensional gamma matrices can be expressed in terms of $\sigma^{\ul a} = (1, \vec\sigma)$ and $\ov\sigma^{\ul a} = (-1, \vec\sigma)$ with $\sigma^{\ul i}$ being the Pauli matrices in the following way
\eq{
  \Gamma^{\ul a} = \left( \begin{array}{cc} 0 & \ov\sigma^{\ul a} \\ \sigma^{\ul a} & 0 \end{array} \right)
  \;,\hspace{40pt} \ul a=\{\ul t, \ul i\}\ .
}

\end{itemize}


\section{Details on the near-boundary asymptotics}
\label{app:asymp}


\subsection{Dirac equation}

The Dirac equation for the fermion field $\Psi$ can be obtained from varying the action \eqref{totac} with respect to $\ov \Psi$, and imposing $\delta \Psi_+=0$ at $r=r_0$ (see \cite{Gursoy:2011gz} for more details). We find the usual expression $(\slashed{\mathcal{D}}-M)\Psi = 0$, which can be written as follows
\eq{
\lab{de1}
\biggl[ \:rV\, \Gamma^{\ul r}\, \6_r + \frac{i}{r}\, \Gamma\cdot
  \tk +\half\, \Gamma^{\ul r}\, p_z(r) -M\: \biggr] \psi(r) =0 \;.
}
Here we defined the function
$p_z(r) = r^{1-z} \partial_r [ r^z V ] +(d-1)V$, and used the notation
$\Gamma\cdot \tk = \Gamma^{\ul t}\, \tw + \Gamma^{\ul i}\, k_i$
together with the {\it generalized momenta}
\eq{\lab{gmom}
\tk = \bigl( \tw, k_i \bigr)\;, \hspace{60pt}
\tw =  -\frac{\omega}{r^{z-1}V}\;.
}
Next, we can simplify  equation \eqref{de1} by introducing
\eq{
\lab{cov}
\psi(r)= \frac{1}{\sqrt{r^{d-1+z}V}} \: \phi(r) \;,
}
which leads to
\eq{
\lab{de2}
\biggl[\: rV\,\Gamma^{\ul r} \6_r + \frac{i}{r}\, \Gamma\cdot
  \tk  -M\: \biggr] \phi(r) =0\;.
}
In terms of the chiral components $\phi_{\pm}(r)$, the Dirac equation \eqref{de2}, both in even and odd dimensions $d$, then reads
\eq{
\lab{de3}
\phi_\pm(r) = \mp \frac{i}{\tk^2}\: (\gamma\cdot \tk) \: \mathcal A( \mp M ) \, \phi_\mp(r) \;,
\hspace{60pt}
\mathcal A(M)\equiv r\bigl(r V \6_r - M \bigr)\;,
}
where we employed the notation for the $\Gamma$ matrices given in appendix \ref{app:spinors}.
Finally, we can also derive a {\em second-order} equation
by applying $\mathcal A(\pm M)(\gamma\cdot\tk)$ to both sides of the Dirac equation in \eqref{de3},
which leads to
\eq{
\lab{de4}
\tk^2 \phi_{\pm}(r)  = \mathcal A(\mp M)\mathcal A(\pm M) \phi_{\pm}(r) -
r^2V\: \frac{\partial_r \tw}{\tk^2}\: \gamma^{\ul 0}\, (\gamma\cdot \tk)
\mathcal A(\pm M)\phi_{\pm}(r)\;.
}
Note that for  the case of interest in this paper, that is $d=4$, half of the components in equations \eqref{de3} and \eqref{de4}
are trivial, because we can use the Weyl representation of the gamma matrices so that $\phi_\pm$ correspond to the upper and lower two components of $\phi$.


\subsection{Asymptotic solutions near the boundary}
\label{app_asymps}

We now determine the form of the asymptotic solutions near the boundary at $r=\infty$.
In contrast to the AdS case with $z=1$, for $z\neq 1$, the asymptotics are different for vanishing and non-vanishing spatial momentum $\vec{k}$. We therefore consider both cases in turn.


\subsubsection*{Asymptotics for $\vec{k}=0$}

In the case of vanishing spatial momentum, we make a power-law ansatz
for the second-order differential equation of the chiral components.
The asymptotic expansion near the boundary then comprises two independent solutions and is given by
\eq{
\lab{asymphi1}
\phi_\pm = r^{\pm M} \le( 1+  r^{-2z}c_\pm + \ldots \ri)A_\pm + r^{\mp M-z}\le(1 +  r^{-2z}d_\pm+\ldots \ri) B_\pm \; ,
}
where we
included subleading corrections in the  two separate branches.
Here, $A_\pm$ and $B_\pm$ are spinors in the Clifford algebra, and the coefficients $c_\pm$ and $d_\pm$ are matrices therein. We also observe that \eqref{asymphi1} agrees with the AdS case \cite{Liu2} for $z=1$,
and we note that a subleading term in the
$A$ (or $B$) branch can be more dominant over the leading term in the $B$ (or $A$) branch for certain ranges of $M$ and $z$.
Employing the expansion \eqref{asymphi1} in the first order Dirac equation \eqref{de3}, we  derive a local relation between $B_\mp$ and $A_\pm$ of the form
\eq{
\lab{ABrel11}
B_\pm = -i
\frac{\o\gamma^0}{2M \pm z}A_\mp\;,
}
which fixes half of the integration constants.
The coefficients of the subleading terms are found as
\eq{
\lab{c1}
c_\pm = \frac{\o^2}{2z (\pm 2M-z)}\ , \hspace{40pt} d_\pm = \frac{\o^2}{2z (\mp 2M-3z)}\;.
}
However, here we need to assume that the subleading terms in the expansion \eqref{asymphi1} are still leading over the subleading terms in the Dirac equation \eqref{de3} which appear in the function $V$. A careful  analysis translates this assumption into the condition
\begin{equation}
z < d-1\ .
\end{equation}

As we can see from \eqref{ABrel11} and \eqref{c1},  the above analysis breaks down for $M=\pm z/2$. Moreover, in that case the powers in the asymptotic expansion \eqref{asymphi1} can become equal and so the spinor coefficients cannot be disentangled. The way to treat the case $M=\pm z/2$ is to allow for logarithmic terms in the asymptotic expansion, which lift the degeneracy in the two branches $A$ and $B$.
In particular, for $M=-z/2$ we have
\eq{
\lab{asymphicp}
\phi_+  & = r^{-\frac{z}{2}} A_+ + r^{-\frac{z}{2}} \log(r) B_+ +\ldots \;, \\
\phi_- &= r^{+\frac{z}{2}} \Big(1+c_-r^{-2z}\log(r)+\ldots\Big)A_- + r^{-3\frac{z}{2}}\Big(1+\ldots\Big)B_-
\;,
}
with the coefficients of the form
\eq{
\lab{ABrelmc1}
B_+ =  i \o \gamma^0 A_- \;,\hspace{40pt}
B_-=\frac{i\omega\gamma^0}{2z}A_+  \;, \hspace{40pt}
c_-=\frac{\omega^2}{2z} \;.
}
For $M=+z/2$, we find
\eq{
\lab{asymphic}
\phi_+ &= r^{+\frac{z}{2}} A_+ + r^{-3\frac{z}{2}} B_+ +\ldots \;, \\
\phi_- &= r^{-\frac{z}{2}} A_- + r^{-\frac{z}{2}} \log(r) B_- +\ldots \;,
}
together with
\eq{\lab{ABrelc}
B_- = - i
\o \gamma^0 A_+\;, \hspace{40pt}
B_+ = -\frac{i\o \gamma^0}{2z} A_- \;.
}


\subsubsection*{Asymptotics for $\vec{k}\neq 0$}

For non-zero spatial  momenta $\vec k$ and generic values of the Dirac mass $M$, the asymptotic solution
to the second-order differential equation \eqref{de4} reads
\eq{
\lab{asymphi2}
\phi_\pm = r^{\pm M} \le( 1+  r^{-2}c_\pm + \ldots \ri)A_\pm
+ r^{\mp M-1}\le(1 +  r^{-\delta_\pm}d_\pm+\ldots \ri) B_\pm \;.
}
Note the important difference compared to
\eqref{asymphi1}, namely
that  the exponent of the leading order term does not depend on $z$.
The relation between $A_\pm$ and $B_\mp$ can again be derived via the first-order Dirac equation \eqref{de3}. Together with the subleading coefficient $c_{\pm}$, their form reads
\eq{
\lab{ABrel2}
B_\pm = i \frac{\vec{k}\cdot\vec{\gamma}}{2M\pm 1}A_\mp \;, \hspace{40pt}
c_\pm = - \frac{\vec{k}^2}{2(\pm 2M-1)} \;.
}
The subleading behavior in the B-branch turns out to be more complicated.
Both the exponent $\delta_\pm$ and the coefficient $d_\pm$ depend on the value of $z$. We have to distinguish between
three separate cases
\eq{
\lab{dd1}
  \begin{array}{r@{\hspace{25pt}}l@{\hspace{25pt}}l}
  1 < z < 3 \; :& \delta_{\pm} = z-1 \;,  &
     \displaystyle d_\pm = \frac{\o}{\vec{k}^2} \le(\frac{\pm 2M+ 1}{\pm 2M+ z}\ri)\vec{\gamma}\cdot\vec{k} \;, \\[4mm]
  z = 3 \; :& \delta_{\pm} = 2 \;,  &
     \displaystyle d_\pm= \frac{1}{3\pm 2M} \le( \frac{\vec{k}^2}{2} +\frac{\o}{\vec{k}^2} ( 2M \pm 1)\,\vec{\gamma}\cdot\vec{k}\ri) \;, \\[4mm]
  3<z \; :& \delta_{\pm} = 2 \;,  &
     \displaystyle     d_\pm = \frac{\vec{k}^2}{2(3\pm 2M)} \;.
  \end{array}
}
As is clear from \eqref{ABrel2}, the case of $M=\pm1/2$ is again special and needs to be treated separately. Along similar lines as for vanishing momenta, logarithmic terms will appear in the expansion.


\subsection{Hydrodynamic limit}
\lab{hydro}

We can determine the Green's function analytically in the strict  hydrodynamic limit $\o\to 0$, $\vec{k}\to 0$ for arbitrary temperatures $T$ in the region $|M|<z/2$.
However, one has to be careful in defining this limit because the $r\to r_h$ and $\o\to 0$ limits do not commute in the differential equation \eqref{xieq}.
Here, we define the limiting procedure as first setting $\vec{k}=0$, and then
taking $\o\to 0$ while keeping $T$ finite. Thus, we should use the same boundary condition $\xi(r_h)= i$ as before.
The differential equation \eqref{xieq}  for $\xi(r)$ then becomes
\eq{\lab{xieqhydro}
r^2V \partial_r \xi_{\pm} + 2Mr \xi_{\pm} = 0\;,
}
whose solution with the boundary condition $\xi_+(r_h) = i$ reads
\eq{
\lab{solhydro1}
\xi_{\pm} (r) = i \hspace{1pt} e^{-2M\int_{r_h}^r \frac{{\rm d}t}{t\, V(t)}} \;.
}
Substituting then the expression for $V(r)$ given in \eqref{metric_02} and performing the integral, we find
\eq{
\lab{solhydro2}
\xi_{\pm}(r) = i \le(\frac{r}{r_h}\ri)^{-2M} \bigl[1+ V(r)\bigr]^{-\frac{4M}{d+z-1}} \;,
}
where we have left the dimension $d$ and the dynamical exponent $z$ unspecified.
The self-energy appearing in the Green's function is given by \eqref{defsig},
and so we obtain
\eq{\lab{GRhydro2}
G_R(\vec 0,0) = \frac{i}{g}\:  2^{\frac{4M}{d+z-1}} \le( \frac{d+z-1}{4\pi }\ri)^{\frac{2M}{z}} T^{-\frac{2M}{z}}\;,
}
where we employed  \eqref{T} to related $r_h$ to the temperature $T$. Note that this expression is valid for $d+z-1\ne 0$, which includes the case of interest in this paper, namely $d=4$ and $z=2$.


\section{Symmetry properties of the Green's function}
\lab{app:symm}

In this appendix, we discuss symmetries of the self-energy $\Sigma(p)$ defined in \eqref{defsig} and of the full Green's function \eqref{GR2}. This analysis utilizes the properties of the differential equation \eqref{xieq} and boundary condition \eqref{bc}, which we recall for convenience
\eq{
\lab{xieq2}
r^2V \dau_r\xi_\pm + 2Mr \xi_\pm = \frac{\omega}{r^{z-1}V} \mp k_3 + \left( \frac{\omega}{r^{z-1}V} \pm k_3\right)
\xi_\pm^2 \;,
\hspace{40pt}
\xi_{\pm}(r_h) = i \;.
}


\subsection{Scale-invariant variables}
\lab{app:sca}

We start our discussion with the scaling properties of \eqref{xieq2}. To simplify our notation, we  define the constant $\kappa = \frac{d+z-1}{4\pi}$, and we can distinguish three cases which will be considered in turn.
\begin{itemize}

\item We define new variables in the following way
\eq{
  \ov\omega = \frac{\omega}{k_3^{z}} \;, \hspace{40pt}
  \ov T = \frac{T}{k_3^{z}} \;, \hspace{40pt}
  x = \frac{r}{k_3} \;,
}
which  leads to
\eq{
  V = \sqrt{1- \left( \frac{x_h}{x} \right)^{d+z-1}} \;, \hspace{40pt}
  x_h = \left( \frac{\ov T}{\kappa} \right)^{\frac{1}{z}} \;.
}
The differential equations \eqref{xieq2} then takes the form
\eq{
x^2V \dau_x\xi_\pm + 2Mx \xi_\pm = \frac{\ov \omega}{x^{z-1}V} \mp 1 + \left( \frac{\ov\omega}{x^{z-1}V} \pm 1 \right)
\xi_\pm^2 \;,
\hspace{30pt}
\xi_{\pm}(x_h) = i \;,
}
and we obtain
\eq{
\xi_{\pm} \bigl(\ov x_0, \ov\omega, \ov T \bigr)  = k_3^{-2M} \,\xi_{\pm} \bigl( r_0,\omega, k_3, T \bigr) \;.
}

\item Along similar lines, we can define another set of variables as follows
\eq{
  \ov k = \frac{k_3}{\omega^{\frac{1}{z}}} \;, \hspace{40pt}
  \ov T = \frac{T}{\omega} \;, \hspace{40pt}
  y = \frac{r}{\omega^{\frac{1}{z}}} \;,
}
leading to
\eq{
  V = \sqrt{1- \left( \frac{y_h}{y} \right)^{d+z-1}} \;, \hspace{50pt}
  y_h = \left( \frac{\ov T}{\kappa} \right)^{\frac{1}{z}} \;,
}
as well as
\eq{
y^2V \dau_y\xi_\pm + 2My \xi_\pm = \frac{1}{y^{z-1}V} \mp \ov k + \left( \frac{1}{y^{z-1}V} \pm \ov k\right)
\xi_\pm^2 \;,
\hspace{30pt}
\xi_{\pm}(y_h) = i \;.
}
We then find
\eq{
\xi_{\pm} \bigl( \ov y_0,\ov k, \ov T \bigr)  = \omega^{-\frac{2M}{z}}\, \xi_{\pm} \bigl( r_0,\omega, k_3, T \bigr) \;.
}

\item Finally, we can define variables as
\eq{
  \ov\omega = \frac{\kappa}{T} \:\omega \;, \hspace{40pt}
  \ov k = \left(\frac{\kappa}{T}\right)^{^{\frac{1}{z}}} k_3 \;, \hspace{40pt}
  z = \frac{r}{r_h} \;,
}
where we recall that $r_h = ( T / \kappa)^{\frac{1}{z}}$. The above choice implies
\eq{
  V = \sqrt{1- \left( \frac{z_h}{z} \right)^{d+z-1}} \;, \hspace{50pt}
  z_h = 1  \;,
}
and the differential equations become
\eq{
z^2V \dau_z\xi_\pm + 2Mz \xi_\pm = \frac{\ov \omega}{r^{z-1}V} \mp \ov k + \left( \frac{\ov \omega}{r^{z-1}V} \pm \ov k\right)
\xi_\pm^2 \;,
\hspace{30pt}
\xi_{\pm}(z_h) = i \;.
}
We then have
\eq{
\xi_{\pm} \bigl( \ov z_0, \ov \omega, \ov k \bigr)  = \left( \frac{T}{\kappa} \right)^{-\frac{2M}{z}} \xi_{\pm}\bigl( r_0, \omega, k_3, T \bigr) \;.
}

\end{itemize}


\subsection{Chirality}

From equation \eqref{xieq2} we observe that  the spin-up
and spin-down  components $\xi_+$ and $\xi_-$  are related by a reflection of the momentum
\eq{
   \xi_{\pm}(r, k_3,\omega)  =  \xi_{\mp}(r,- k_3 ,\omega) \;.
}
Furthermore,
equations \eqref{defsig} and \eqref{GR2} which relate $\xi$ to the self-energy $\Sigma(p)$ and to the Green's function $\Sigma$ to $G_R$ do not spoil this symmetry. We thus obtain the following relation for the components of the full Green's function
\eq{
G_R^{\pm}(\kv,\om) = G_R^{\mp}(-\kv,\om) \;.
}
This symmetry can easily be understood in the relativistic case of $z=1$:
since $\psi_\pm$ correspond to
left- and right-handed spinors in the boundary theory, the helicity $h$ given by \mbox{$h=\vec{\sigma}\cdot\kv/|\vec{\sigma}\cdot\kv|$} is conserved.
In the non-relativistic case $z\ne 1$, this may be viewed as a generalization of chirality.


\subsection{Particle-hole symmetry}

From equation \eqref{xieq2} we also observe that the components $\xi_+$ and $\xi_-$ enjoy the additional symmetry
\eq{\lab{symPH}
  \xi_\pm(r, k_3 ,\omega) = - \xi_\pm^*(r,-k_3,-\omega) \;,
}
where $\xi^*$ denotes the complex conjugate of $\xi$.
Noting then again that  \eqref{defsig} and \eqref{GR2} preserve this property as well, we see that also the full Green's function satisfies
\eq{
 \label{phsymm}
 G_R^{\pm}(\kv,\o) = -\le(G_R^{\pm}(-\kv,-\om)\ri)^* = - \le(G_R^{\mp}(\kv,-\om)\ri)^* \;,
}
where we used the chirality symmetry in the last equality. This symmetry can be seen as a {\em particle-hole} symmetry since it relates components with frequency $\om$ to components with frequency $-\om$.


\subsection{A symmetry relating $M$ to $-M$}
\label{app:Msym}

In the case when the operator under consideration is right handed, that is $\cO_+$, the source is identified with the boundary value of $\psi_-$ instead. Hence,
the transfer matrix $\xi(r,p)$ defined in \eqref{xi1} is replaced with
\eq{
\lab{defxin}
\zeta_+ = i\,\frac{\psi_{+,1}}{\psi_{-,1}} \;,\hspace{40pt}
 \zeta_-= i\,\frac{\psi_{+,2}}{\psi_{-,2}} \;,
}
where $\psi_{1,2}$ are the up and down components of the spinors defined in \eqref{psipm} and \eqref{FTspinor}.
Let us now consider $\zeta_+$ for definiteness, the discussion for $\zeta_-$ is analogous. First, we see that $\zeta_+ = -1/\xi_+$, and we easily verify that the equation for $\zeta_+$
is exactly the same equation \eqref{xieq} that $\xi_+$ satisfies, but with the replacement $k_3\to -k_3$ and $M\to -M$.
In addition,  the boundary conditions  \eqref{bc} for $\zeta_+$ and $\xi_+$ are the same, and therefore we have the following symmetry property:
\eq{
\lab{invxi}
\zeta_+(r,M,\o,k_3) = -\xi_+^{-1}(r,M,\o,k_3) = \xi_+(r,-M,\o,-k_3) \;.
}
On the other hand, equation \eqref{defsig}
is now replaced with
\eq{
 \Sigma(p) = -g \lim_{r_0\to\infty} r_0^{-2M} \zeta (r_0,p)\;,
 }
and so we arrive at
\eq{
\lab{symGR}
\Sigma(-M,-\vec k,\o) = -g^2 \, \Sigma^{-1}(+M,+\vec k,\o)  \;.
}
However, note that strictly speaking this derivation is valid only within the range $-z/2<M<z/2$ for $\vec{k}=0$ and within  $-\half<M<\half$ for $\vec{k}\ne 0$.  Outside this region one needs
to subtract divergent terms  to obtain the Green's functions from $\xi$ and $\zeta$, which can be shown by observing that the counter terms that are needed in order to renormalize $G_R$ also respect this symmetry.

As a consequence of this symmetry, conditions on the functions $s_{1,M}$ and $s_{2,M}$ from \eqref{Gzko} can be obtained as
\eq{
\lab{cond1}
&s_{1,-M}(u) s_{2,+M}(u) =s_{1,+M}(u)s_{2,-M}(u)\;,\\
&s_{1,+M}(u) s_{1,-M}(u) - s_{2,+M}(u)s_{2,-M}(u) = -1 \;.
}
Solving \eqref{cond1} in favor of $s_{2,M}(u)$, we find for non-degenerate values of $s_{1,M}$ and $s_{2,M}$ that
\eq{
\lab{f2}
s_{2,M}(u) = s_{1,M}(u)\sqrt{1+ \frac{1}{ s_{1,M}(u)\,s_{1,-M}(u)}  } \;,
}
and by using \eqref{f2} in \eqref{Gzko} we  arrive at
\eq{
 \lab{Gzko2}
G_{R}( \vec k,\o) = -\frac{1}{\omega -\frac{1}{\lambda}\, \sik\; k^{z-1} - g\,k^{2M} s_{1,M}(u)
\le( 1 +\frac{\sik}{k}  \sqrt{1+\frac{1}{s_{1,M}(u)s_{1,-M}(u)}}  \ri)}\;.
}
Thus, we have reduced the problem to the determination of a single function $s_{1,M}(u)$ of a single variable $u=\omega/k^z$ for two values of $M$.


\subsection{A symmetry of the momentum distribution}

Using the above symmetries of the Green's function in combination with the sum rule, we can easily derive a convenient identity for the momentum distribution \eqref{mom_dis_4}. Namely, we have
\eq{ \lab{nksymm}
N^{\pm}_{\kv} + N^{\pm}_{-\kv} & = \frac{1}{\pi} \int_{-\infty}^{+\infty}\;\textrm{d}\om \;\left[ \I G_R^{\pm}(\kv,\om) n_F(\om) + \I G_R^{\pm}(-\kv,\om) n_F(\om)\ri]\\
& = \frac{1}{\pi} \int_{-\infty}^{+\infty}\;\textrm{d}\om \; \I G_R^{\pm}(\kv,\om) \le[n_F(\om) + n_F(-\om) \ri] \\
& = 1\;,
}
where we used \eqref{phsymm} in the second line, and the sum rule \eqref{sumrule} and $n_F(\om) + n_F(-\om) = 1$ in the third line.


\section{Pole structure and sum rules}
\label{app:sum_rule}

In this appendix,  we analyze the pole structure of the Green's function, and we show analytically for  the case of vanishing temperature that the sum rule is satisfied. The situation of vanishing momenta can be analyzed for all values of $z$, however,  analytic expressions for the Green's function for $k\neq0$ are only available for $z=1$. We therefore restrict the latter computations to the relativistic case.


\subsection{Pole structure}
\label{app:sum3}

We now consider the pole structure of the Green's function in the case of vanishing momentum and zero temperature. Employing \eqref{Gz}, we see that there is a trivial pole at $\omega=0$ which corresponds to the free particle. A non-trivial pole is determined by the equation
\eq{
  \label{poles01}
  0 = 1 -  g \hspace{1pt}c_z\:  \omega^{\frac{2M}{z}-1} e^{-i \pi (\frac{M}{z}+\half)} \;.
}
Using then polar coordinates $\omega = |\omega| e^{i \theta}$, we find for the imaginary part of \eqref{poles01} the equation
\eq{
  0 = \sin \left[ \frac{2M-z}{z} \,\theta - \frac{2M+z}{z} \, \frac{\pi}{2}\right] \;,
}
with solutions
\eq{
  \label{poles02}
  \theta_n = \frac{\pi}{2} - \frac{n+1}{z-2M}\, \pi \hspace{1pt}z \;,
  \hspace{40pt} n \in \mathbb{Z} \;.
}
If we want to satisfy the Kramers-Kronig relations, that is the requirement of no poles in the upper half-plane, we have to make sure that there is no solution to \eqref{poles02} in the range $(0,+\pi)$ for all values of $n$, that is
\eq{
\lab{nosol}
  \t_n\notin (0,+\pi) \hspace{40pt}  \forall n \;.
}
By noting that $0<z-2M<2z$ for $|M|<\frac{z}{2}$, we see that the only possible solution in the forbidden  range is given by $n=-1$ with $\theta_{-1} = \frac{\pi}{2}$.
All the other solutions are already out of the forbidden range, but yield poles on the non-principal sheets of the complex plane. For example,
for $n=-2$ the pole resides in $\t_{-2} \in (\pi, +\infty)$ depending on the value of $M$, while for $n=0$ it is in the range $\t_{0} \in (0,-\infty)$.

Let us now turn to the real part of \eqref{poles01}. Employing again polar coordinates and using
\eqref{poles02}, we arrive at the equation
\eq{
 0 = 1 - g \hspace{1pt}c_z\:  |\omega|^{\frac{2M}{z}-1} (-1)^n\;,
}
which is solved by
\eq{
  \label{poles03}
  |\omega| = |g \hspace{1pt} c_z |^{\frac{z}{z-2M}} \;, \hspace{50pt}
  {\rm sign}\, (g \hspace{1pt} c_z) = (-1)^{n} \;.
}
It is now clear that in order to avoid the pole $\theta_{-1} = \frac{\pi}{2}$ on the principal sheet, we have to require
\eq{
  \label{poles04}
  g \hspace{1pt} c_z >0 \;.
}
Furthermore, all other possible poles of the Green's function (specified by \eqref{poles02} and \eqref{poles03}) with even $n$
are either on the lower-half of the principal sheet or on non-principal sheets of the complex plane.
Therefore, for the choice of sign \eqref{poles04} the Kramers-Kronig relations are satisfied.

For the case of vanishing temperature and $z=1$, analytic results also for $\kv\neq0$  are available. As it has been explained in \cite{Gursoy:2011gz}, the analytic continuation of the Green's function for non-vanishing momenta is given by  replacing $\omega \to \sqrt{\omega^2 - k^2}$. This implies that there are poles at $\omega = \pm |\vec k|$, with the pole in the lower half-plane corresponding to the retarded Green's function. Furthermore, for the choice of sign \eqref{poles04}, there are no additional poles in the upper half-plane of the principal sheet. Therefore, we have shown that in the relativistic case the Kramers-Kronig relations are satisfied also for non-vanishing momenta.


\subsection{Sum rule for $\vec k=0$}
\label{app:sum1}

The Green's function in the case of vanishing temperature and momenta can be found in equation \eqref{Gz}. The spectral density \eqref{imspec} then reads
\eq{
  \rho(\kv=0,\om) = \frac{1}{2\pi} \I \Tr \bigl[G_R(\kv=0,\om) \bigr]
  =
 -\frac{1}{\pi} \I \left[ \frac{1}{\omega -  g \hspace{1pt}c_z\:  \omega^{\frac{2M}{z}} e^{-i \pi (\frac{M}{z}+\half)}} \right]
 \;.
}
Introducing the following constants
\eq{
  \mathcal B = g \hspace{1pt} c_z \cos \left[ \pi \left( \frac{M}{z} + \frac{1}{2} \right) \right] \;,
  \hspace{ 40pt}
  \mathcal C = g \hspace{1pt} c_z \sin \left[ \pi \left( \frac{M}{z} + \frac{1}{2} \right) \right] \;,
}
we  express the spectral density as
\eq{
  \rho(\kv=0,\om)
  =
 +\frac{1}{\pi} \, \frac{ \mathcal C\, \omega^{\frac{2M}{z}}}{
 \left( \omega - \mathcal B \,\omega^{\frac{2M}{z}} \right)^2 +
 \left( \mathcal C\, \omega^{\frac{2M}{z}} \right)^2 } \;.
}
Next, to compute the sum rule we recall the symmetry $\rho(\vec k,-\omega) = \rho(\vec k,+\omega)$. We then have
\eq{
  \int_{-\infty}^{+\infty} {\rm d}\omega\, \rho (\vec k=0, \omega)
  &=  2 \int_{0}^{+\infty} {\rm d}\omega\, \rho (\vec k=0, \omega) \\
  & = \frac{2}{\pi} \,\frac{z}{2M -z} \,\arctan \left[
  -\frac{\mathcal B}{\mathcal C} + \omega^{\frac{2M}{z}} \, \frac{\mathcal B^2+\mathcal C^2}
  {\mathcal C}   \right]_{\omega =0}^{\omega=\infty} \\
  & = \frac{2M}{2M-z} - \frac{z}{2M-z} \:{\rm sign}\hspace{1pt}( g \hspace{1pt}c_z) \;.
}
Therefore, for the choice of sign \eqref{posgot} we indeed find that the sum rule \eqref{sumrule} for vanishing momenta and vanishing temperature is satisfied
\eq{
  \int_{-\infty}^{+\infty} {\rm d}\omega\, \rho (\vec k=0, \omega) = 1 \;.
}


\subsection{Sum rule for $\vec k\neq0$}
\label{app:sum2}

In the case of $z=1$ and vanishing temperature, the expression for the Green's function
was given in equation \eqref{GRAdS2} together with \eqref{realline}.
To compute the sum rule in this situation, we write
\eq{
  \int_{-\infty}^{+\infty} {\rm d}\omega \, \rho(\vec k,\omega)
  =  \frac{2}{2\pi} \int_{0}^{+\infty} {\rm d}\omega \,{\rm Im}\,
  \frac{-2\omega}{p^2\le(1-g \hspace{1pt}c_1\, e^{-i\pi(M+\half)}p^{2M-1}\ri)} \;,
}
with $p$ as in \eqref{realline}.
Next, we take \eqref{realline} into account to obtain
\eq{
    \int_{-\infty}^{+\infty} {\rm d}\omega \, \rho(\vec k,\omega)
    = &+\frac{1}{\pi} \int_{0}^{+|\vec k|} {\rm d}\omega\,
    {\rm Im}\,
  \frac{-2\omega}{(\omega^2 - k^2) \le(1+g \hspace{1pt} c_1\, (k^2-\omega^2)^{2M-1}\ri)} \\
  & + \frac{1}{\pi} \int_{+|\vec k|}^{+\infty}
  {\rm d}\omega \,{\rm Im}\,
  \frac{-2\omega}{(\omega^2 - k^2)^2\le(1-g \hspace{1pt}c_1\, e^{-i\pi(M+\half)}(\omega^2-k^2)^{2M-1}\ri)} \;,
}
where we note that the first term on the right-hand side vanishes due to taking the imaginary part.
Performing then the change of variables $x = \omega^2 - k^2$, we find
\eq{
    \int_{-\infty}^{+\infty} {\rm d}\omega \, \rho(\vec k,\omega)
     = - \frac{1}{\pi} \int_0^{\infty} {\rm d}x \,
     \I \left[ \frac{1}{x^2-  g \hspace{1pt}c_1\:  x^{2M+1} e^{-i \pi (M+\half)}} \right] \;.
}
After a further change of variables $y=x^2$, we arrive at
\eq{
    \int_{-\infty}^{+\infty} {\rm d}\omega \, \rho(\vec k,\omega)
     = - \frac{2}{\pi} \int_0^{\infty} {\rm d}y \,
     \I \left[ \frac{1}{y-  g \hspace{1pt}c_1\:  y^{2M} e^{-i \pi (M+\half)}} \right] \;,
}
which is a special case of the computation in appendix \ref{app:sum1}. We therefore find
\eq{
      \int_{-\infty}^{+\infty} {\rm d}\omega \, \rho(\vec k,\omega) = 1 \;.
}


\section{WKB calculation}\label{app:WKB}

In this appendix, we present details of the calculation leading to equation (\ref{WKBres}). We consider the case $M>0$, $T=0$ and without loss of generality make the assumptions $\o>0$, $\vec k=(0,0,k_3)$ and $k_3>0$. Furthermore, we consider the spin-up component of the Green's function, the other cases can be worked out  using the symmetries outlined in appendix \ref{app:symm}.

The parameter which we take large in the WKB approximation is $\bk \equiv k/\sqrt{\o}$, and the fluctuation equations of $u_{\pm}$ are given by (\ref{de4}) which yield
\begin{align}
\lab{wkb1}
0 &= \6_r^2 u_\pm + \le[\frac{2}{r}  + \frac{\o}{r^2\tilde{k}^2}\left(-\frac{\o}{r}\pm k\right)\ri]\6_r u_\pm +  \le[-\frac{\tilde{k}^2}{r^4}-\frac{M\o}{r^3\tilde{k}^2}\left(\mp\frac{\o}{r} + k\right) - \frac{M(M\pm1)}{r^2}\ri]u_\pm \nonumber
\\
&\equiv  \6_r^2 u_\pm + f_1^\pm(r) \6_r u_\pm + f_2^\pm(r) u_\pm \;,
\end{align}
where $\tilde k$ had been defined in \eqref{gmom}.
This expression can be put in the Schr\"odinger form by the transformation
\eq{
\lab{wkb2}
u_\pm = h_\pm(r) y_{\pm}\;,
\hspace{40pt}
 h_{\pm}(r) = \exp\le(-\half\int^r {\rm d}t \,f_1^\pm(t)  \ri) \;.
}
A suitable parametrization of the extra spatial coordinate in the limit $\bk \gg1$ is given in terms of $x=r/\sqrt{\o}$.
In this variable, the  transformation \eqref{wkb2} reads as follows
\eq{
\lab{wkb3}
h_\pm(x)  =  \frac{\sqrt{x \bk \pm 1}}{x^{3/2}} \;.
}
Then, the Schr\"odinger equations become
\eq{
\lab{wkb31}
\frac{1}{\bk^2} \,\frac{d^2}{dx^2}y_\pm(x) - \bV_\pm(x)\, y_\pm(x)=0 \;,
}
with the effective Schr\"odinger potentials
\eq{
\lab{wkb4}
\bV_{\pm}(x)=\frac{1}{x^4}-\frac{1}{\bk^2 x^6} +\frac{(\half\pm M)(\frac32\pm M)}{\bk^2x^2} +\frac34 \frac{1}{(x\bk \pm 1)^2}
-\frac{\frac32\pm M}{x\bk(x\bk \pm 1)}\;.
}
In addition to solving the above Schr\"odinger equation for $y_+$ analytically using the WKB approximation as described below, we have also solved it numerically. The numerical results are presented in figure \ref{figWKBwfnum}, which we discuss as we proceed in our computation.

\begin{figure}[p]
\centering
\vskip6pt
\subfigure[Double-logarithmic plot of $|V_+|$. The turning point is approximately $\xtp=1/10$.]
{
\includegraphics[width=0.46\textwidth]{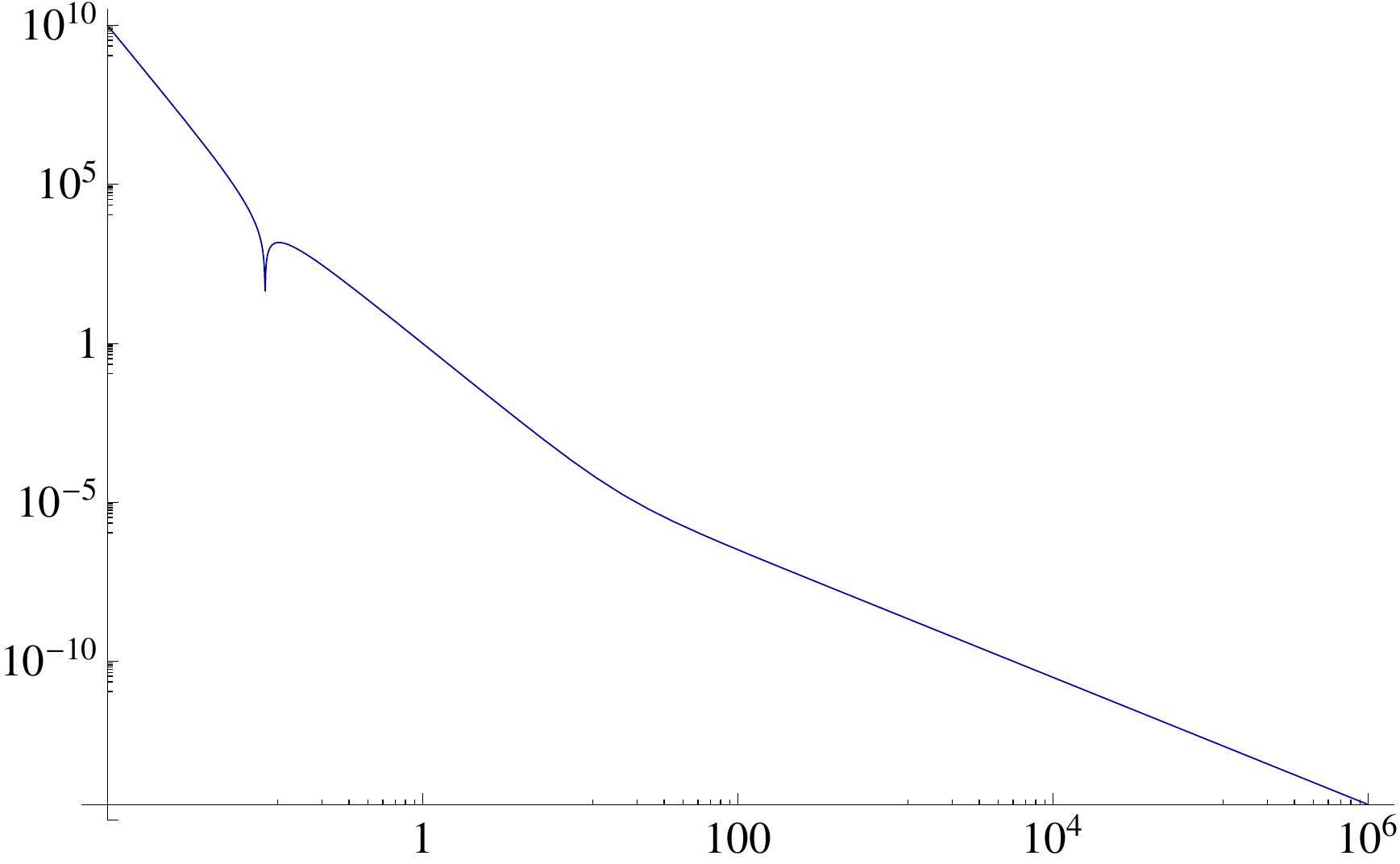}
\begin{picture}(0,0)
\put(-1,7){\scriptsize $x$}
\put(-190,126){\scriptsize $|V_+|$}
\end{picture}
}
\hspace{\stretch{1}}
\subfigure[Double-logarithmic plot of $\R y_+$. The imaginary part is similar but with a minus sign. Left of the turning point there is oscillatory behavior, the other regions exhibit exponential behavior.]{
\includegraphics[width=0.46\textwidth]{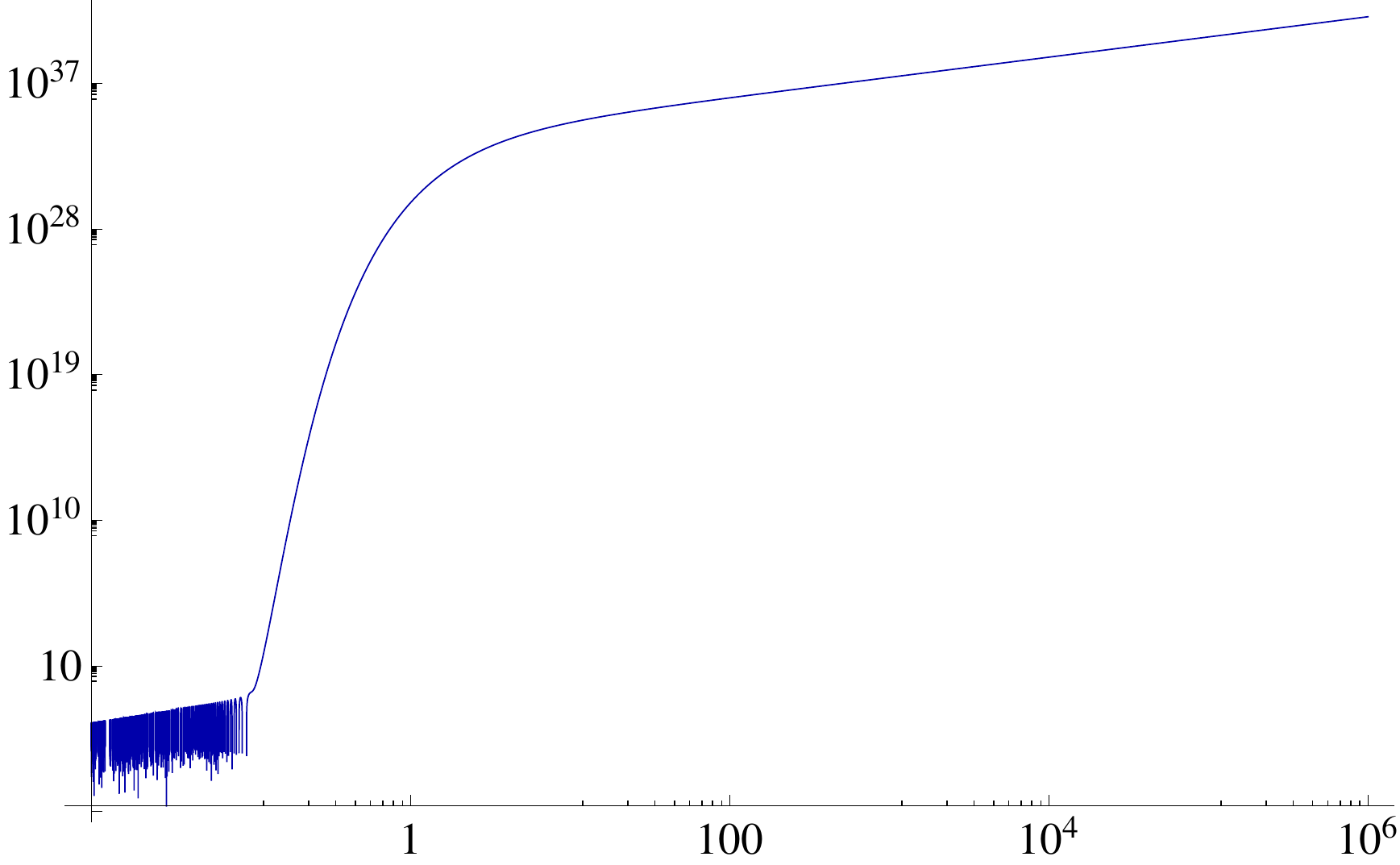}
\begin{picture}(0,0)
\put(-1,7){\scriptsize $x$}
\put(-193,125){\scriptsize $\R y_+$}
\end{picture}
}
 \\[5mm]
\subfigure[Near-horizon oscillatory behavior. Until the turning point, there is good agreement between the numerical (blue) and WKB (red) result. After that, the WKB solution is no longer valid.]{
\includegraphics[width=0.56\textwidth]{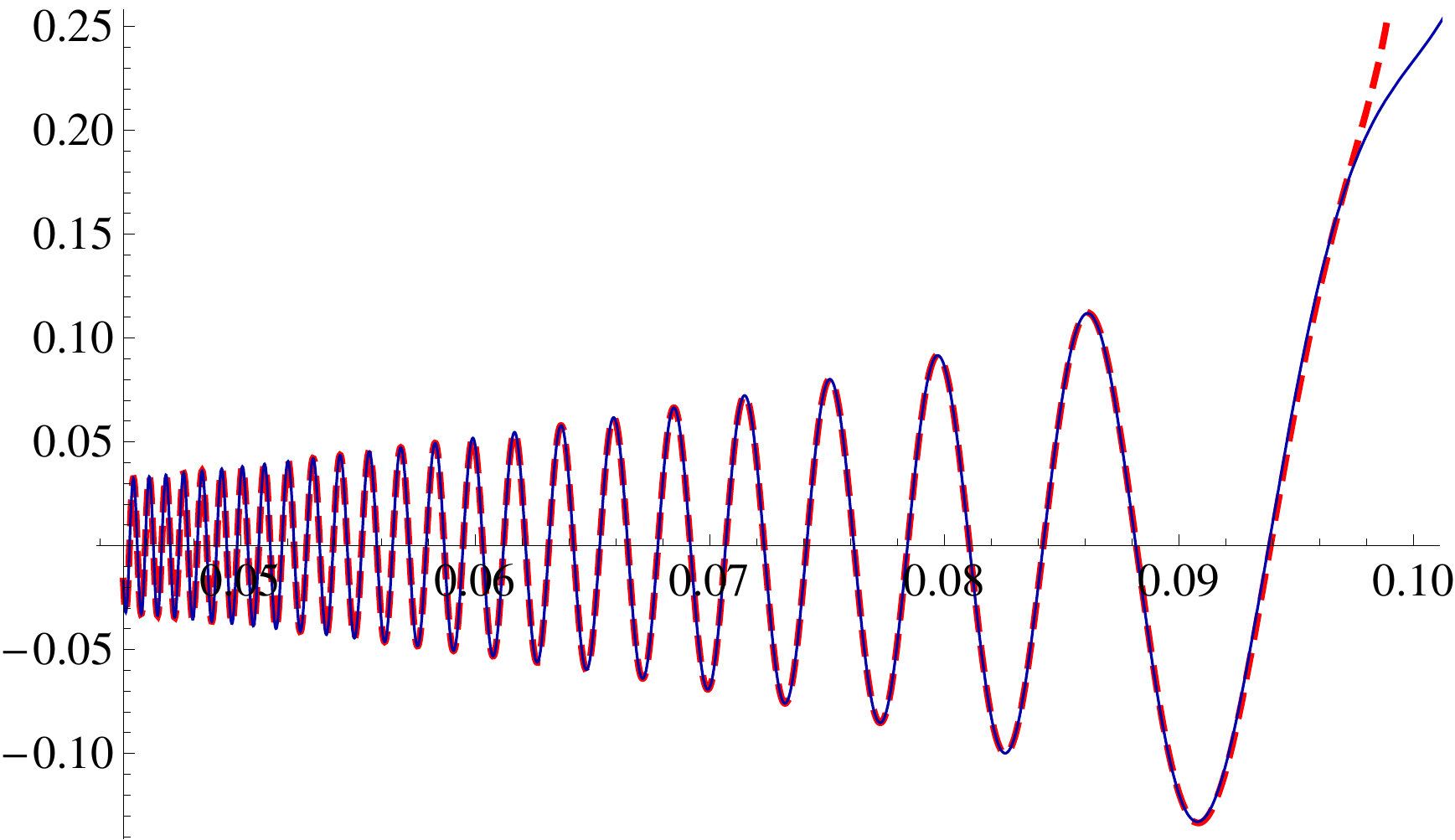}
\begin{picture}(0,0)
\put(0,45){\scriptsize $x$}
\put(-228,140){\scriptsize $\R y_+$}
\end{picture}
}
\\[5mm]
\subfigure[Matching procedure near the classical turning point.
The dashed red curves are the WKB results, the solid blue curve is the numerical solution, and the solid green curve is the Airy function solution needed in the matching formulae which agrees with the numerics around the turning point.]{
\includegraphics[width=0.56\textwidth]{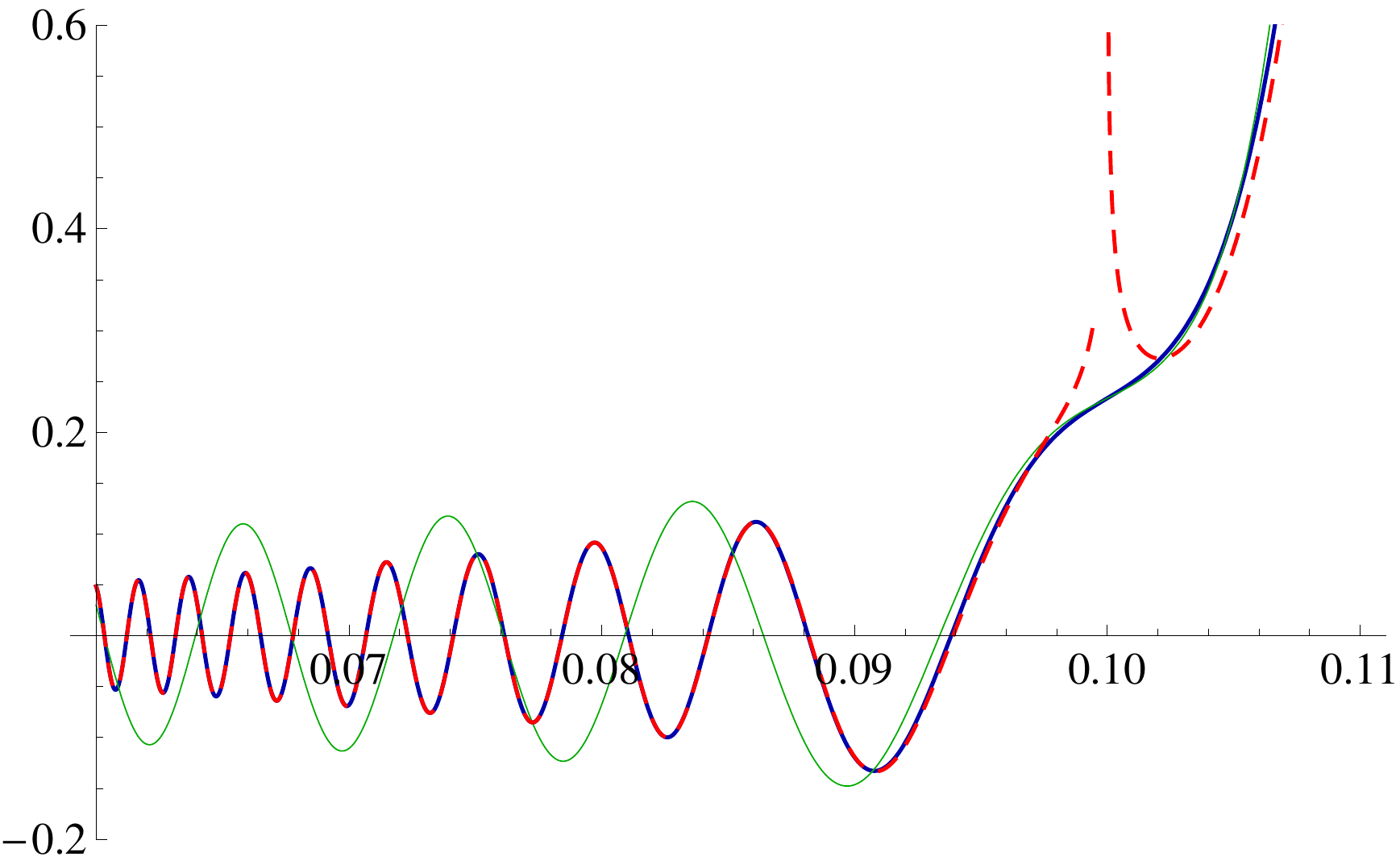}
\begin{picture}(0,0)
\put(0,36){\scriptsize $x$}
\put(-232,148){\scriptsize $\R y_+$}
\end{picture}
}
\end{figure}
\begin{figure}[p]
\ContinuedFloat
\centering
\vspace{\stretch{1}}
\subfigure[Double-logarithmic plot of $\R y_+$ in the classically disallowed region.
The solid blue curve is the numerical solution, the dashed lines are the analytic solution \eqref{wkb7} for exponents $\alpha=\pi/2$ (red) and $\alpha=5/3$ (yellow). The former agrees with the numerical result, the latter deviates from it. For large $x$, the WKB results are not valid anymore but are taken over by the Bessel-function solutions  plotted on the far right, which overlap with the WKB solution for smaller $x$.]{
\includegraphics[width=0.65\textwidth]{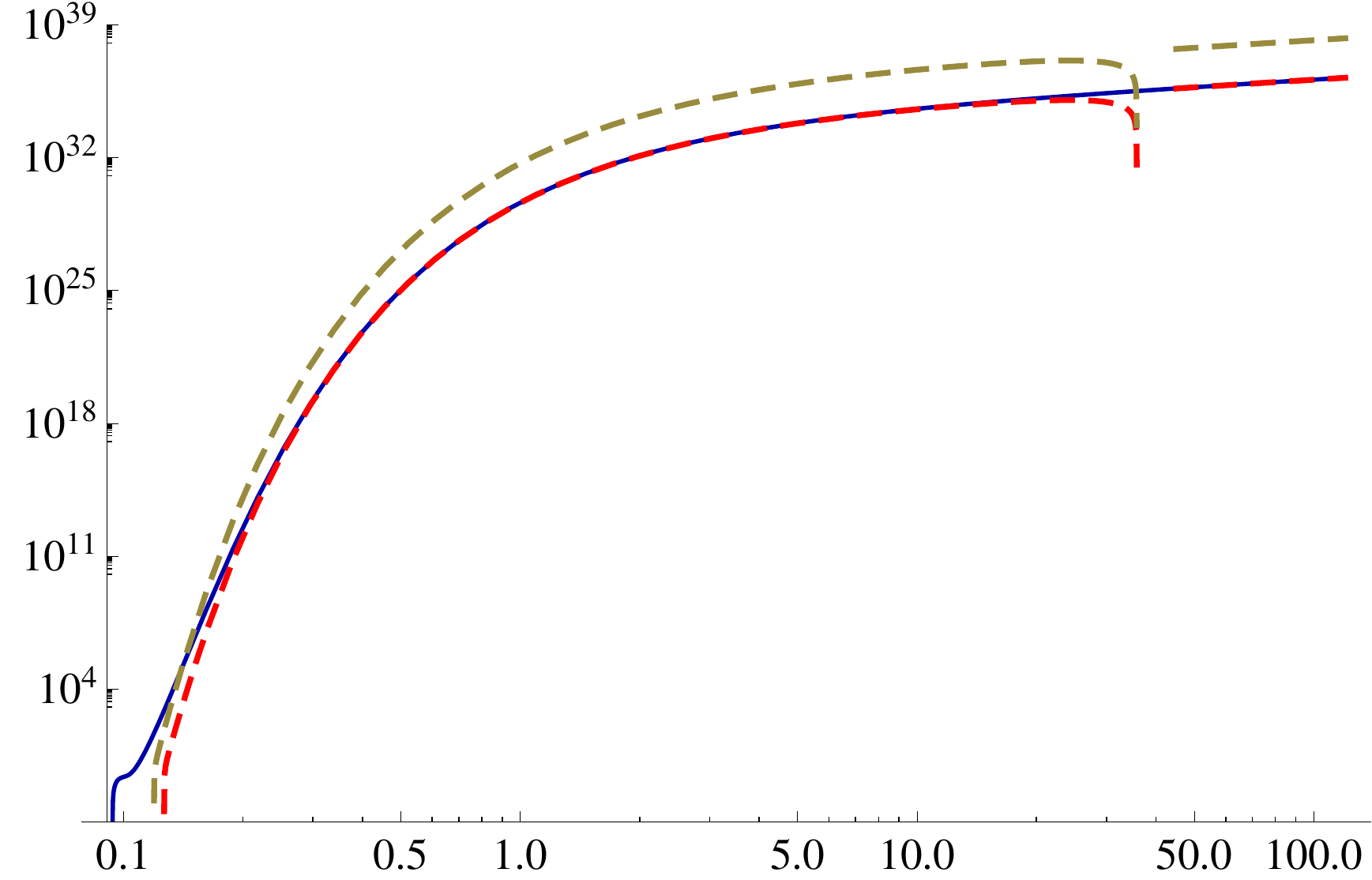}
\begin{picture}(0,0)
\put(-1,9.5){\small $x$}
\put(-267,180){\small $\R y_+$}
\end{picture}
}
\\[8mm]
\subfigure[Double-logarithmic plot of the near-boundary region, where the Bessel-function solution is valid. The blue curve is the numerical result, the dashed  curves are the analytic result \eqref{wkb9} for exponents $\alpha=\pi/2$ (red) $\alpha=5/3$ (yellow).]{
\includegraphics[width=0.65\textwidth]{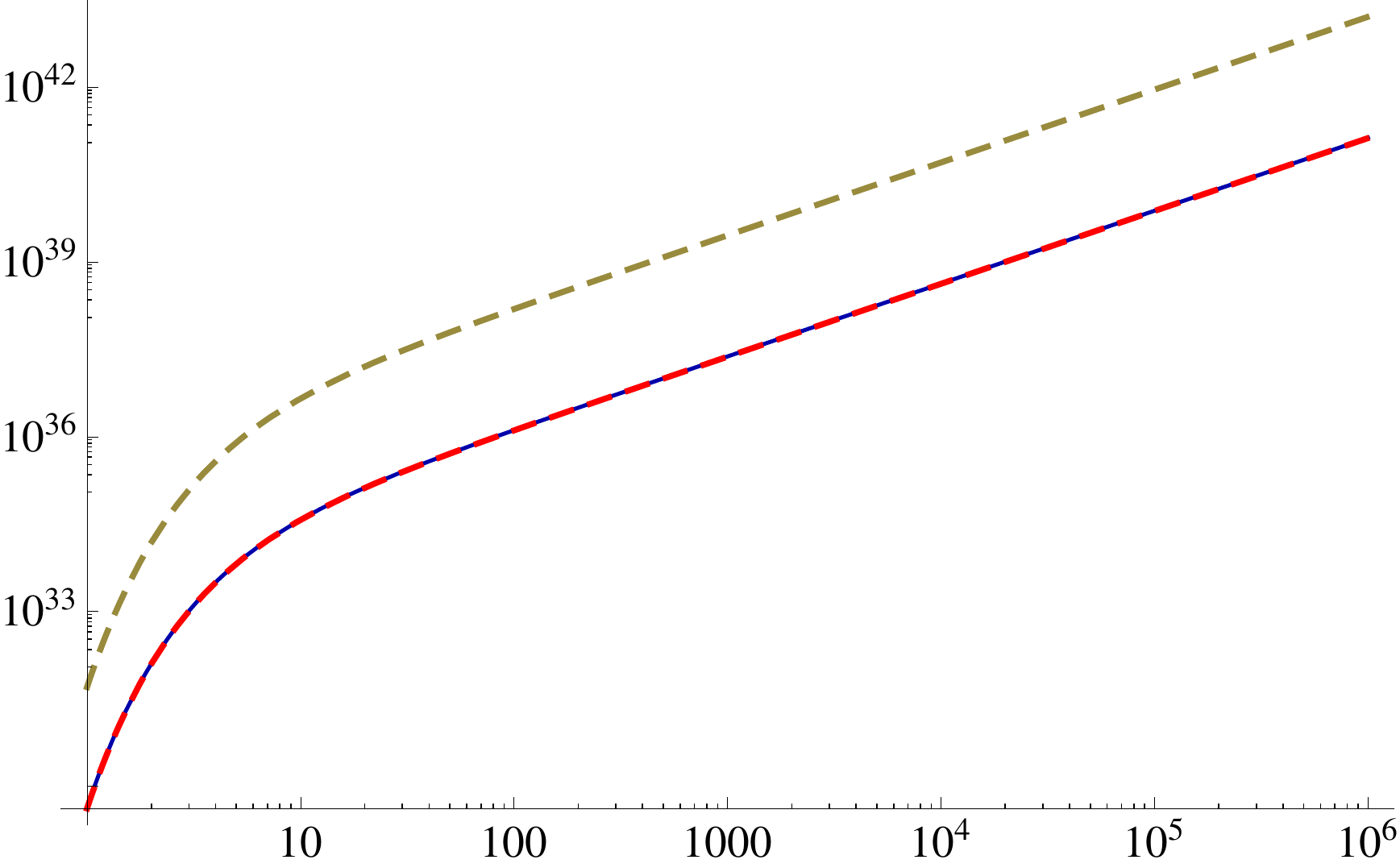}
\begin{picture}(0,0)
\put(0,9.5){\small $x$}
\put(-270,175){\small $\R y_+$}
\end{picture}
}
\hspace{5pt} \caption{Illustration of the numerical WKB calculation for $\bk=10$ and $M=1/4$.}
\label{figWKBwfnum}
\end{figure}

We now focus on the plus component of the fluctuations. In the region close to the horizon at $x \ll 1$, and in the limit $\bk \gg 1$, the relevant piece of the potential for  $y_+$
 reduces to
\eq{
\lab{wkb5}
\bV_+ \approx \frac{1}{x^4}-\frac{1}{\bk^2 x^6} \;,
}
from which we see that there is a turning point located at approximately $\xtp = 1/\bk$. In the region $x< \xtp$ the WKB wavefunction is given by
\eq{
\lab{wkb51}
y_+ (x) = \frac{A_1}{(-\bV_+)^{1/4}}\, e^{+ i\bk \int_x^{\xtp} {\rm d}t \sqrt{-\bV_+(t)} } +   \frac{A_2}{(-\bV_+)^{1/4}}\, e^{- i \bk\int_x^{\xtp} {\rm d}t \sqrt{-\bV_+(t)}} \;,
}
with $\bV_+$ obtained from (\ref{wkb5}). The solution for $x<\xtp$ is compared with the numerics in figure \ref{figWKBwfnum}(c).
This solution should be connected to the solution near the horizon at $x\to 0$, which  yields the condition that $A_2=0$. We will not need to fix the coefficient $A_1$ because, as shown below, it will cancel out in the final answer.
The next step is to continue the solution from the region $x<\xtp$ to the region $x>\xtp$ by using the well-known WKB connection formula, see for example \cite{Iizuka,HHV}. This procedure is illustrated in figure \ref{figWKBwfnum}(d). One important subtlety here is that in the final answer we will need second-order corrections to the WKB formula as well. Noting that the parameter that
plays the role of $\hbar$ here is $1/\bk$, we find,
\eq{
\lab{wkb6}
y_+ (x) = \frac{A_1 e^{-i\frac{\pi}{4}}}{\bV_+(x)^\frac14} \le[ e^{+  \bk \int_{\xtp}^x {\rm d}t\sqrt{\bV_+(t)}}\le(1+ \frac{\sigma_+(x)}{\bk}\ri) +   \frac{i}{2} e^{-  \bk \int_{\xtp}^x {\rm d}t\sqrt{\bV_+(t)}}\le(1+ \frac{\sigma_-(x)}{\bk}\ri)\ri] \;,
}
where the second-order WKB coefficients can be found as (see for example \cite{Landau})
\eq{
\lab{wkb61}
\sigma_\pm = \pm \int_{\xtp}^x {\rm d}t \le( \frac18 \frac{\bV_+''(t)}{\bV_+(t)^\frac32} - \frac{5}{32}\frac{\bV_+'(t)^2}{\bV_+(t)^\frac52}\ri) \;.
}
This is our WKB solution in the classically disallowed region $\xtp\ll x<\infty$, where we can now approximate the potential by\footnote{Note that we keep the leading $1/\bk^2$ corrections in order to be consistent with keeping the $\sigma_\pm$ corrections in (\ref{wkb6}).}
\eq{
\lab{wkb8}
\bV_+\to \frac{1}{x^4}+\frac{M(M+1)}{\bk^2x^2} + \frac{M}{x^3\bk^3} - \frac{1}{x^6 \bk^2} \;.
}
To the desired order in $\bk$, (\ref{wkb61}) then yields
\eq{\lab{wkb62}
\pm \sigma_\pm = \frac{1}{4\bk^2 x} + \frac{M(M+1)x^3}{4\bk^2}+\frac{M x^2}{8\bk^3} -\frac{1}{4\bk} + \ldots  \;,
}
and with this approximation we find from (\ref{wkb6}) the following solution, which is illustrated in figure \ref{figWKBwfnum}(e),
\eq{\lab{wkb7}
y_+ (x) = A_1\,x\,e^{-i\frac{\pi}{4}} \le( e^{-\frac{\bk}{x}} e^{\bk^2}\kappa_+(x) +\frac{i}{2}  e^{+\frac{\bk}{x}} e^{-\bk^2}\kappa_-(x)\ri),
}
where we employed
\eq{\lab{wkb71}
\xtp = \frac{1}{\bk} - \frac{M(M+1)}{\bk^5} + \ldots \;,
}
and have defined the functions
\eq{\lab{wkb72}
\kappa_\pm \approx \le[1\pm \frac{M(M+1)x}{2\bk} \pm \frac{M}{2\bk^2}\log(\bk x)\ri]
\!\!
\le[1-\frac{M(M+1)x^2}{4\bk^2}+\frac{1}{4x^2\bk^2}\ri]
\!\!
\le[1\mp \frac{1}{4\bk^2} \pm \frac{1}{4\bk^3 x}\ri]\!.
}
Note that here we only show terms up to the desired $1/\bk^2$ order. The terms in the first brackets come from the corrections to the exponential term in (\ref{wkb6}), the second terms come from the expansion of the $\bV_+^{-1/4}$  and the last ones from the $\sigma_\pm$ corrections in (\ref{wkb6}).

The next step is to connect this solution to the region near the boundary at $x\to\infty$ and read off the Green's function from the coefficients of the near-boundary expansion. Note that the WKB approximation in the near-boundary limit $x\to\infty$ fails, because in this region we have an
inverse-square potential which does not satisfy the WKB condition $|V_+'(x)/V^\frac32(x)|\ll 1$. Thus, one has to solve the Schr\"odinger equation with the potential (\ref{wkb8}) exactly and connect it to the solution (\ref{wkb7}) in the overlapping region $\bk/x \gg 1$. In the latter, we can approximate the potential as
\eq{\lab{wkb73}
\bV_+(x) \approx \frac{1}{x^4} + \frac{M(M+1)}{x^2\bk^2} \;,
}
and the exact solution to the Schr\"odinger equation with this potential is found in terms of the Bessel functions as
\eq{\lab{wkb9}
y_+ (x) = \sqrt{x} \le[ B_1\, I_{-\half-M} \le(\frac{\bk}{x}\ri) +B_2\, I_{\half+M} \le(\frac{\bk}{x}\ri) \ri],
}
where $B_{1,2}$ are coefficients so far undetermined.
This solution is also shown in figure \ref{figWKBwfnum}(e). In order to connect it to (\ref{wkb7}), we use the well-known asymptotic formula for the Bessel functions for large values of the variable $\bk/x$ and we find
\eq{\lab{wkb10}
y_+ (\bk/x) \to  \frac{x}{\sqrt{2\pi \bk}} \le[ B_1 \le(e^{\frac{\bk}{x}} +  e^{-i \pi M} e^{-\frac{\bk}{x}}\ri) +B_2 \le(e^{\frac{\bk}{x}} -  e^{i \pi M} e^{-\frac{\bk}{x}}\ri) \ri] +\ldots \;,
}
from which we obtain
\eq{\lab{wkb11}
B_1 = \frac{A_1e^{-i\frac{\pi}{4}}\sqrt{2\pi\bk}}{2\cos(\pi M)}\le(\frac{i}{2}  e^{i\pi M} e^{-\bk^2} \epsilon_- +  e^{\bk^2} \epsilon_+\ri)\, .
}
We shall not need the expression for $B_2$, and the constants $\epsilon_\pm$ are defined as follows
\eq{\lab{wkb111}
\epsilon_\pm = 1\pm \frac{M}{2\bk^2}\log\bk \mp \frac{1}{4\bk^2}\, .
}
On the other hand, using equation \eqref{wkb3} together with the series expansion of the Bessel functions near $x\to\infty$ given in (\ref{wkb9}),
we find that
\eq{\lab{wkb12}
u_+(x) \approx  \bar{A}^u_+ x^M +\ldots \;,
}
where the superscript $u$ denotes that we consider the spin-up component of the spinor $\bar{A}_+$.
This is in accord with the asymptotic solution given in appendix \ref{app_asymps} with
\eq{\lab{wkb13}
\bar{A}^u_+  = B_1 \frac{\bk^{-M} 2^{M+\half}}{\Gamma(\half-M)}\, ,
}
where $B_1$ is determined by (\ref{wkb11}).

\medskip
The next step in the calculation of the Green's function is to obtain the WKB solution for the $u_-$ component in \eqref{wkb1}. One way would be to apply the same steps as above in the WKB calculation, however this process is more difficult than the $u_+$ case, since the potential for $u_-$  shown in (\ref{wkb4}) has extra poles at $x=1/\bk$. Fortunately, $u_-$ is related to $u_+$ by the first  order differential equation
(\ref{de3}), and so we can use the result for $u_+$ to obtain $u_-$ directly.  Equation (\ref{de3}), specified to the spin-up component and in the rescaled variables $x$ and $\bk$, reads
\eq{
\lab{wkb14}
u_-=\frac{i}{\bk +\frac{1}{x}} \le( x^2 \6_x -Mx\ri)u_+\, ,
}
which for the solution (\ref{wkb6}) becomes
\eq{
\lab{wkb15}
u_-=\frac{i}{\bk +\frac{1}{x}} \le[ x^2 \le(\frac{h_+'(x)}{h_+(x)}-\frac14 \frac{\bV'(x)}{\bV(x)}\ri) -Mx \ri]u_+ +A(x) \, ,
}
where the function $A(x)$ is given by
\eq{
\lab{wkb16}
A(x) = \frac{ix^2 h_+ }{\bk +\frac{1}{x}}\:
\frac{A_1 e^{-i\pi/4}}{\bV_+^{1/4}(x)}
\biggl[ \hspace{24pt}
&e^{+  \bk \int_{\xtp}^x {\rm d}t\sqrt{\bV_+(t)} }\le(\bk\bV_+^\half(x)\left(1+ \frac{\sigma_+(x)}{\bk}\right) +\frac{\sigma'_+(x)}{\bk}\ri)    \\
-\frac{i}{2} \;&e^{-  \bk \int_{\xtp}^x {\rm d}t\sqrt{\bV_+(t)} }\le(\bk\bV_+^\half(x)\left(1+ \frac{\sigma_-(x)}{\bk}\right) -\frac{\sigma'_-(x)}{\bk}\ri)\bigg] .
}
On the other hand, very close to the boundary, we can approximate the Schr\"odinger potential as
\eq{\lab{wkb17}
\bV_-(x) \approx \frac{1}{x^4} + \frac{M(M-1)}{x^2\bk^2},
}
and again, using this potential in the Schr\"odinger-like equation for $y_-$, the exact solution reads
\eq{\lab{wkb18}
y_- (x) = \sqrt{x} \le[ C_1\, I_{-\half+M} \le(\frac{\bk}{x}\ri) +C_2\, I_{\half-M} \le(\frac{\bk}{x}\ri) \ri] \;,
}
with $C_{1,2}$ some coefficients to be determined.
Similarly as above, expanding the Bessel functions in this solution for large $x/\bk$,  and using $u_- = h_-(x) y_-(x)$ with (\ref{wkb3}),
we find
\eq{\lab{wkb19}
u_-(x) \approx  \bar{A}^u_- x^{-M} +\ldots \;,
}
with
\eq{\lab{wkb20}
\bar{A}^u_-  = C_1 \frac{\bk^{M} \,2^{-M+\half}}{\Gamma(\half+M)}\, .
}
Next, we employ the asymptotic formula for the Bessel functions  in (\ref{wkb18}) for large  $\bk/x$ to
match to the WKB result (\ref{wkb15}). In this matching, we also need to use (\ref{wkb16}), (\ref{wkb7})  and the relation $u_\pm(x) = h_\pm(x)y_\pm(x)$ with (\ref{wkb3}). The result is
\eq{\lab{wkb21}
C_1 = \frac{A_1e^{-i\frac{\pi}{4}}\sqrt{2\pi\bk}}{2\cos(\pi M)}\le( \half e^{-i\pi M} e^{-\bk^2} \g_- + i e^{\bk^2} \g_+\ri)\;,
}
and we shall not need the expression for $C_2$. The constants $\g_\pm$ read
\eq{\lab{wkb23}
\g_\pm  = 1\pm \frac{M}{2\bk^2}\log\bk \mp \frac{3}{4\bk^2}\; .
}

\medskip
Finally, as can be inferred for instance from (\ref{xi1}), the spin-up eigenvalue of the matrix $\xi$ is proportional to the ratio of $u_-$ and $u_+$. Thus, using its definition from (\ref{defsig}), the spin-up component of the self-energy is proportional to the ratio of the constants $\bar{A}^u_-$ and $\bar{A}^u_+$. To be more precise,
\eq{
\Sigma^+  = - i\, g\, \o^{M} \frac{\bar{A}^u_-}{\bar{A}^u_+}
 = - i \, g\, 2^{-2M} k^{2M}\,\frac{\Gamma\le(\tfrac{1}{2}-M\ri)}{\Gamma\le(\tfrac{1}{2}+M\ri)} \frac{C_1}{B_1}\;.
}
If we then use (\ref{wkb11}), (\ref{wkb13}), (\ref{wkb20}) and  (\ref{wkb21}), we obtain the final result for the spin-up component of the self-energy as it was given in (\ref{WKBresSigma}).

Before closing this appendix, let us remark the following.
As explained at the beginning of section \ref{sec:WKBmain}, the numerical prefactors both of the terms in $\epsilon_{\pm}$ and $\gamma_{\pm}$ and the one in the exponent, i.e., the number $    \alpha$ in the end result $\I \Sigma \sim \exp(-\alpha k^2/\o)$, are difficult to determine analytically. This is because the integral in(\ref{wkb6}) leading to $\kappa_{\pm}$ cannot be done analytically. In the calculation presented here, we have approximated the potential $\bV_+$ by taking into account only the three most relevant terms in the region $x>\xtp$, which leads to $\alpha=2$ as in (\ref{wkb7}). This is reasonable away from the turning point, but close to  the lower integration limit $\xtp$ the $x^{-6}$ term has a considerable contribution  not taken into account. Furthermore, the turning point itself is determined by setting to zero an approximation to the effective potential.
To do better, we should use the full potential everywhere, and expand the integrand as a series in $1/\bk$. However, this series is not uniformly convergent because the integration limit $\xtp$ depends on $1/\bk$. In practice we can terminate the series at some point, which leads to a different approximation of the numerical prefactors. For example, if the full effective potential is taken into account and the integrand $\sqrt{\bV_+}$ is expanded in $1/\bk$, the next contribution of order $\bk^2$ is $-\bk^2/6$, which results in $\alpha = 5/3$, and thus in a decay rate of $\Gamma \sim \exp(-5k^2/3 \om)$. Furthermore, we can do the same for the integral in the $\sigma_{\pm}$ from (\ref{wkb61}). Also, in the prefactor $\bV_+^{-1/4}$ the full potential can be taken into account.
Then, the $\mathcal{O}(1/\bk^{2})$ terms in $\g_{\pm}$ and $\epsilon_{\pm}$ also receive extra corrections. Still, we are left with an approximation of the exact numbers, and it is not clear whether there occur cancellations between terms later on in the $1/\bk$ expansion.

\begin{figure}[t]
\centering
\vskip 10pt
\includegraphics[width=0.7\textwidth]{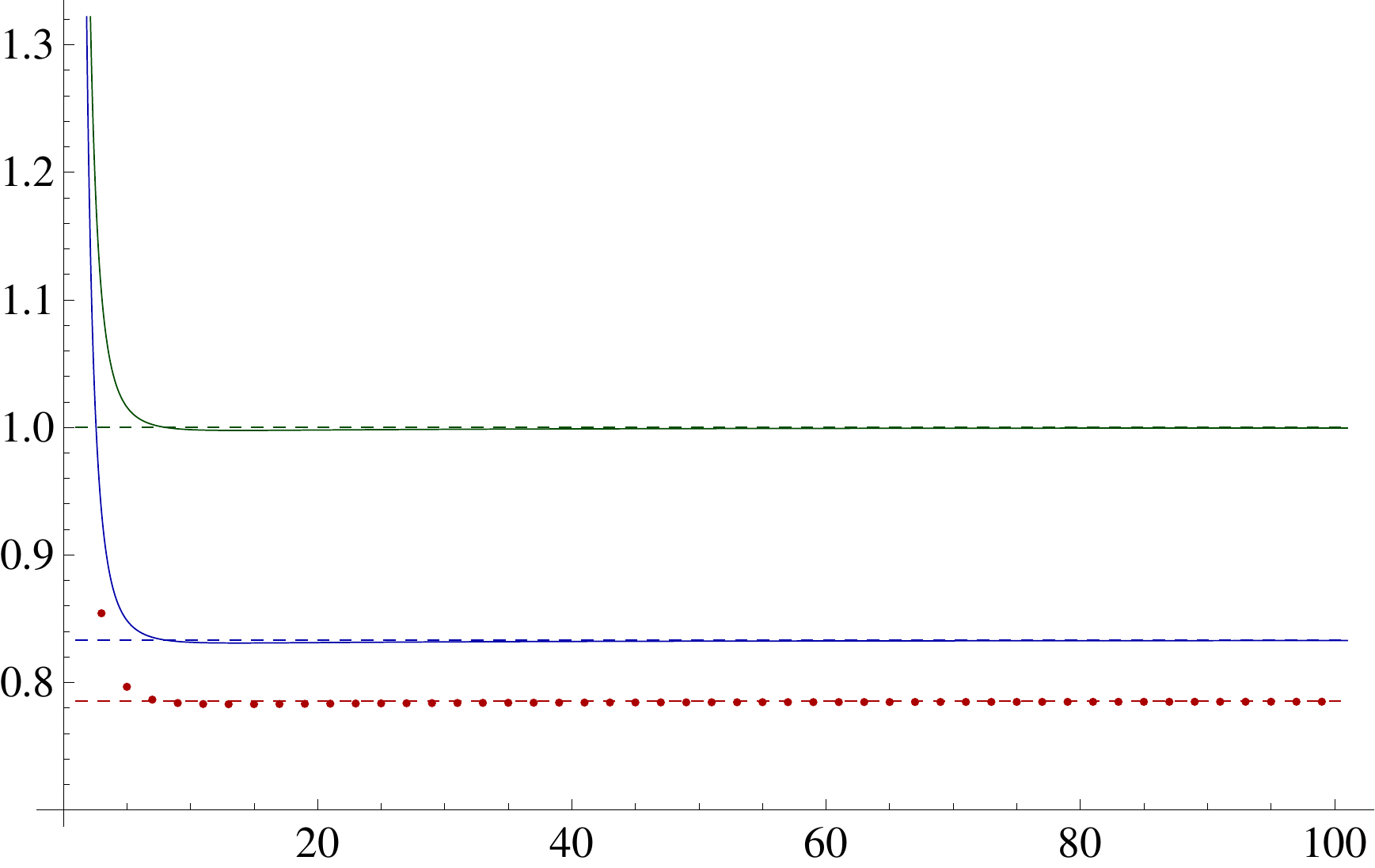}
\begin{picture}(0,0)
\put(-318,195){\small $\frac{1}{\bk}\int_{\xtp}^x {\rm d}t\sqrt{V_+(t)}$}
\put(0,8){\small $\bk$}
\end{picture}
\caption{Plot of $\frac{1}{\bk}\int_{\xtp}^x {\rm d}t \sqrt{V_+(t)} $ as a function of $\bk$, which appears in the exponents of (\ref{wkb6}) that determine the quasi-particle decay rate. For large $\bk$, the $\bk^2$ term in the exponent is dominant and therefore the curves converge to the numerical prefactor of this term. In this way, the correct factor can be determined numerically.
Here, we have taken $M=+1/4$ and $x=20$ as the upper integration limit, which is an intermediate value of $x$. The value of the integral is independent of $M$ for $\bk \rightarrow \infty$.
\newline
The solid green curve is the result when the two most relevant terms in the potential are taken into account,  leading  to $\alpha = 2$ (dashed green line).
The solid blue curve is the resulting expression when the integration over the full potential is done analytically before terminating the series in $\bk$, leading to the $\alpha=5/3$ prefactor of the $\bk^2$ term (the dashed blue line is the number $5/6$).
The red dots show the value of the integral when it is performed numerically, and the dashed red line is the number $\pi/4$. The former converge to the latter curve, which is evidence that the value of the integral is approximately $\pi/4 \bk^2 + \ldots$. Thus, the exponent that determines the quasi-particle decay rate in the final WKB  result will be $\exp(-\pi k^2/2\om)$.}
\label{figpi4}
\end{figure}

The integral over $\sqrt{\bV_+}$ can however be performed numerically. Figure \ref{figpi4} compares the numerical and analytic values of the number $\alpha/2$ that appears in the exponent. The numerical result is approximately $\pi/4$, whereas our improved approximated analytic calculation yields $5/6$. When we use $\alpha=\pi/2$ in the exponent, the resulting self-energy  agrees more accurately with the numerical results, as is shown in figures \ref{figImSWKB} and \ref{figWKBwfnum}.



\clearpage
\begin{thebibliography}{00}

\bibitem{Hartnoll:2009sz}
  S.~A.~Hartnoll,
  {\it Lectures on holographic methods for condensed matter physics},
  Class.\ Quant.\ Grav.\  {\bf 26} (2009) 224002
  [arXiv:0903.3246 [hep-th]].

\bibitem{Herzog:2009xv}
  C.~P.~Herzog,
  {\it Lectures on Holographic Superfluidity and Superconductivity},
  J.\ Phys.\ A A {\bf 42} (2009) 343001
  [arXiv:0904.1975 [hep-th]].

\bibitem{McGreevy:2009xe}
  J.~McGreevy,
  {\it Holographic duality with a view toward many-body physics},
  Adv.\ High Energy Phys.\  {\bf 2010} (2010) 723105
  [arXiv:0909.0518 [hep-th]].

\bibitem{Sachdev:2010ch}
  S.~Sachdev,
  {\it Condensed Matter and AdS/CFT},
  [arXiv:1002.2947 [hep-th]].

\bibitem{Hartnoll:2011fn}
  S.~A.~Hartnoll,
  {\it Horizons, holography and condensed matter},
  [arXiv:1106.4324 [hep-th]].


\bibitem{Lee:2008xf}
  S.~-S.~Lee,
  {\it A Non-Fermi Liquid from a Charged Black Hole: A Critical Fermi Ball},
  Phys.\ Rev.\ D {\bf 79} (2009) 086006
  [arXiv:0809.3402 [hep-th]].

\bibitem{Faulkner:2009wj}
  T.~Faulkner, H.~Liu, J.~McGreevy, D.~Vegh,
  {\it Emergent quantum criticality, Fermi surfaces, and AdS(2)},
  Phys.\ Rev.\ D {\bf 83} (2011) 125002
  [arXiv:0907.2694 [hep-th]].

\bibitem{Liu:2009dm}
  H.~Liu, J.~McGreevy, D.~Vegh,
  {\it Non-Fermi liquids from holography},
  Phys.\ Rev.\  D {\bf 83} (2011) 065029
  [arXiv:0903.2477 [hep-th]].

\bibitem{Zaanen}
  M.~Cubrovic, J.~Zaanen, K.~Schalm,
  {\it String Theory, Quantum Phase Transitions and the Emergent Fermi-Liquid},
  Science {\bf 325} (2009) 439
  [arXiv:0904.1993 [hep-th]].

\bibitem{Cubrovic:2010bf}
  M.~Cubrovic, J.~Zaanen, K.~Schalm,
  {\it Constructing the AdS Dual of a Fermi Liquid: AdS Black Holes with Dirac Hair},
  JHEP {\bf 1110} (2011) 017
  [arXiv:1012.5681 [hep-th]].

\bibitem{Pomarol}
  R.~Contino, A.~Pomarol,
  {\it Holography for fermions},
  JHEP {\bf 0411} (2004) 058
  [arXiv:hep-th/0406257].

\bibitem{Faulkner:2010tq}
  T.~Faulkner, J.~Polchinski,
  {\it Semi-Holographic Fermi Liquids},
  JHEP {\bf 1106} (2011) 012
  [arXiv:1001.5049 [hep-th]].

\bibitem{Hartnoll:2009ns}
  S.~A.~Hartnoll, J.~Polchinski, E.~Silverstein, D.~Tong,
  {\it Towards strange metallic holography},
  JHEP {\bf 1004} (2010) 120
  [arXiv:0912.1061 [hep-th]].

\bibitem{Gursoy:2011gz}
  U.~Gursoy, E.~Plauschinn, H.~Stoof, S.~Vandoren,
  {\it Holography and ARPES sum-rules},
  JHEP {\bf 1205} (2012) 018
   [arXiv:1112.5074 [hep-th]].



\bibitem{Kachru}
  S.~Kachru, X.~Liu, M.~Mulligan,
  {\it Gravity Duals of Lifshitz-like Fixed Points},
  Phys.\ Rev.\  D {\bf 78} (2008) 106005
  [arXiv:0808.1725 [hep-th]].

\bibitem{Taylor}
  M.~Taylor,
  {\it Non-relativistic holography},
  [arXiv:0812.0530 [hep-th]].

\bibitem{Keranen:2012mx}
  V.~Keranen, L.~Thorlacius,
  {\it Thermal Correlators in Holographic Models with Lifshitz scaling},
  [arXiv:1204.0360 [hep-th]].

\bibitem{Horowitz:2011gh}
  G.~T.~Horowitz, B.~Way,
 {\it Lifshitz Singularities},
  Phys.\ Rev.\ D {\bf 85} (2012) 046008
 [arXiv:1111.1243 [hep-th]].

\bibitem{Harrison:2012vy}
  S.~Harrison, S.~Kachru, H.~Wang,
  {\it Resolving Lifshitz Horizons},
  [arXiv:1202.6635 [hep-th]].

\bibitem{Bao:2012yt}
  N.~Bao, X.~Dong, S.~Harrison, E.~Silverstein,
  {\it The Benefits of Stress: Resolution of the Lifshitz Singularity},
  [arXiv:1207.0171 [hep-th]].

\bibitem{Weyl1}
  A.~A.~Burkov, L.~Balents,
  {\it Weyl Semimetal in a Topological Insulator Multilayer},
  Phys.\ Rev.\ Lett.\ {\bf 107} (2011) 127205
  [arXiv:1105.5138 [cond-mat.mes-hall]].

\bibitem{Weyl2}
  X.~Wan, A.~M.~Turner, A.~Vishwanath, S.~Y.~Savrasov,
  {\it Topological semimetal and Fermi-arc surface states in the electronic structure of pyrochlore iridates},
  Phys.\ Rev.\ B {\bf 83} (2011) 205101
  [arXiv:1007.0016 [cond-mat.str-el]].



\bibitem{Korovin:2011kw}
  Y.~Korovin,
  {\it Holographic Renormalization for Fermions in Real Time},
  [arXiv:1107.0558 [hep-th]].

\bibitem{Alishahiha:2012nm}
  M.~Alishahiha, M.~R.~Mohammadi Mozaffar, A.~Mollabashi,
  {\it Fermions on Lifshitz Background},
  Phys.\ Rev.\  D {\bf 86} (2012) 026002
  [arXiv:1201.1764 [hep-th]].



\bibitem{Fang:2012pw}
  L.~Q.~Fang, X.~-H.~Ge, X.~-M.~Kuang,
  {\it Holographic fermions in charged Lifshitz theory},
  [arXiv:1201.3832 [hep-th]].

\bibitem{Liu1}
  N.~Iqbal, H.~Liu,
  {\it Universality of the hydrodynamic limit in AdS/CFT and the membrane
  paradigm},
  Phys.\ Rev.\  D {\bf 79} (2009) 025023
  [arXiv:0809.3808 [hep-th]].



\bibitem{Iizuka}
  N.~Iizuka, N.~Kundu, P.~Narayan, S.~P.~Trivedi,
  {\it Holographic Fermi and Non-Fermi Liquids with Transitions in Dilaton Gravity},
  JHEP {\bf 1201} (2012) 094
  [arXiv:1105.1162 [hep-th]].


\bibitem{HHV}
  S.~A.~Hartnoll, D.~M.~Hofman, D.~Vegh,
  {\it Stellar spectroscopy: Fermions and holographic Lifshitz criticality},
  JHEP {\bf 1108} (2011) 096
  [arXiv:1105.3197 [hep-th]].

\bibitem{Meng}
  T.~Meng, L.~Balents,
  {\it Weyl superconductors,}
  Phys.\ Rev.\  B {\bf83 } (2012)  054504
  [arXiv: 1205.5202v2 [cond-mat.mes-hall]].


\bibitem{Burkov}
  A.~A.~Zyuzin, A.~A.~Burkov,
  {\it Topological response in Weyl semimetals and the chiral anomaly},
  [arXiv: 1206.1868v2 [cond-mat.mes-hall]].

\bibitem{Spivak}
  D.~T.~Son, B.~Z.~Spivak,
  {\it Chiral Anomaly and Classical Negative Magnetoresistance of Weyl Metals},
  [arXiv:1206.1627v1 [cond-mat.mes-hall]].

\bibitem{Zhang}
  Z.~Wang, S.~-C.~Zhang,
  {\it Charge Density Waves and Axion Strings from Weyl Semimetals},
  [arXiv:1207.5234v1 [cond-mat.str-el]].

\bibitem{Hosur}
  P.~Hosur,
  {\it Friedel oscillations due to Fermi arcs in Weyl semimetals},
  [arXiv:1208.0027v2 [cond-mat.str-el]].

\bibitem{Voskresensky}
  D.~N.~Voskresensky,
  {\it Screening and anti-screening in QED and in Weyl semimetals},
  [arXiv:1208.5163v1 [cond-mat.mes-hall]].


\bibitem{Hartnoll:2007ih}
  S.~A.~Hartnoll, P.~K.~Kovtun, M.~Muller, S.~Sachdev,
 {\it Theory of the Nernst effect near quantum phase transitions in condensed
  matter, and in dyonic black holes},
  Phys.\ Rev.\  B {\bf 76} (2007) 144502
  [arXiv:0706.3215 [cond-mat.str-el]].


\bibitem{Tarrio:2011de}
  J.~Tarrio, S.~Vandoren,
  {\it Black holes and black branes in Lifshitz spacetimes},
   JHEP {\bf 1109} (2011) 017
  [arXiv:1105.6335 [hep-th]].

\bibitem{ssprivate}
  S.~Sachdev,
  private communication.


\bibitem{Liu2}
  N.~Iqbal, H.~Liu,
  {\it Real-time response in AdS/CFT with application to spinors},
  Fortsch.\ Phys.\  {\bf 57} (2009) 367
  [arXiv:0903.2596 [hep-th]].


\bibitem{Landau}
L.~D.~Landau, E.~M.~Lifshitz, Course of Theoretical Physics, Vol. 3, Quantum Mechanics,
Pergamon Press, 1977.


\end{thebibliography}
\end{document}